\documentclass[amsfonts]{article}
\usepackage{graphicx}
\pdfoutput=1
\usepackage{subcaption}

\newcommand{\be}{\begin{equation}}
\newcommand{\ee}{\end{equation}}
\newcommand{\bea}{\begin{array}}
\newcommand{\ea}{\end{array}}
\newcommand{\beqa}{\begin{eqnarray}}
\newcommand{\eeqa}{\end{eqnarray}}

\newcommand{\eean}{\end{eqnarray*}}

\def\up#1{\leavevmode \raise.16ex\hbox{#1}}

\newcommand{\gapproxeq}{\lower
 .7ex\hbox{$\;\stackrel{\textstyle >}{\sim}\;$}}
\newcommand{\lapproxeq}{\lower .7ex\hbox{$\;\stackrel
{\textstyle <}{\sim}\;$}}

\newcounter{appendice}

\def\thebibliography#1{{\bf REFERENCES\markboth
 {REFERENCES}{REFERENCES}}\list
 {[\arabic{enumi}]}{\settowidth\labelwidth{[#1]}\leftmargin\labelwidth
 \advance\leftmargin\labelsep
 \usecounter{enumi}}
 \def\newblock{\hskip .11em plus .33em minus -.07em}
 \sloppy
 \sfcode`\.=1000\relax}

\begin{document}
\centerline{ \LARGE  Possible Evidence of  Thermodynamic Equilibrium in Dark Matter  Haloes   }

\vskip 2cm

\centerline{ Joshua Davidson\footnote{jadavidson@crimson.ua.edu},   Sanjoy K.  Sarker\footnote{  ssarker@ua.edu} and Allen Stern\footnote{astern@ua.edu}   }

\vskip 1cm
\begin{center}
  { Department of Physics, University of Alabama,\\ Tuscaloosa,
Alabama 35487, USA\\}

\end{center}
\vskip 2cm

\vspace*{5mm} 

\normalsize
\centerline{\bf ABSTRACT}

After deducing the density profiles and   gravitational  potential functions  of  eight galaxies from the rotation velocity data from THINGS, we  find that the density decreases exponentially with the potential in substantial regions of the haloes.  Such behavior is 
  in agreement with that of a single-component isothermal Boltzmann gas, and suggests that  an  effective description in terms of a  Boltzmann gas  is possible for dark matter  in these regions. This could be an indication that dark matter self-interactions are sufficient in strength and number to lead to thermal equilibrium in these regions. We write down the dynamics and boundary conditions for a Boltzmann gas description and  examine some of its  qualitative and quantitative  consequences.
 Solutions to the dynamical system are determined by three  dimensionfull parameters, and  provide reasonable fits to the rotational velocity data in the regions where the Boltzmann-like behavior was found.    Unlike in  the usual approach to  curve fitting, we do not assume a specific form for the dark matter density profile and we  do not require a detailed knowledge of the baryonic content of the galaxy.

\bigskip
\bigskip

\newpage
\section{Introduction and Outline}

The paradigm of collisionless cold dark matter has had much success in describing  the large scale structure of the universe.  Nevertheless, inconsistencies persist with observations   at  smaller scales.  
One example is the  cusp singularity in the dark matter density profile, which is predicted by simulations of collisionless particles,\cite{Navarro:1996gj}  but is in contrast  with observations,  in particular, for dwarf galaxies.\cite{Oh:2010ea},\cite{Frusciante:2012jg}
The possibility that  such  inconsistencies may be cured with the inclusion of  non gravitational self-interactions was suggested a number of years ago.\cite{Spergel:1999mh}  The proposal of self-interacting dark matter (having interaction times  less than the Hubble time) is  not currently excluded, despite  existing constraints  on the interactions coming from halo shapes and cluster interactions.  Numerous models for dark matter self-interactions have  been considered, and their role in curing  observational anomalies has been explored.\cite{Mohapatra:2001sx},\cite{Foot:2004wz},\cite{Ackerman:2008gi},\cite{Buckley:2009in},\cite{Feng:2009hw},\cite{Chang:2009yt},\cite{Loeb:2010gj},\cite{Tulin:2012wi},\cite{Destri:2012yn}
Since self-interactions among dark matter particles are  not  excluded,  it is then reasonable to ask whether  conditions may be suitable in galaxies for a gas of dark matter particles to reach a state of thermodynamic equilibrium. From the theoretical side,  it is a non trivial problem to know precisely what conditions should be satisfied for this to occur since the nature of the interaction is unknown.  However,  it is possible to search for indications  of thermal equilibrium in regions of galactic haloes from direct observations. This can be done 
 using the rotation curve data, which is a central theme  of the article.

Our paper contains two parts.   In the first part (section 2), we  search for regions in galactic halos that show behavior  analogous  to what one would expect from a classical (Boltzmann) gas in thermal equilibrium.  In  addressing this issue,  we   only need to require  that  the region of interest in the  halo is approximately  spherically symmetric and that  the source of the rotation curves are circular orbits. Both of these assumptions are commonly utilized in curve fittings of the rotation velocity data.
With these two assumptions  one can simultaneously determine the halo density $\rho$ and gravitational potential $\phi$ from  orbital speeds of HI gas  in the halo.  The assumption of spherical symmetry means that  these quantities are  functions of a single parameter.  Rather than taking the parameter to be the radial coordinate $r$, we can write $\rho$ as a function of $\phi$.  (This is a single-valued function because $\phi$  monotonically increases with $r$.)   It  is then of interest to search for some  universal behavior for $\rho(\phi)$.  In particular, an exponentially decreasing function would indicate agreement with a self-interacting classical gas in thermodynamic equilibrium. 
To be more specific, the determination of $\phi(r)$ and   $\rho(r)$ from the circular  orbital speed $v(r)$ follows from simple classical physics considerations.  The former is obtained  by integrating Newton's law,
\be\frac{d\phi}{dr} =\frac{v^2(r)}r\label{dphidr}\;,\ee  while the latter comes  from the  Poisson equation,
$ \nabla^2\phi=  4\pi G\rho(r),$  $G$ being the gravitational constant.  Upon    assuming spherical symmetry and then substituting  (\ref{dphidr}) one gets 
\be \rho(r)= \frac 1{ 4\pi G r^2}\frac d{dr} \Bigl(r{v^2(r)}
\Bigr)\;.\qquad\label{1.3}\ee  From $\phi(r) $ and $\rho(r)$, we  then obtain  $\rho(\phi)$. 

Before discussing  analysis of  the rotation curve data,  it is useful to make a few remarks about   the ideal case of  {\it exactly} flat rotation curves, i.e., $v(r)=\,$constant.   $\phi$ varies logarithmically in $r$ in this case, while from (\ref{1.3}) one gets that
$ \rho(r)\propto 1/{r^2}$.  It is a well known curiosity that the
  combination of these two results gives an exponential behavior for $\rho(\phi)$, i.e., 
$ \rho(\phi) \propto e^{-\phi/\phi_0}$, where  $\phi_0$ is a constant associated with the so-called velocity dispersion.  
 Therefore, the source of the gravitational potential  responsible for exactly flat rotation curves 
behaves in an identical manner to a single component  isothermal Boltzmann gas.

Of course, a galactic system is not a single component system, and the density and gravitational potential get contributions from both dark matter and baryonic matter.  Moreover, the above  scenario is an oversimplification, since in nature, rotation curves are not exactly flat.  On the other hand,  the absence of exactly flat rotation curves from galaxies, does not  rule out an exponential behavior for $\rho(\phi)$.   For that reason it is of interest to make a more accurate  determination of $ \rho(\phi) $  from the rotation curves in galactic haloes.  We do so  using the data from THINGS\cite{deBlok:2008wp}.  
A local maximum can be identified for most of the rotation curves.  So for them, there is always a small  neighborhood where $v $ is approximately constant, and it would come as no  surprise to find that $\rho$ decreases exponentially with $\phi$ in that neighborhood.  A more significant result would be to find   exponential behavior for $\rho(\phi)$ in a region that extends  far beyond a local maximum.   We present evidence that this is the case for  eight  galaxies (out of 19) in the THINGS survey. They are  NGC 2841,
NGC 5055, NGC 7331, NGC 2403, NGC 2903, NGC 3521, NGC 3198 and DD0 154.  
The  results suggest that an  effective Boltzmann gas description of dark matter is possible  for  certain regions of galactic haloes, more precisely,  spherical shells  $R_{\tiny {\tt G}}\le r\le R_{\tiny {\tt max}}$ .  The constant $1/\phi_0$ can be approximately determined for these regions, which can then be identified with the ratio of particle mass $m$ with some effective temperature $T$ (times Boltzmann's constant $k_B$). In section 2  we  estimate $m/T$ for the eight galaxies in the relevant regions.

In the second part of the paper,  beginning with section 3,  we make a  model utilizing the effective Boltzmann description.  The dynamics of the model is taken  to be that of a single-component self-gravitating isothermal  gas, which is assumed to be valid for the  spherical shell, $R_{\tiny {\tt G}}\le r\le R_{\tiny {\tt max}}$.  Along with spherical symmetry, the model makes the simplifying assumption that  only dark matter is present at distances larger than  $ R_{\tiny {\tt G}}$, and that all  baryonic matter is contained within a sphere of radius $R_{\tiny {\tt G}}$.  (Dark matter is also assumed to be present in the region $r<R_{\tiny {\tt G}}.)$ The baryonic component nevertheless strongly influences  the region  of interest $R_{\tiny {\tt G}}\le r\le R_{\tiny {\tt max}}$, as it contributes to the  gravitational acceleration $\frac{d\phi}{dr}$, and thus is responsible for a tidal force on the isothermal gas.  
The pressure balance  implicitly holds everywhere within and beyond the isothermal region, with the pressure within the isothermal region given by the ideal gas equation.

 The relevant dynamical equation for such a system has been known for a long time.\cite{Emden}   It is the Emden equation, which is just the Poisson equation with an isothermal  Boltzmann gas source.   
   It has a well known solution, found by Emden, 
	which satisfies the condition that it be nonsingular for {\it all} values of the radial coordinate, $0\le r<\infty$.  The solution results from the boundary condition that states that the gravitational force vanishes  at the origin. Emden's solution gives a unique density profile, and  yields a flat rotation curve in the asymptotic  limit $r\rightarrow \infty$.  

Since for us the relevant domain   is a spherical shell, the condition  that  the solution be nonsingular for all $r$ is too restrictive, and we must seek alternative  boundary conditions, and hence alternative solutions.
The  appropriate  conditions  for our case should be  imposed at the inner boundary, 
  $r=R_{\tiny {\tt G}}$.   The derivative of $\phi$ at this point gives the gravitational attraction to the  matter in the interior region $r<R_{\tiny {\tt G}}$.  The baryonic component dominates  the interior region, and so it is largely responsible for the boundary conditions at $r=R_{\tiny {\tt G}}$, and it contributes to the gravitational potential  of the Boltzmann gas. In section 3, 
we show that a family of spherically symmetric solutions to the Emden equations results from such boundary conditions.  They then lead to a family of density profiles that are parameterized by two dimensionless constants, which we  denote by $\kappa$ and $\tau$.  The parameters can be expressed as functions of the total mass $M_{\tiny {\tt G}}$ in the interior region $r<R_{\tiny {\tt G}}$, as well as $R_{\tiny {\tt G}}$ and the effective temperature of the gas.  The family of solutions contains Emden's solution as a special case.  
  Another special solution is  one  which gives an exactly flat rotation curve for all of  $R_{\tiny {\tt G}}\le r\le R_{\tiny {\tt max}}$.\footnote{ This solution has often been referred to as the 'singular' isothermal sphere because it is singular in the limit  $r=0$.  For us, however, the solution is {\it nonsingular} because the origin is excluded from the domain.} 
 It  results from particular values of the two dimensionless parameters  $\kappa$ and $\tau$. Since the solution  yields an exactly  flat rotation curve  it should serve as a crude fit to observations.  From the crude fit one can determine the ratio $m/T$ in terms of the orbital speed in the halo.    For instance, from typical orbital speeds of $\sim 200$ km/s, one gets $m/T\sim 400$ eV per  Kelvin.  If one   takes the analogy with the Boltzmann gas further and regards $T$ as the temperature of a gas in equilibrium,   the equipartition theorem for this system implies that the mean speed of the dark matter particles in the gas is of the same order of magnitude  as the rotational speed for HI in the halo.  So if the orbital speed is  a few hundred  km/s, the speed of the particles in the gas is also a few hundred  km/s, and this result is independent of the dark matter mass.  The details  of this argument are given in Section 3.

We go beyond the crude approximation in section 4.  There we   perform fits of the rotation curve data to the solutions of the self-gravitating isothermal gas for the eight galaxies examined in section 2.  The fits are applied only to the regions  $R_{\tiny {\tt G}}\le r\le R_{\tiny {\tt max}}$ of the haloes where $\rho(\phi)$ exhibited  exponential behavior.   Our approach
is quite different from  the usual approach to  curve fitting, such as Navarro-Frenk-White (NFW) fits.\cite{Navarro:1996gj} 
 We do not assume any specific form for the dark matter density profile. Rather, the density $\rho(r)$  is inferred from the fitted solution. We also do not require any detailed knowledge of the baryonic content of the galaxy.  
These details are often difficult to accurately determine, yet they play a crucial role in the usual approach to curve fitting.  Moreover, results differ significantly from one mass model to another.  In the usual approach to curve fitting, the matter contributions from the  disk, bulge and HI gas must be subtracted from the rotation curve data. 
 Here, on the other hand, it would be a mistake to subtract off the baryonic contribution to rotation velocity before doing the curve fitting.  This is because the contribution of baryonic matter from the inner region  $r\le R_{\tiny {\tt G}}$ is already included in the  gravitational potential  $\phi(r)$. Moreover,  the Boltzmann equation determines the actual rotational velocity, not individual contributions from  the different components.  Finally, while fits to rotational  curve data are usually  performed for the entire galaxy,  here it makes sense to perform the fits to the Boltzmann gas only for the  spherical shell,  $R_{\tiny {\tt G}}\le r\le R_{\tiny {\tt max}}$, where evidence for such a behavior has been found from the data.

 While only two dimensionless parameters are needed to specify solutions to the Boltzmann equation, three   {\it dimensionfull } parameters are required to fit the rotational velocity data.  They are   $m/T$, the density $\rho(R_{\tiny {\tt G}})$  at   $r=R_{\tiny {\tt G}}$ and the total mass $M(R_{\tiny {\tt G}})$ contained inside the sphere of radius  $R_{\tiny {\tt G}}$.   Not too surprisingly, the result for $m/T$ obtained from the fit  are reasonably consistent with the  corresponding values  obtained in section 2. We use the fit to estimate the mass $ M_{(R_{\tiny {\tt G}}, R_{\tiny {\tt max}})}$  of the region $R_{\tiny {\tt G}}\le r\le R_{\tiny {\tt max}}$ exhibiting the Boltzmann behavior.  In most cases, we find that this mass is larger than that of the inner region  $r\le R_{\tiny {\tt G}}$.  We  also find that  in most cases, $ M_{(R_{\tiny {\tt G}}, R_{\tiny {\tt max}})}$ is larger than previous estimates of the total baryonic mass of the galaxy, obtained from mass models.\cite{deBlok:2008wp}  This lends support to the conclusion that the region $R_{\tiny {\tt G}}\le r\le R_{\tiny {\tt max}}$ exhibiting Boltzmann-like behavior contains a significant fraction of the total mass of the galaxy, and that it consists primarily of dark matter.

A summary of the results for the eight galaxies, including  two tables, is given   in section 5, along with concluding remarks.

\section{Determination  of $\rho(\phi)$ from rotation curve data}

As stated in the introduction, the  density function  $\rho(r)$  and  gravitational potential  $\phi(r)$ in the halo can be numerically determined from the rotation velocity data under certain simplifying  assumptions.  The assumptions are that the system is spherically symmetric and that the rotation velocity data are attributed to circular orbits.  $\phi(r)$ and  $\rho(r)$ are obtained from    $v(r)$ using (\ref{dphidr}) and (\ref{1.3}), respectively.  The two functions can then be combined to determine $\rho(\phi)$.  In order to search for Boltzmann-like behavior, it is useful to plot  ${\log}\,\rho$ versus $\phi$.  We  do this for the eight mentioned  galaxies in the THINGS survey.  Boltzmann-like behavior would correspond to a straight line with negative slope in the plot of ${\log}\,\rho$ versus $\phi$, the slope being $\phi_0$ or $T/m$.  We also plot   $-\frac d{d\phi}{\log \rho}\,$ versus $r$ for the eight galaxies.   Boltzmann-like behavior would yield a constant in the latter graph, corresponding to the value of $m/T$ for the gas.

The first step in the analysis is to fit the rotation velocity data to a smooth curve $v(r)$. We should mention that the results for $\rho(\phi)$ are somewhat sensitive to how closely one fits the  data to a smooth curve.   In order to reduce the subjectivity of the analysis, we fit all eight galaxies in the sample to  the same eight-parameter series, i.e., $\sum_{n=-3}^4a_n r^{n}$. Our analysis reveals an exponential behavior  for regions of the haloes.  This  behavior becomes less evident if we make a cruder  fit to the data, using only a few  parameters.
  On the other side, additional features of the halo obscure the exponential behavior if one includes too many parameters.

 The eight galaxies are organized into four categories: unbarred galaxies (NGC 2841, NGC 5055 and NGC 7331), weakly barred galaxies (NGC 2403,  NGC 2903 and NGC 3521), a barred spiral galaxy (NGC 3198), and finally, a  dwarf galaxy (DD0 154). 

We begin with the unbarred galaxies. After first listing some  relevant information for these galaxies, we then determine $\log\rho(\phi)$  from the data.

\subsection{NGC 2841}

The morphology classification of NGC 2841 is   SA(r)b.  It is a flocculent, unbarred  giant spiral galaxy,  with a classical bulge and a significant  population of blue stars.  A certain mass model gives the disk and bulge masses for  NGC 2841 of   $1.096\times 10^{11}\; M_{\odot}$ and  $2.51\times 10^{10}\; M_{\odot}$, respectively.\footnote{More specifically, these values were obtained using models with fixed values of the mass to luminosity ratio and the diet Salpeter stellar initial mass function. (They were obtained from Table 3 of \cite{deBlok:2008wp}.) As we indicated previously,  the baryonic masses for any given galaxy differ significantly from one model to another.  These details  play no direct role in the analyses we do here. }
Using  \cite{Mannheim:2010xw} the  distance scale of  the disk is $\sim3.5$ kpc.
  The HI gas mass for  NGC 2841 is $8.6\times 10^{9} \;M_{\odot}$.\cite{Walter:2008wy} This is approximately $6\%$ of the disk plus bulge mass, and so one would expect it to make only a small contribution to the total density in the halo.  More background information about  NGC 2841 can be found in \cite{deBlok:2008wp}. 

The 
THINGS rotation velocity data  for NGC 2841   is available at distances up to $51.6$ kpc, or  almost $15$ times the distance scale of the disk. 
Figure 1  shows an eight-parameter  fit of the rotation velocity data to the series  $\sum_{n=-3}^4a_n r^{n}$  from $r= 3.8$ to $ 51.6$ kpc.  The fit is  done to the midpoints [corresponding to the
dots] of the error bars. 
   From the fitted function for $v(r)$,  we obtain the gravitational potential $\phi(r)$ by numerically integrating (\ref{dphidr}).  The result is  shown in figure 2.  The density profile $\rho(r)$ is found using (\ref{1.3}), and is shown in figure 3.\footnote{Figure 3 exhibits a region beyond $40$ kpc where the density increases with increasing $r$.  Similar results are seen for very distant regions of NGC 7331, NGC 2903, NGC 3521 and NGC  3198. This behavior is presumably unphysical, and so if one can trust the fits in these regions, it may indicate that the two assumptions (spherical symmetry and circular orbits) used in obtaining $\rho(r)$ are invalid for these particular regions. In any case, these regions are far beyond the domains of interest of our paper. }
  The latter two plots are combined   in figure 4 to give $\log{({\rm \rho(r)/  \rho_0})}$ versus $\phi(r)$, where $\rho_0$ is a constant with units of density.  In the region $7<r<22$ kpc, it approximately coincides with a straight line with negative slope.   From this figure it  thus appears that $\rho(\phi)$ has an exponential behavior in the region  $7<r<22$ kpc,  or approximately from $2$ to $6$ times the distance scale of the disk.  
Finally, in figure 5 we plot  minus the derivative of  $\log{{\rm \rho}}$ with $\phi$ versus $r$.    An approximately constant  value of  $\sim 190$ eV/Kel is obtained  for $\;\;-\frac{d\log{{\rm \rho}}}{d\phi}$ over the range  $7<r<22$ kpc, which is shown in the inset. Therefore this region is consistent with the description of a Boltzmann gas with $m/T\sim 190$ eV/Kel.  $\;\;-\frac{d\log{{\rm \rho}}}{d\phi}$ gradually  increases to $\sim 250$ eV/Kel over the range $20<r<30$ kpc, before rapidly decreasing beyond $30$ kpc.

\begin{figure}[placement h]
\begin{center}
\includegraphics[height=2in,width=3in,angle=0]{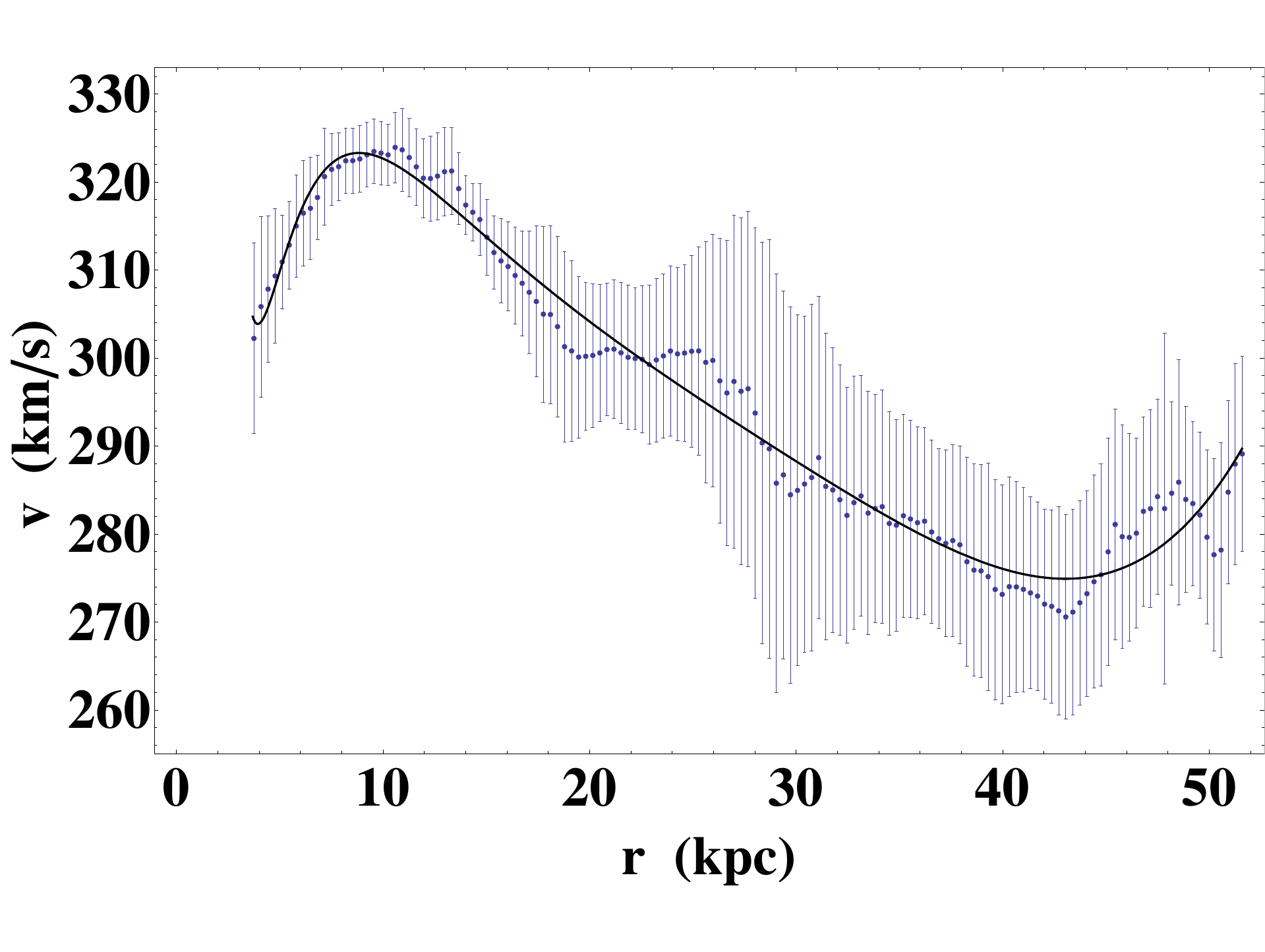}
\caption {Eight-parameter series fit of rotation velocity data for  NGC 2841 from $r=3.8$ to $51.6$ kpc.}
\end{center}

\begin{center}
\includegraphics[height=1.65in,width=2.5in,angle=0]{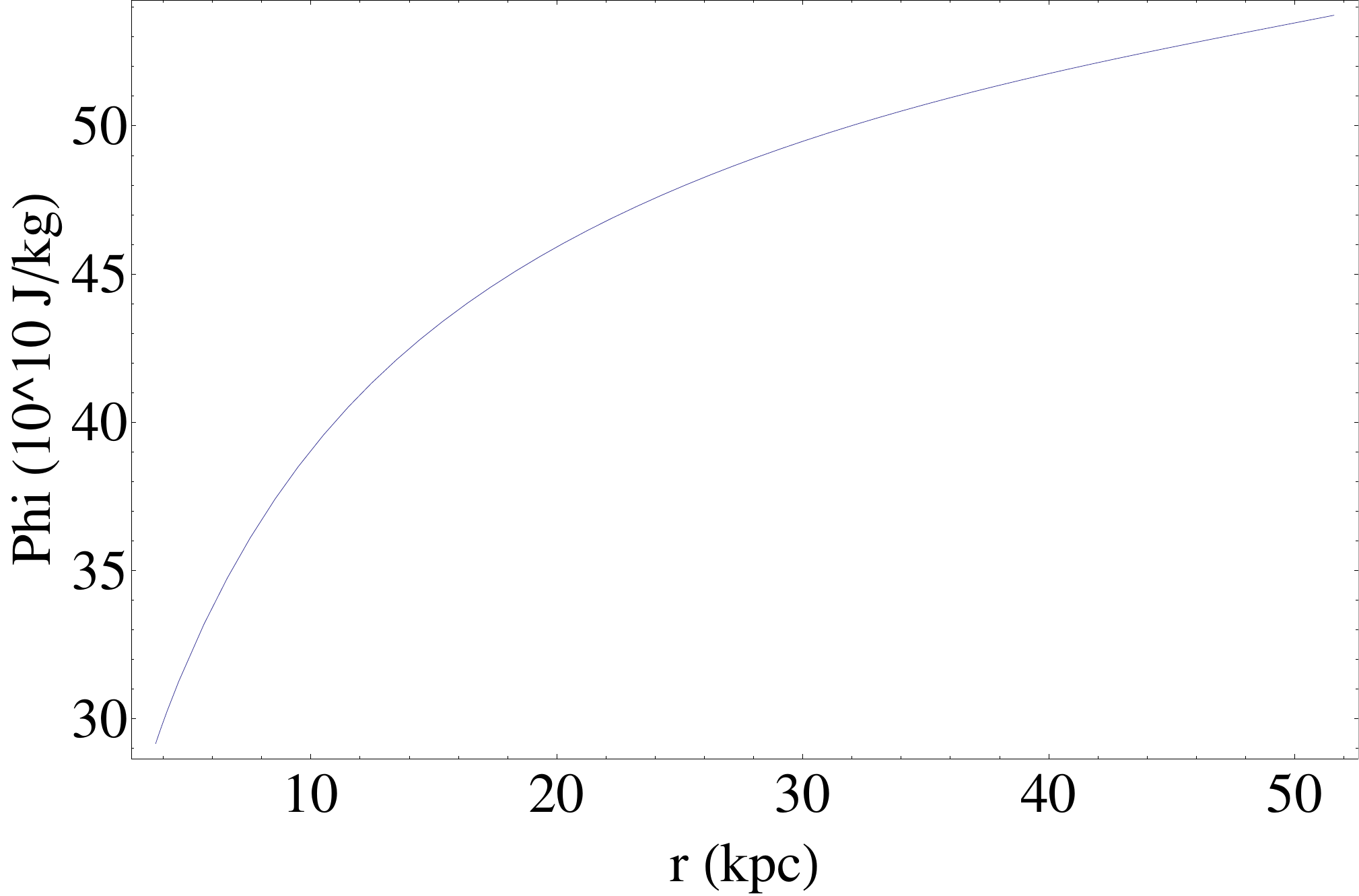}
\caption {Resulting gravitational potential  versus $r$ for  NGC 2841 from $r=3.8$ to $51.6$ kpc.  The zero of the potential is at  $r=3.8$.}
\end{center}

\end{figure}

\begin{figure}[placement h]
\begin{center}
\includegraphics[height=1.65in,width=2.8in,angle=0]{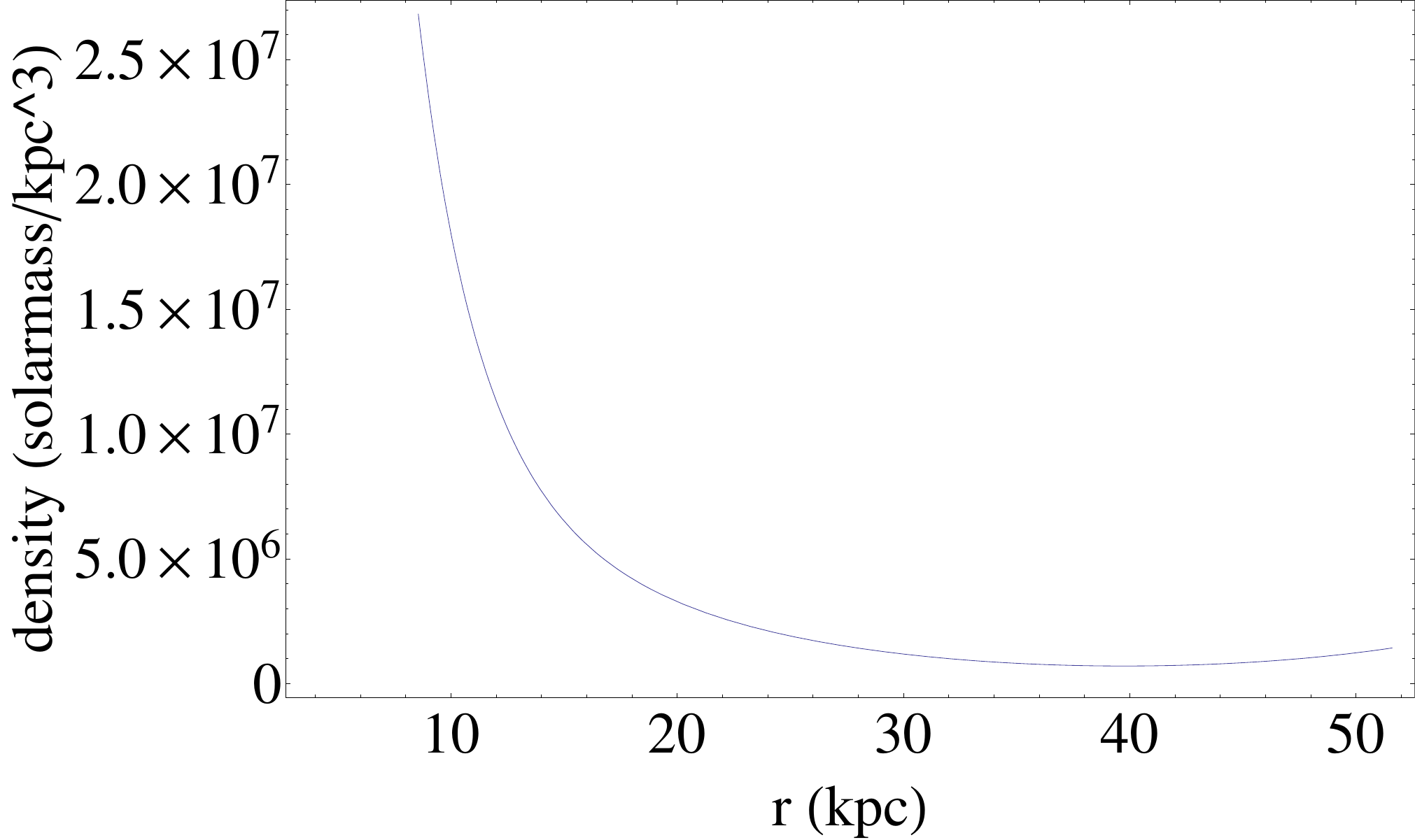}
\caption {Resulting density versus $r$ for  NGC 2841 from $r=3.8$ to $51.6$ kpc. }
\end{center}

\begin{center}
\includegraphics[height=2in,width=3.3in,angle=0]{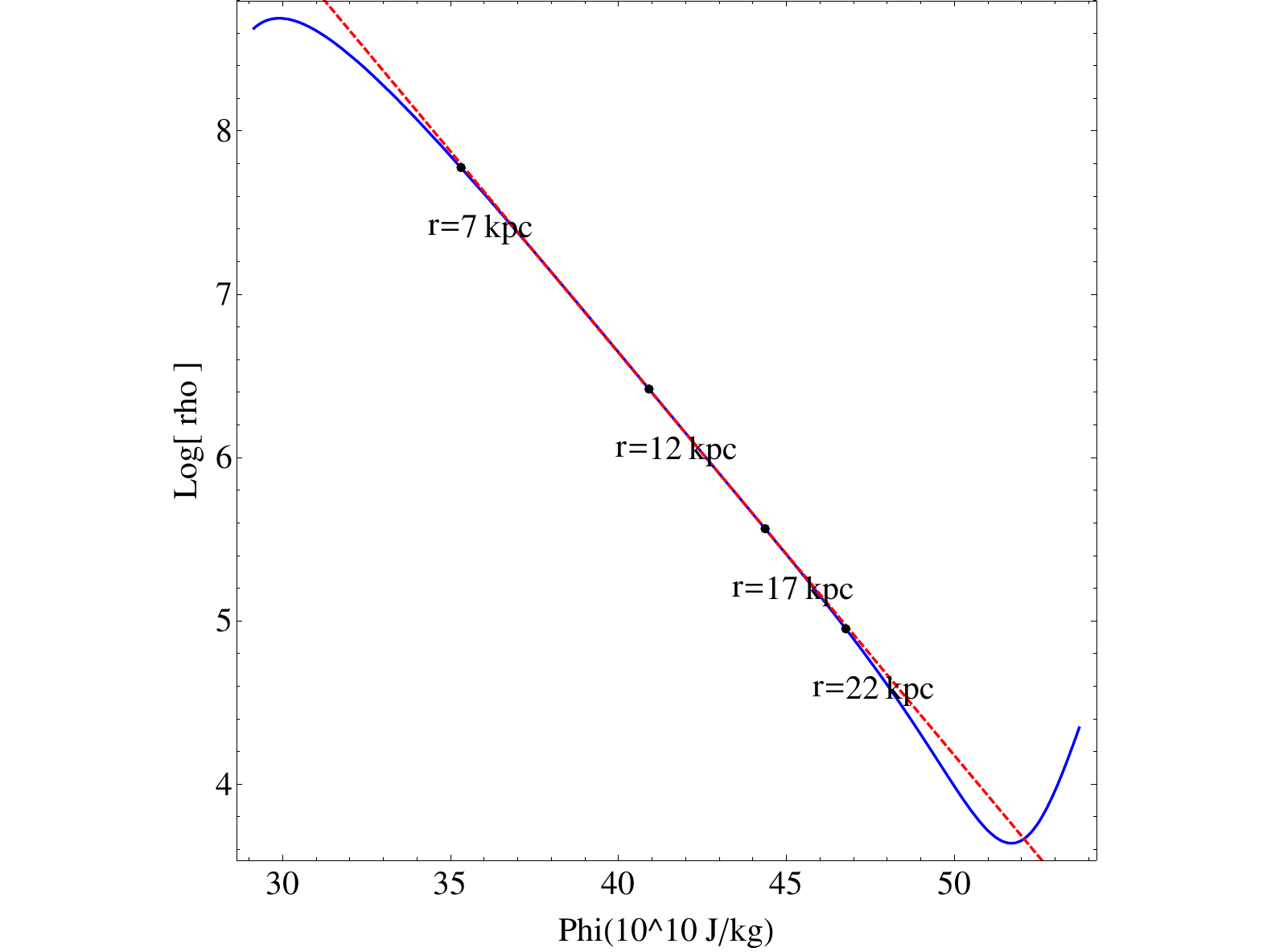}
\caption {$\log{({\rm density/  constant})}$ versus the gravitational potential for  NGC 2841 for $r=3.8$ to $51.6$ kpc.  It is compared to a straight line (red, dashed) with  slope  $\approx -190$ eV/Kel.  The two plots approximately coincide for     $7<r<22$ kpc. }
\end{center}

\begin{center}
\includegraphics[height=2.5in,width=2.5in,angle=0]{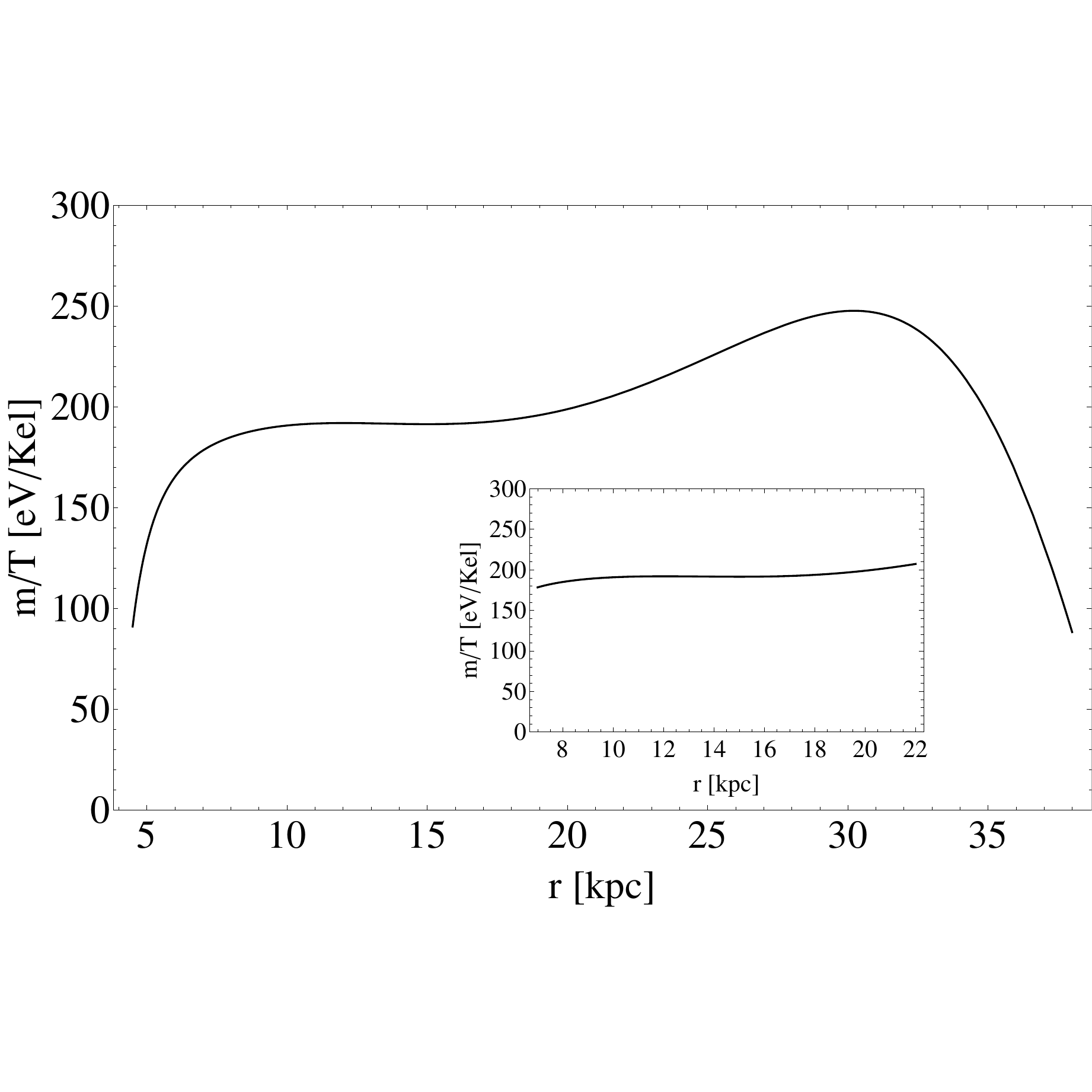}
\caption {$\;\;-\frac{d\log{{\rm \rho}}}{d\phi}$ versus $r$ for  NGC 2841. The plot is approximately constant  for the domain  $7<r<22$ kpc, which is shown in the inset.   }
\end{center}
\end{figure}
\noindent 
\newpage

\subsection{ NGC 5055}

The morphology classification of NGC 5055 is SA(rs)bc. It is an unbarred spiral galaxy, with an extended warp at end of the optical disk.  It has a regular structure with  well defined flocculent spiral arms, and is part of the M51 group of galaxies. 
  Mass models ([see  \cite{deBlok:2008wp})  give the disk and bulge masses for  NGC 5055 of   $1.203\times 10^{11}\; M_{\odot}$ and  $2.09\times 10^{9}\; M_{\odot}$.  The  distance scale of the disk is $\sim 3.622$ kpc. \cite{Mannheim:2010xw}
  The HI gas mass for  NGC 5055 is $.91\times 10^{10} \;M_{\odot}$,\cite{Walter:2008wy} which is approximately $7\%$ of the disk plus bulge mass.

THINGS rotation velocity data are available at distances up to $\sim44.4$ kpc,   around $12$ times the distance scale of the disk.   An eight-parameter series fit, again using   $\sum_{n=-3}^4a_n r^{n}$,  of the  data from $r= .2$ to $ 44.4$ kpc appears in figure 6.  Here we don't include a plot of $\phi(r)$ and $\rho(r)$, but give the  plot of $\log{({\rm \rho(r)/  constant})}$ versus $\phi(r)$  in figure  7.  An approximate straight line with negative slope is recovered upon restricting the radial coordinate to $10<r<25$ kpc.  In figure 8 we plot  minus the derivative of  $\log{{\rm \rho}}$ with $\phi$ versus $r$.   It is approximately constant in the domain $10<r<25$ kpc  (shown in the inset), with an average  value  of $\sim 440$ eV/Kel. $\;\;-\frac{d\log{{\rm \rho}}}{d\phi}$ reaches a minimum value of  $\sim 375$ eV/Kel at  $r\approx 4$ kpc, before rapidly increasing at smaller distances.

 \begin{figure}[placement h]
\begin{center}
\includegraphics[height=2in,width=3in,angle=0]{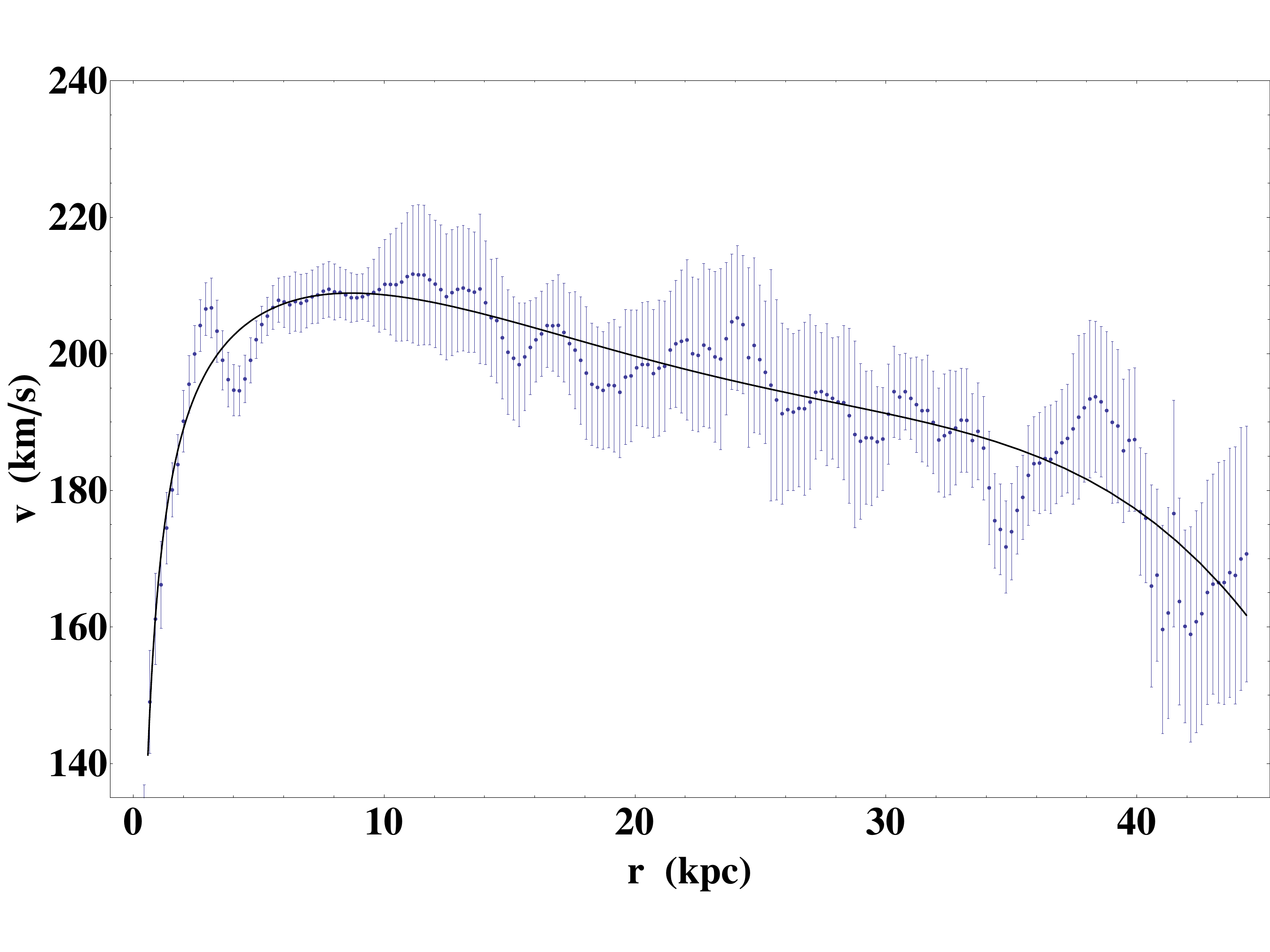}
\caption {Series fit of rotation velocity data for  NGC 5055 from $r=.2$ to $44.4$ kpc.}
\end{center}

 \begin{center}
\includegraphics[height=1.5in,width=3.3in,angle=0]{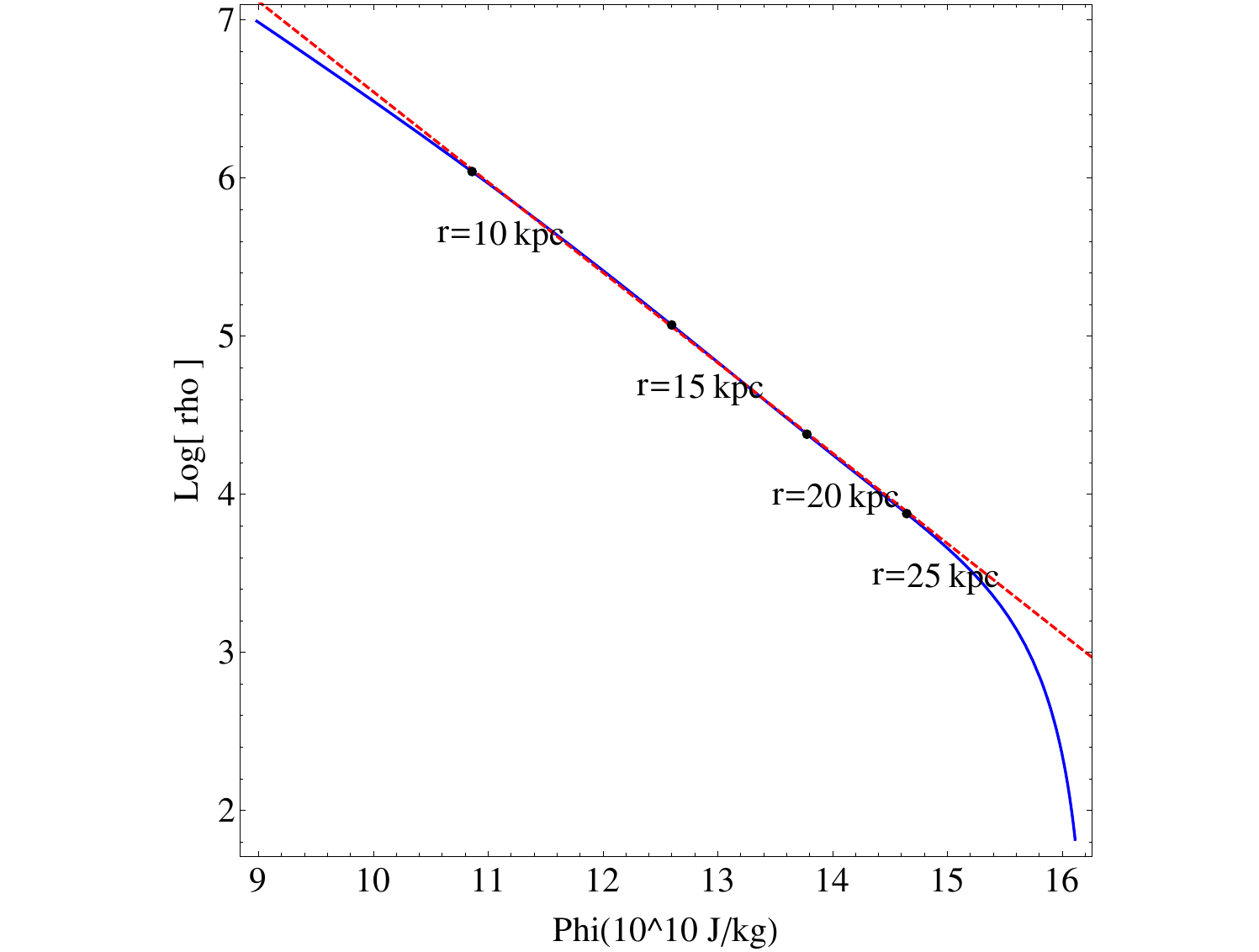}
\caption {$\log{({\rm density/  constant})}$ versus the gravitational potential for  NGC 5055 for $r=6.5$ to $37.5$ kpc.  It is compared to a straight line (red, dashed) with  slope  $\approx - 440$ eV/Kel.  The two plots approximately coincide for     $10<r<25$ kpc. }
\end{center} 
\end{figure}

\begin{figure}[placement h]
 \begin{center}
\includegraphics[height=2.25in,width=2.5in,angle=0]{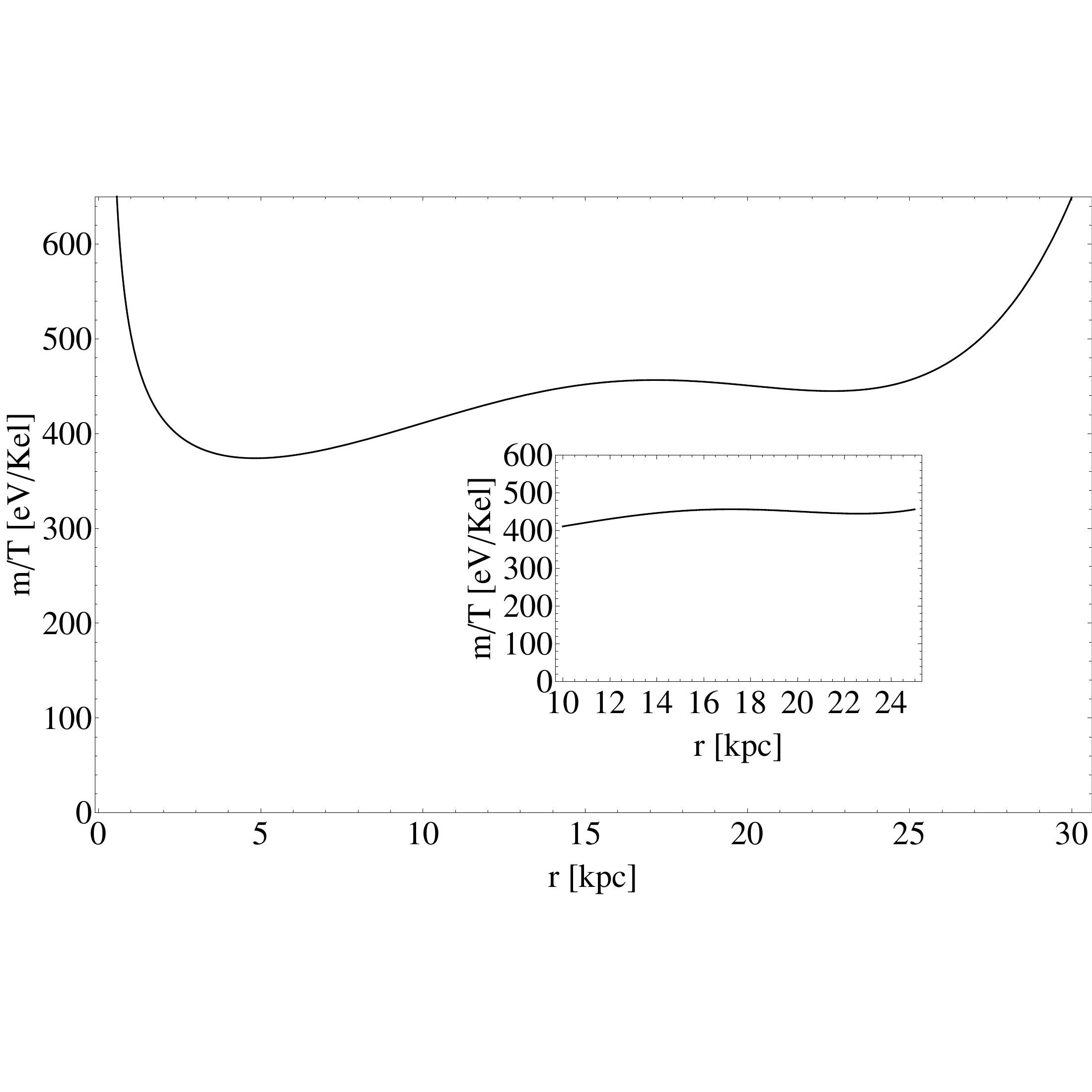}
\caption {$\;\;-\frac{d\log{{\rm \rho}}}{d\phi}$ versus $r$ for  NGC 5055. The plot is approximately constant  for the domain  $10<r<25$ kpc, which is shown in the inset.   }
\end{center}
\end{figure}

\newpage
\bigskip
\noindent 
\subsection{ NGC 7331 }

The morphology classification of NGC 7331 is SA(s)b.  It is regular at large scales, but spiral arms cause wiggles in velocity contours, and there are  large differences in the  rotation curve for the  approaching and receding sides.\cite{deBlok:2008wp}
  Mass models give the disk and bulge masses for  NGC 7331 of   $1.66\times 10^{11}\; M_{\odot}$ and  $1.74\times 10^{10}\; M_{\odot}$.  The  distance scale of the disk is $\sim 3.2$ kpc. \cite{Mannheim:2010xw}
  The HI gas mass for  NGC 7331 is $.91\times 10^{10} \;M_{\odot}$,\cite{Walter:2008wy} which is approximately $5\%$ of the disk plus bulge mass.

THINGS rotation velocity data are available at distances up to $\sim 24.4$ kpc, or over seven times the   distance scale of  the disk. The eight-parameter series fit of rotation velocity data from $r=3.3$ to $ 24.4$ kpc appears in figure 9.  The resulting  plot of $\log{({\rm \rho(r)/  constant})}$ versus $\phi(r)$ is given in figure 10.  An approximate straight line with negative slope is recovered upon restricting the radial coordinate to $8<r<16$ kpc.  In figure 11 we plot  minus the derivative of  $\log{{\rm \rho}}$ with $\phi$ versus $r$.   It is approximately constant in the domain $8<r<16$ kpc  (shown in the inset), with an average  value  of $\sim 260$ eV/Kel. $\;\;-\frac{d\log{{\rm \rho}}}{d\phi}$ reaches a maximum value of  $\sim 340$ eV/Kel at  $r\approx 5.5$ kpc.

 \begin{figure}[placement h]
\begin{center}
\includegraphics[height=2in,width=3in,angle=0]{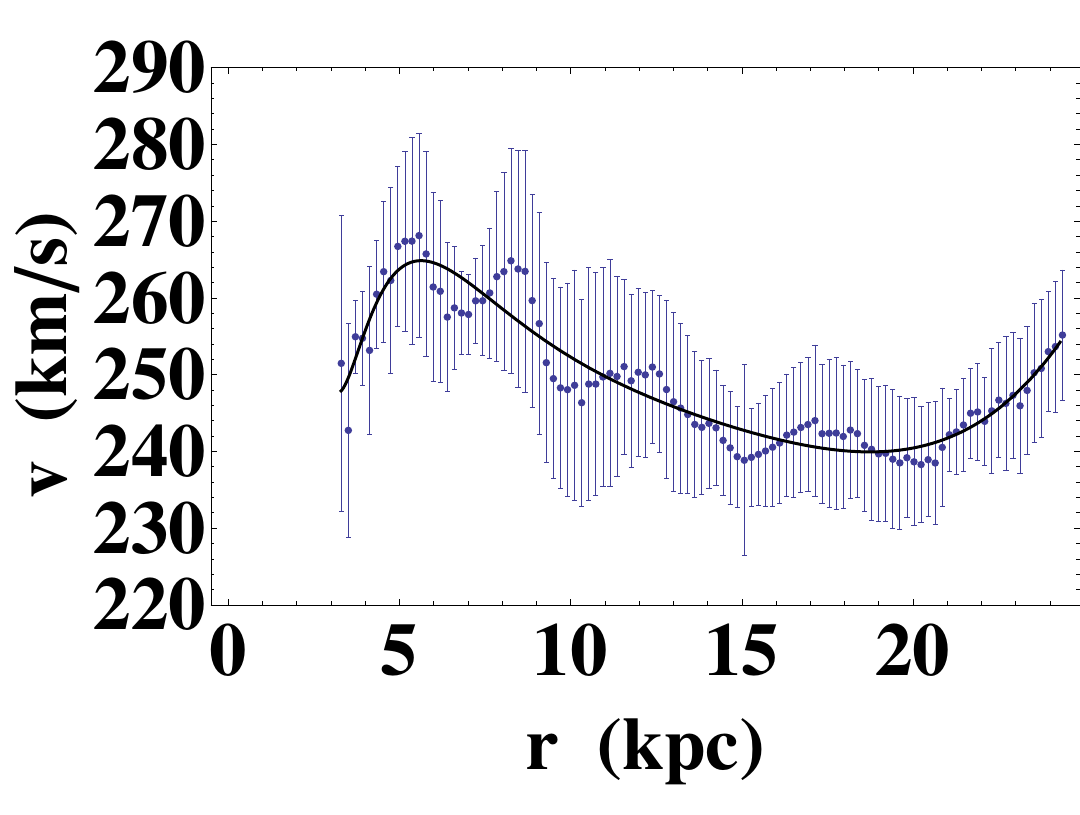}
\caption {Series fit of rotation velocity data for  NGC 7331 from  $r=3.3$ to $ 24.4$ kpc.}
\end{center} \end{figure}
\begin{figure}[placement h]
\begin{center}
\includegraphics[height=1.5in,width=3.3in,angle=0]{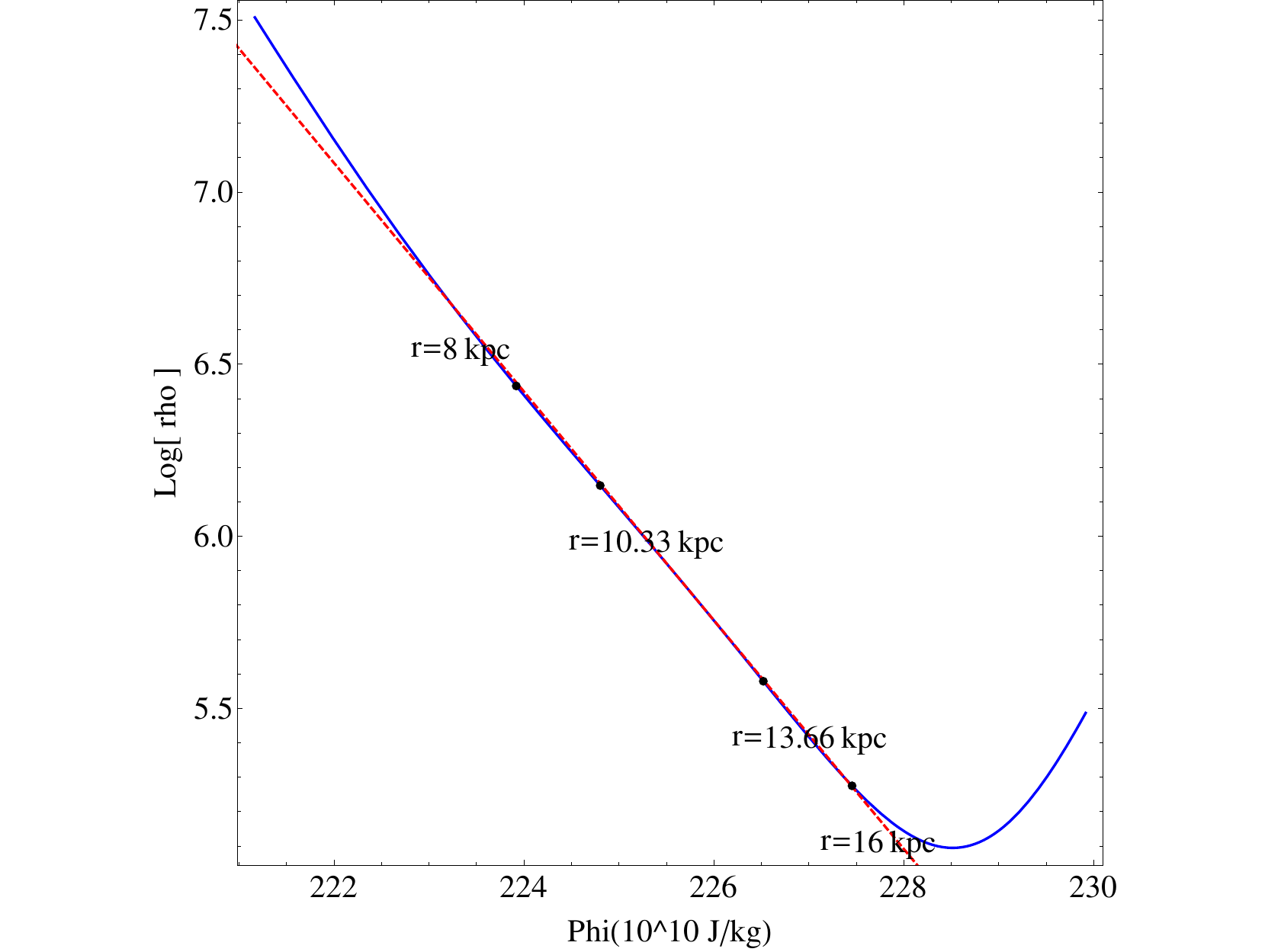}
\caption {$\log{({\rm density/  constant})}$ versus the gravitational potential for  NGC 7331 for $r=6$ to $24.3$ kpc.  It is compared to a straight line (red, dashed) with  slope $\approx -260$ eV/Kel.  The two plots approximately coincide for     $8<r<16$ kpc. }
\end{center}

\begin{center}
\includegraphics[height=2.5in,width=2.5in,angle=0]{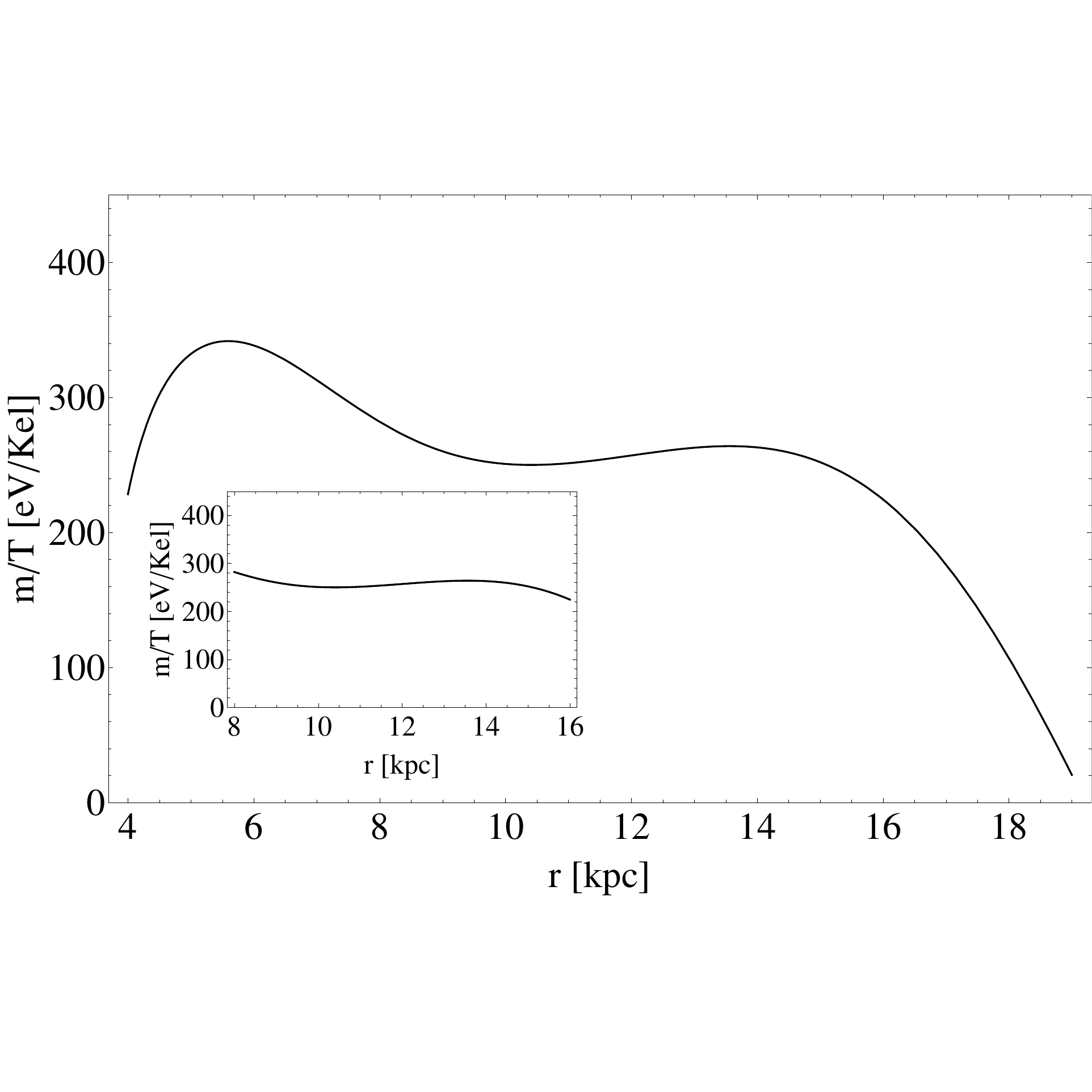}
\caption {$\;\;-\frac{d\log{{\rm \rho}}}{d\phi}$ versus $r$ for  NGC 7331. The plot is approximately constant over the domain  $8<r<16$ kpc, which is shown in the inset.   }
\end{center}

\end{figure}

\newpage

We  next repeat the previous  analysis for weakly barred spiral galaxies:  NGC 2403,  NGC 2903 and NGC 3521.

\subsection{ NGC 2403}

The morphology classification of NGC
2403 is  SAB(s)cd. 
  Mass models  (see  \cite{deBlok:2008wp}) give the disk and bulge masses for  NGC 2403 of   $4.68\times 10^{9}\; M_{\odot}$ and  $4.27\times 10^{8}\; M_{\odot}$, respectively.  The  distance scale of the disk is   $\sim 2.7$ kpc.  \cite{Mannheim:2010xw}
  The HI gas mass for  NGC 2403 is $2.58\times 10^{9} \;M_{\odot}$,\cite{Walter:2008wy} which is approximately $50\%$ of the disk plus bulge mass.

THINGS rotation velocity data are available at distances up to $\sim 24$ kpc, or almost nine times the   distance scale of  the disk.   The eight-parameter series fit of rotation velocity data from $r= .1$ to $ 24$ kpc appears in figure 12.  The resulting  plot of $\log{({\rm \rho(r)/  constant})}$ versus $\phi(r)$ is given in figure 13.  An approximate straight line with negative slope is recovered upon restricting the radial coordinate to $3<r<10$ kpc.  In figure 14 we plot  minus the derivative of  $\log{{\rm \rho}}$ with $\phi$ versus $r$.   It is approximately constant in the domain $3<r<10$ kpc  (shown in the inset), with an average  value  of $ \sim 1100$ eV/Kel.

 \begin{figure}[placement h]
\begin{center}
\includegraphics[height=2in,width=3in,angle=0]{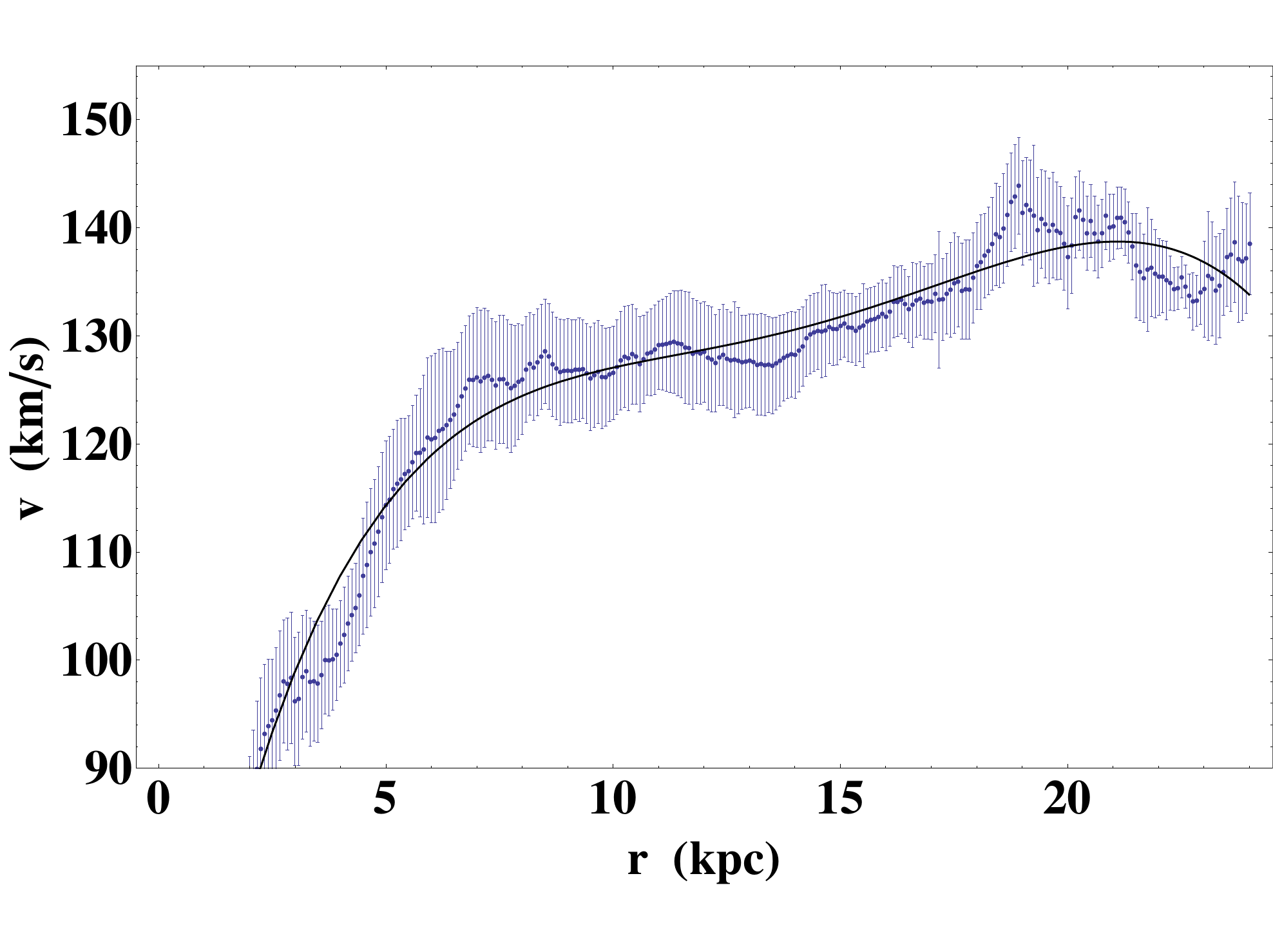}
\caption {Eight-parameter series fit of rotation velocity data for  NGC 2403 from $r=.1$ to $24$ kpc.}
\end{center} 

\begin{center}
\includegraphics[height=1.5in,width=3.3in,angle=0]{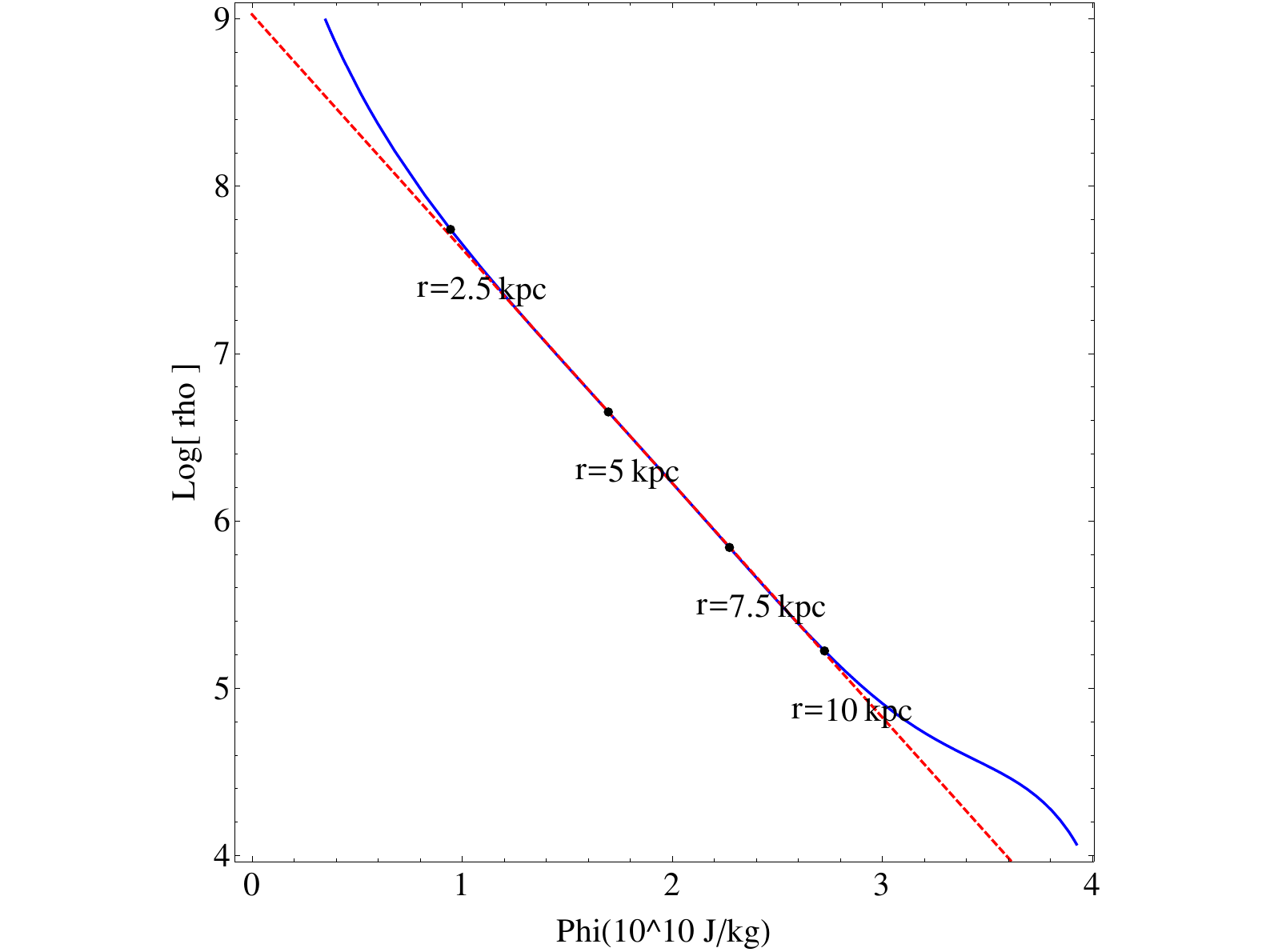}
\caption{$\log{({\rm density/  constant})}$ versus the gravitational potential for  NGC 2403 for $r=1$ to $20$ kpc.  It is compared to a straight line (red, dashed) with  slope $\approx -1100$ eV/Kel.  The two plots approximately coincide for     $3<r<10$ kpc. }
\end{center}
\end{figure}

\begin{figure}[placement h]
\begin{center}
\includegraphics[height=2.5in,width=2.5in,angle=0]{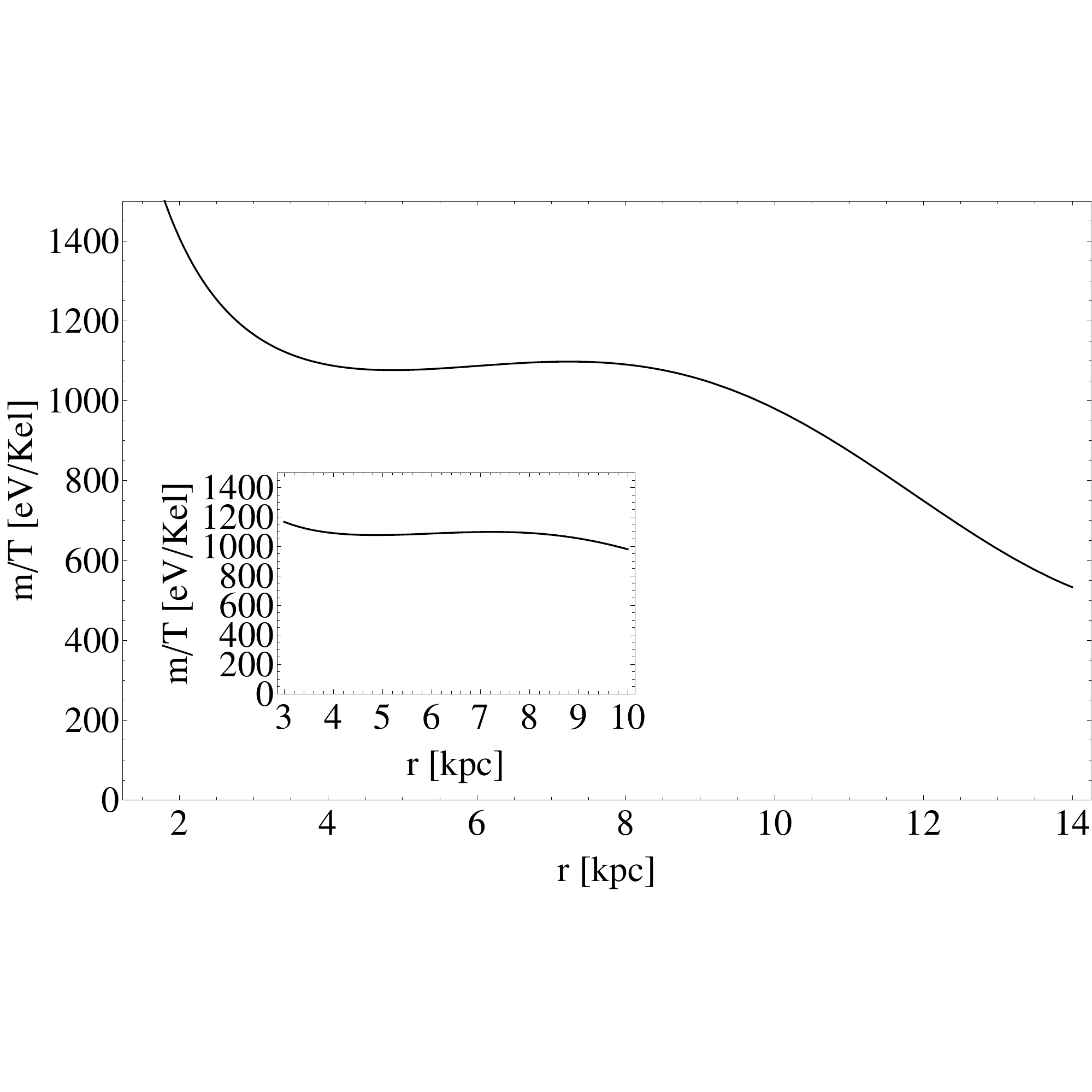}
\caption {$\;\;-\frac{d\log{{\rm \rho}}}{d\phi}$ versus $r$ for  NGC 2403. The plot is approximately constant over the domain  $3<r<10$ kpc, which is shown in the inset.   }
\end{center}

\end{figure}
\newpage 
\subsection{ NGC 2903}

The morphology classification of NGC
2903 is SAB(rs)bc. It has  tightly wound spiral arms.
 Mass models  (see  \cite{deBlok:2008wp}) give the disk and bulge masses for  NGC 2903 of   $1.41\times 10^{10}\; M_{\odot}$ and  $2.14\times 10^{9}\; M_{\odot}$, respectively.  The   distance scale of the disk is  $\sim 3.0$ kpc. \cite{Mannheim:2010xw}
  The HI gas mass for  NGC 2903 is $4.35\times 10^{9} \;M_{\odot}$,\cite{Walter:2008wy} which is approximately $27\%$ of the disk plus bulge mass.

 THINGS rotation velocity data are available at distances up to $31$ kpc, which is over ten times the  distance scale of  the disk.  An eight-parameter series fit of rotation velocity data from $r= .3$ to $ 31$ kpc appears in figure 15.   The resulting  plot of $\log{({\rm \rho(r)/  constant})}$ versus $\phi(r)$ is given in figure 16.  An approximate straight line with negative slope is recovered upon restricting the radial coordinate to $3<r<9$ kpc. In figure 17 we plot  minus the derivative of  $\log{{\rm \rho}}$ with $\phi$ versus $r$.   It is approximately constant in the domain $3<r<9$ kpc  (shown in the inset), with an average  value  of $\sim  490$ eV/Kel.  If one extends the domain up to $25$ kpc, the value of $m/T$ varies from $\sim 350$ to $\sim  650$ eV/Kel.

 \begin{figure}[placement h]
\begin{center}
\includegraphics[height=2in,width=3in,angle=0]{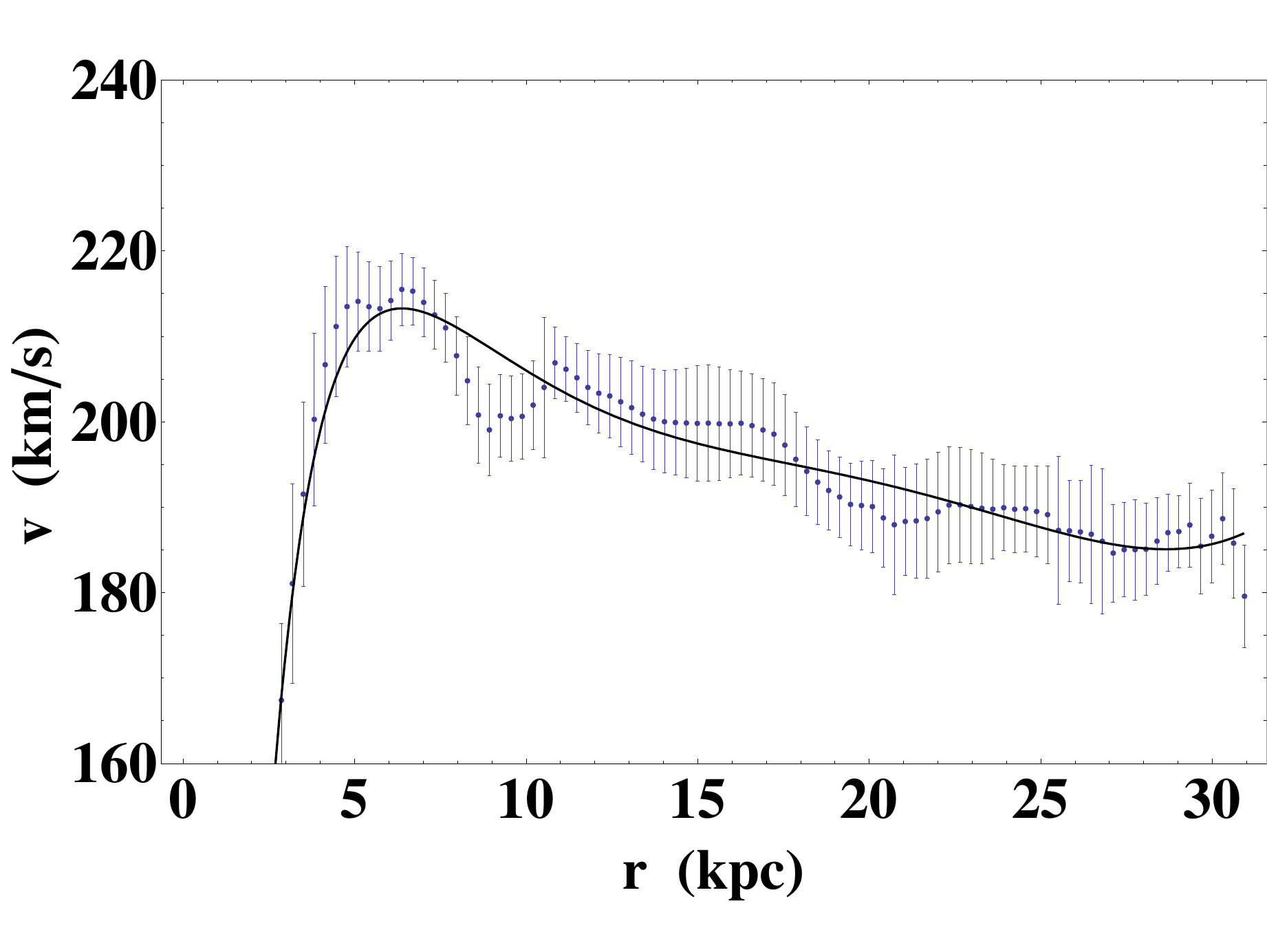}
\caption {Series fit of rotation velocity data for  NGC 2903 from $r= .3$ to $ 31$ kpc .}
\end{center}
 \end{figure}

\begin{figure}[placement h]
\begin{center}
\includegraphics[height=1.5in,width=3.3in,angle=0]{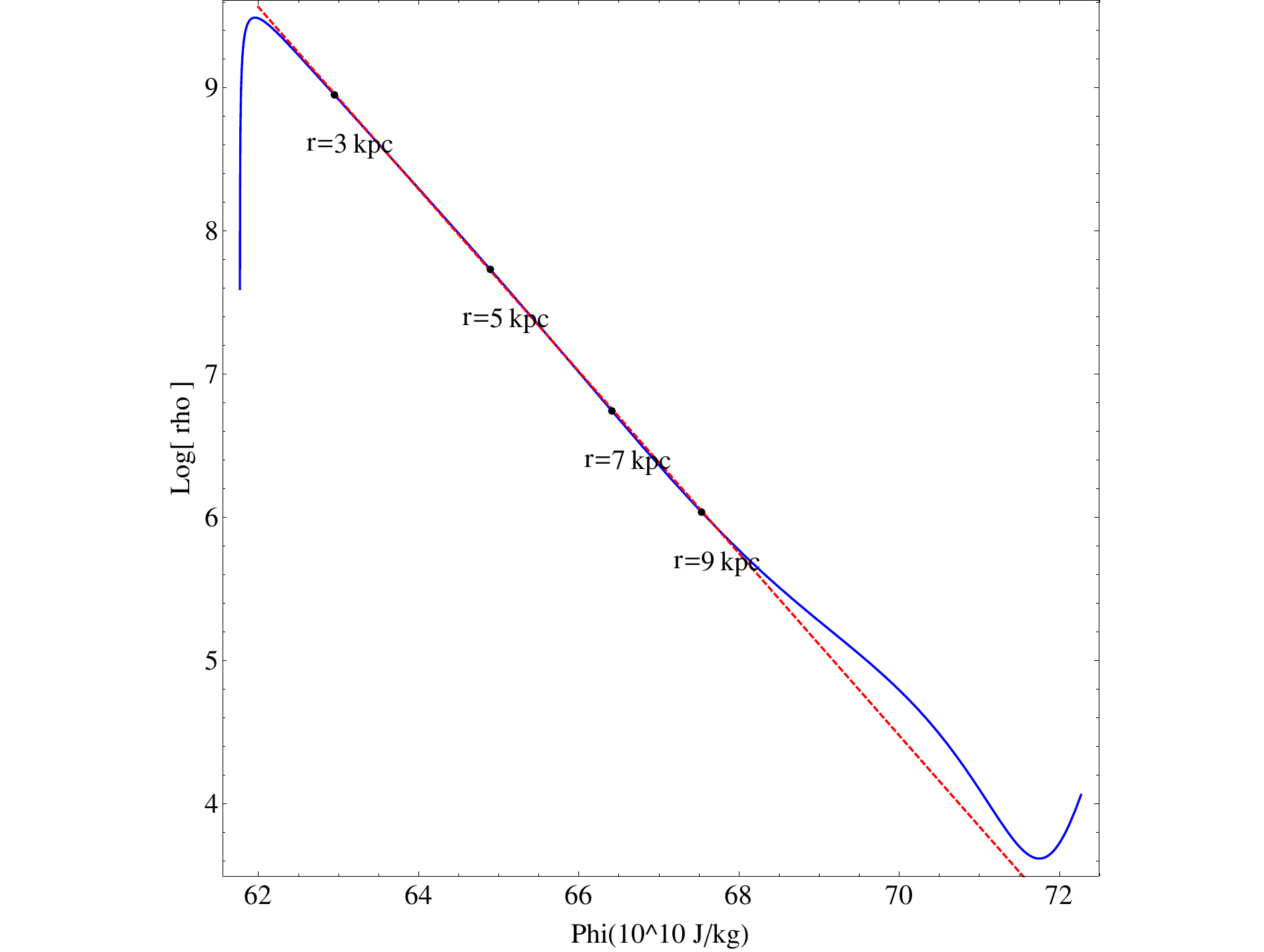}
\caption{$\log{({\rm density/  constant})}$ versus the gravitational potential for  NGC 2903 for $r=1$ to $31$ kpc.  It is compared to a straight line (red, dashed) with  slope $\approx -490$ eV/Kel.  The two plots approximately coincide for     $3<r<9$ kpc. }
\end{center}
\begin{center}
\includegraphics[height=2.5in,width=2.5in,angle=0]{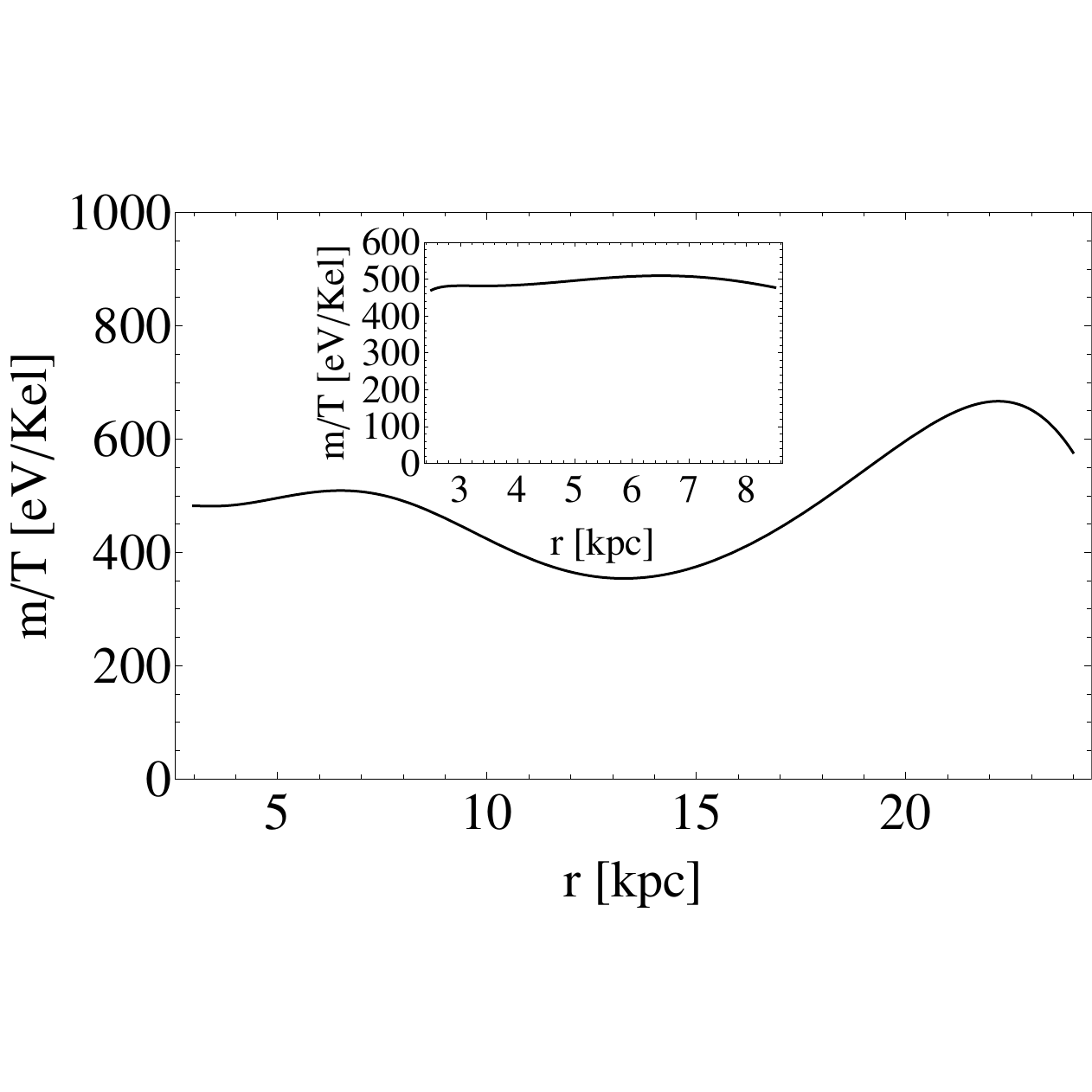}
\caption {$\;\;-\frac{d\log{{\rm \rho}}}{d\phi}$ versus $r$ for  NGC 2903. The plot is approximately constant over the domain  $3<r<9$ kpc, which is shown in the inset.   }
\end{center}

 \end{figure}

\noindent 
\subsection{ NGC 3521}

The morphology classification of NGC 3521 is SAB(rs)bc.   It has a flocculent spiral structure and a negligible bulge.
 Mass models  (see  \cite{deBlok:2008wp}) give the disk  mass for  NGC 3521 of  
 $1.23\times 10^{11}\; M_{\odot}$, with a  corresponding  distance scale of  $\sim 3.3$ kpc. \cite{Mannheim:2010xw}
  The HI gas mass for  NGC 3521 is $.802\times 10^{10} \;M_{\odot}$,\cite{Walter:2008wy} which is approximately $6.5\%$ of the disk  mass.

 THINGS rotation velocity data are available at distances up to $35.5$ kpc, which is over ten times the  distance scale of  the disk.  An eight-parameter series fit of rotation velocity data from $r= 3.2$ to $ 35.5$ kpc appears in figure 18.   The resulting  plot of $\log{({\rm \rho(r)/  constant})}$ versus $\phi(r)$ is given in figure 19.  An approximate straight line with negative slope is recovered upon restricting the radial coordinate to $7<r<15$ kpc.  In figure 20 we plot  minus the derivative of  $\log{{\rm \rho}}$ with $\phi$ versus $r$.   It is approximately constant in the domain $7<r<15$ kpc  (shown in the inset), with an average  value  of $\sim  380$ eV/Kel. A  minimum of  $\sim 230$ eV/Kel occurs at  $\sim 4$ kpc. 

 \begin{figure}[placement h]
\begin{center}
\includegraphics[height=2in,width=3in,angle=0]{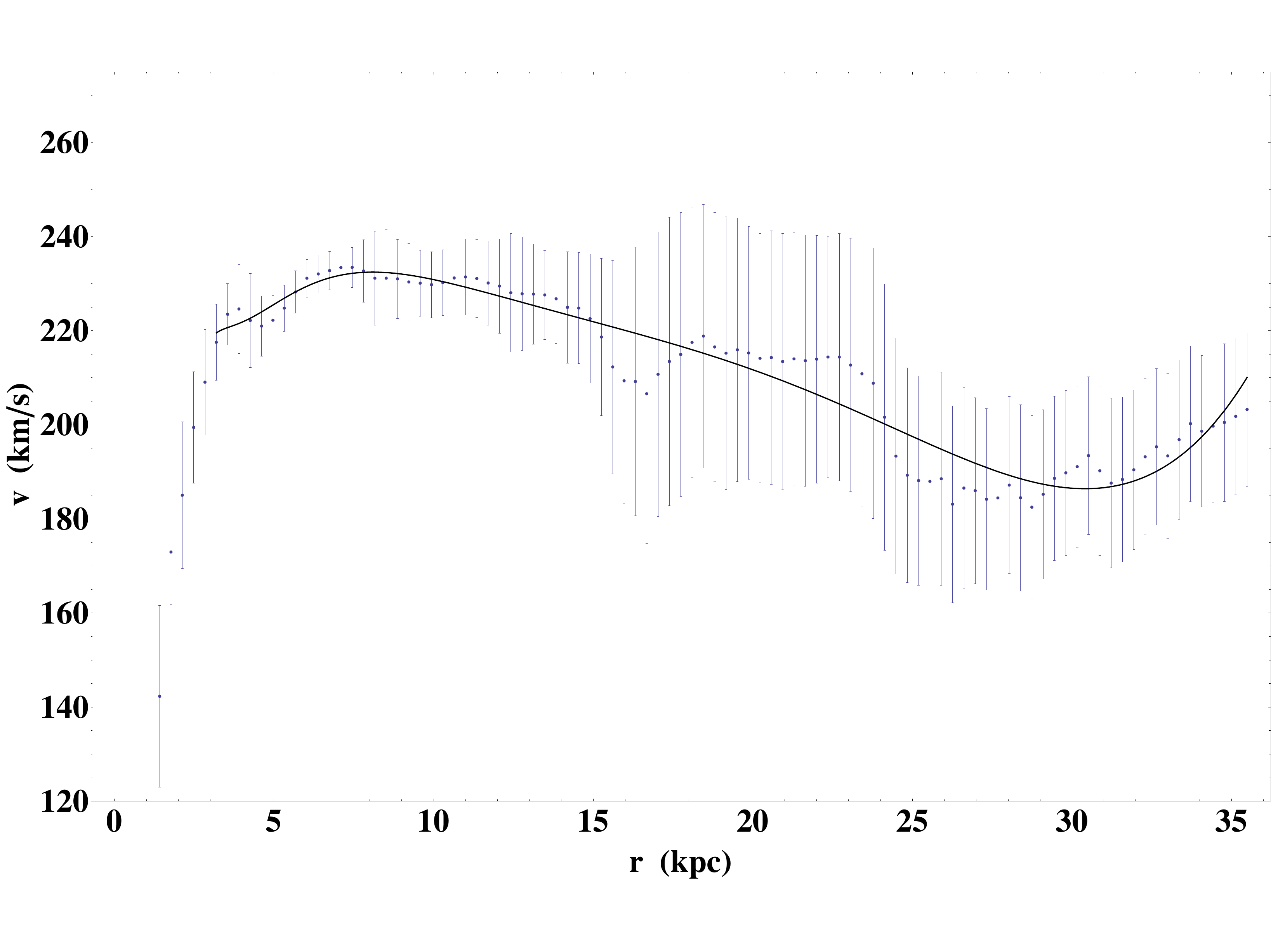}
\caption {Series fit of rotation velocity data for  NGC 3521 from $r=3.2$ to $35.5$ kpc.}
\end{center}

\begin{center}
\includegraphics[height=1.5in,width=3.3in,angle=0]{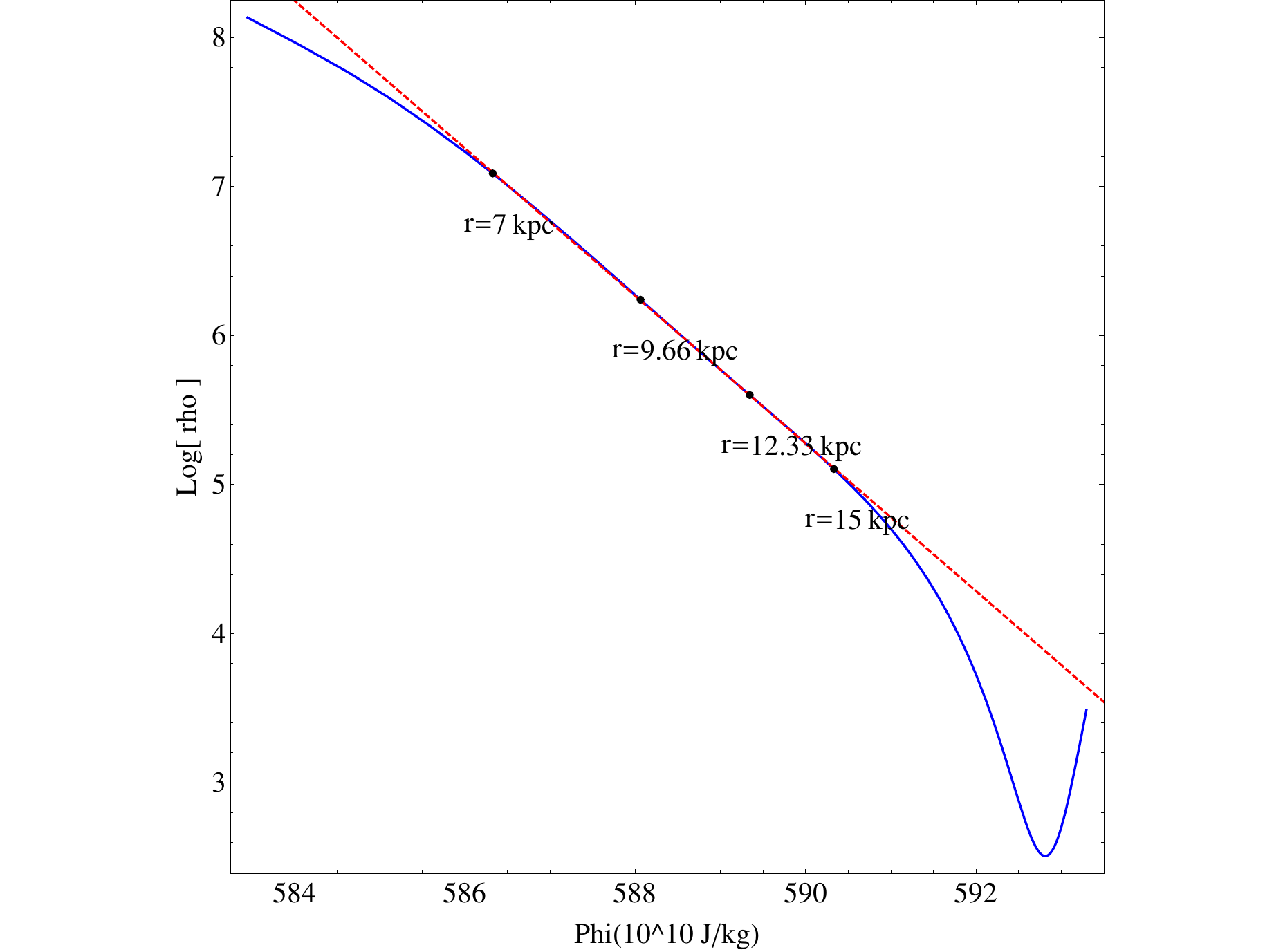}
\caption{$\log{({\rm density/  constant})}$ versus the gravitational potential for  NGC 3521 for $r=4$ to $30$ kpc.  It is compared to a straight line (red, dashed) with  slope  $\approx -380$ eV/Kel .  The two plots approximately coincide for     $7<r<15$ kpc. }
\end{center}

\end{figure}
\newpage
\begin{figure}[placement h]
\begin{center}
\includegraphics[height=2.5in,width=2.5in,angle=0]{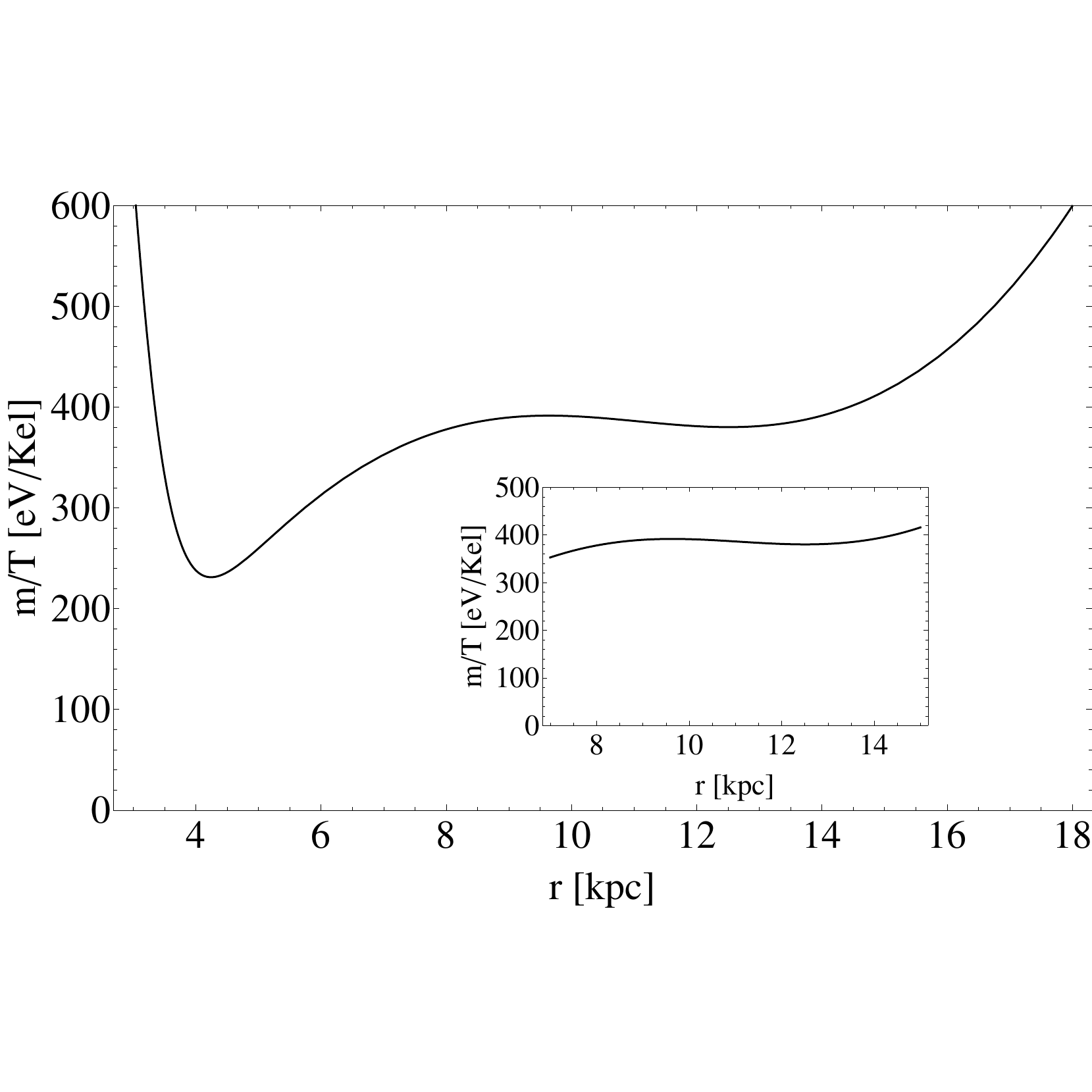}
\caption {$\;\;-\frac{d\log{{\rm \rho}}}{d\phi}$ versus $r$ for  NGC 3521. The plot is approximately constant over the domain  $7<r<15$ kpc, which is shown in the inset.   }
\end{center}
 \end{figure}

We next consider  the barred spiral galaxy NGC 3198.

\noindent 
\subsection{ NGC  3198}

The morphology classification of NGC 3198 is SB(rs)c.  This galaxy has two well-defined  spiral arms, and has indications of a modest warp.
 Mass models (see  \cite{deBlok:2008wp}) give the disk and bulge masses for  NGC 3198 of   $2.82\times 10^{10}\; M_{\odot}$ and  $2.88\times 10^{9}\; M_{\odot}$, respectively.  The  distance scale of the disk is  $\sim 4.0$ kpc. \cite{Mannheim:2010xw}
  The HI gas mass for  NGC 3198 is $1.017\times 10^{10} \;M_{\odot}$,\cite{Walter:2008wy} which is approximately $33\%$ of the disk plus bulge mass.

THINGS rotation velocity data are available at distances up to $38.5$ kpc, which is over  nine times the  distance scale of  the disk.  The eight-parameter series fit of rotation velocity data from $r= 4.1$ to $ 38.5$ kpc appears in figure 21. The resulting  plot of $\log{({\rm \rho(r)/  constant})}$ versus $\phi(r)$ is given in figure 22.  An approximate straight line with negative slope is recovered upon restricting the radial coordinate to $5<r<15$ kpc. In figure 23 we plot  minus the derivative of  $\log{{\rm \rho}}$ with $\phi$ versus $r$.  It ranges between $\sim 750$ eV/Kel  and  $\sim 950$ eV/Kel  and has an average value of $\sim 820$ eV/Kel for the domain  $5<r<25$ kpc.   
 \begin{figure}[placement h]
\begin{center}
\includegraphics[height=2in,width=3in,angle=0]{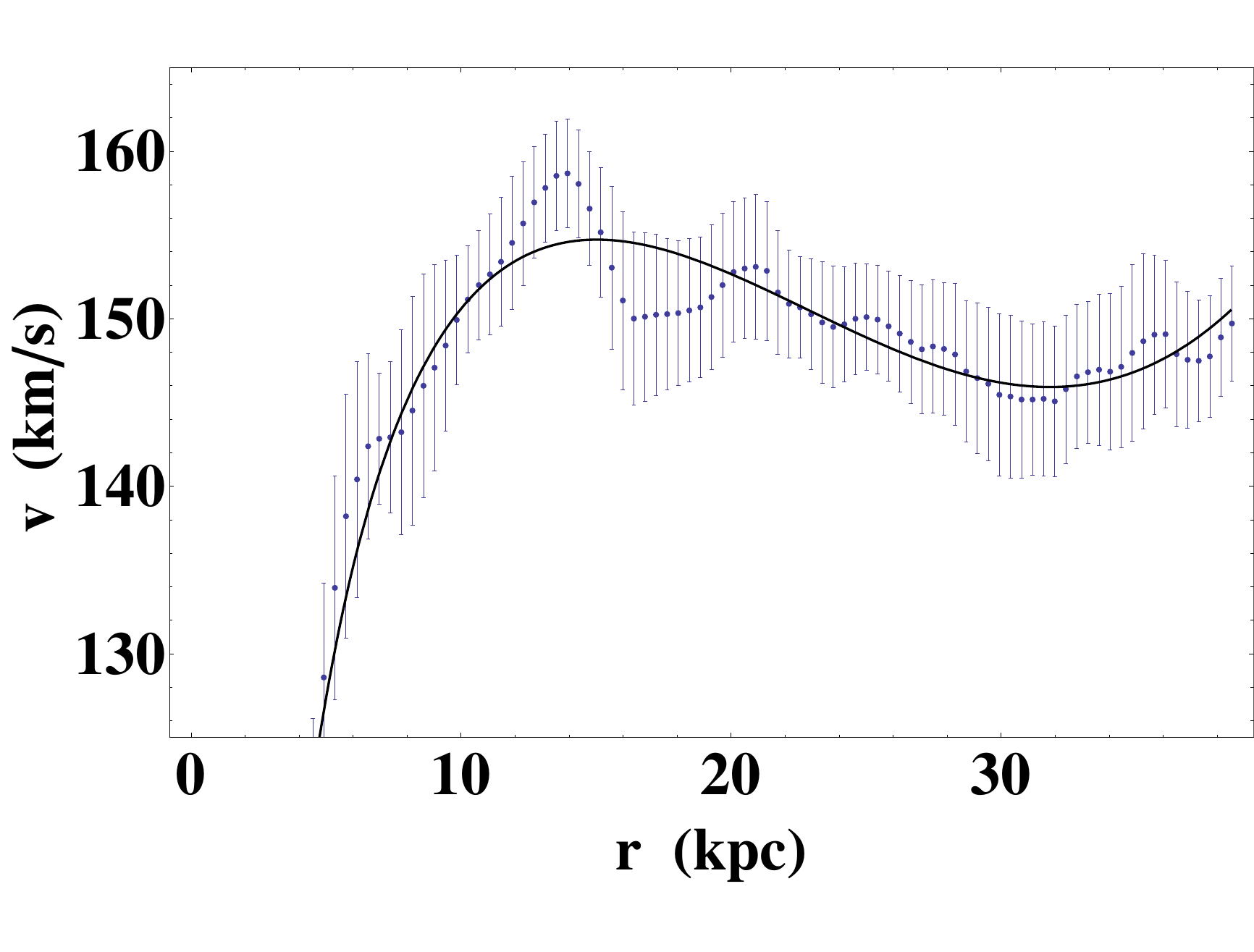}
\caption {Eight-parameter series fit of rotation velocity data for  NGC 3198 from $r= 4.1$ to $ 38.5$ kpc.}
\end{center}

\begin{center}
\includegraphics[height=1.5in,width=3.3in,angle=0]{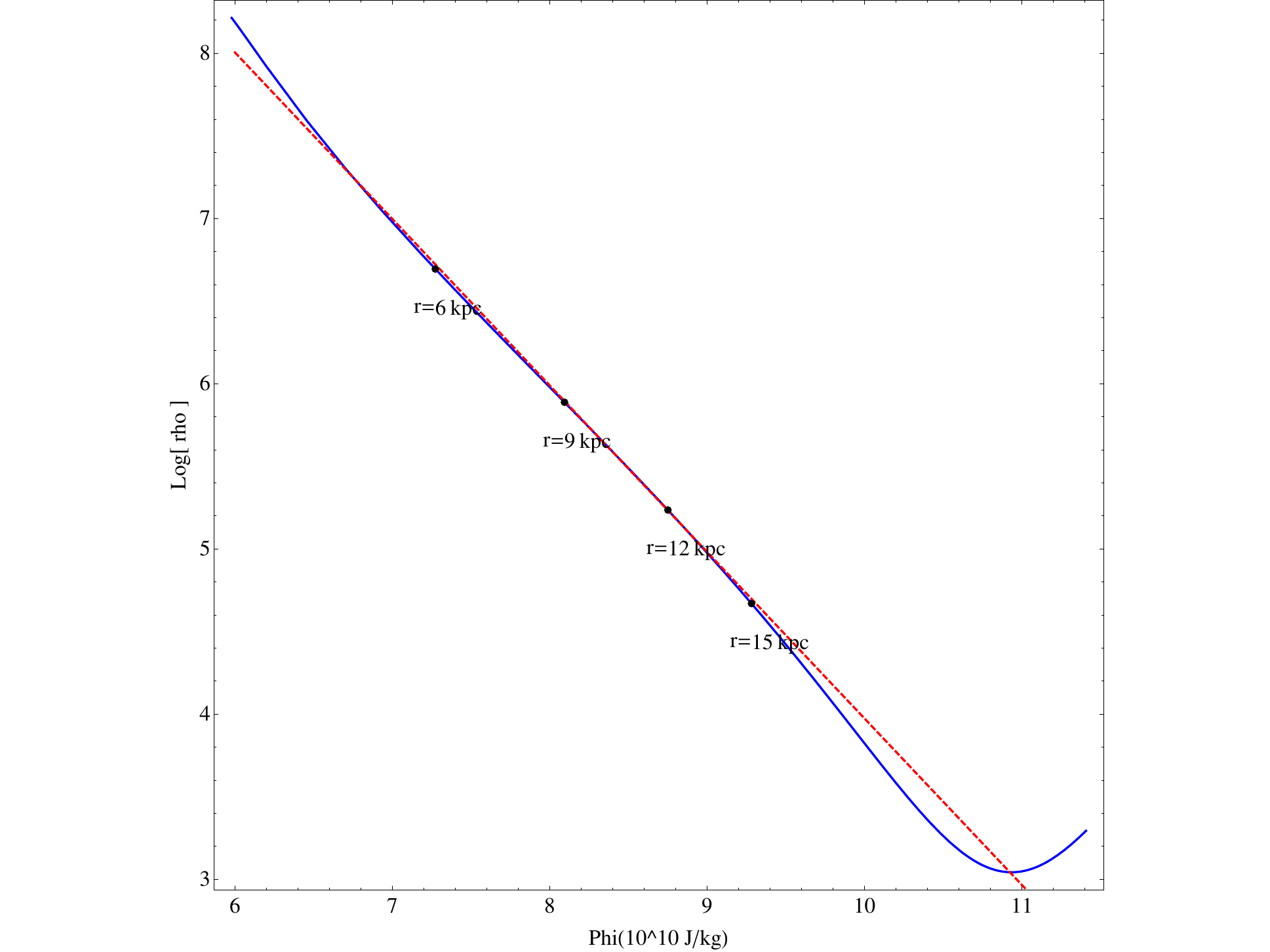}
\caption{$\log{({\rm density/  constant})}$ versus the gravitational potential for  NGC 3198 for $r=2$ to $39.5$ kpc.  It is compared to a straight line (red, dashed) with  slope $\approx - 820$ eV/Kel.  The two plots approximately coincide for     $5<r<15$ kpc. }
\end{center}\end{figure}

\begin{figure}[placement h]
\begin{center}
\includegraphics[height=1.5in,width=2in,angle=0]{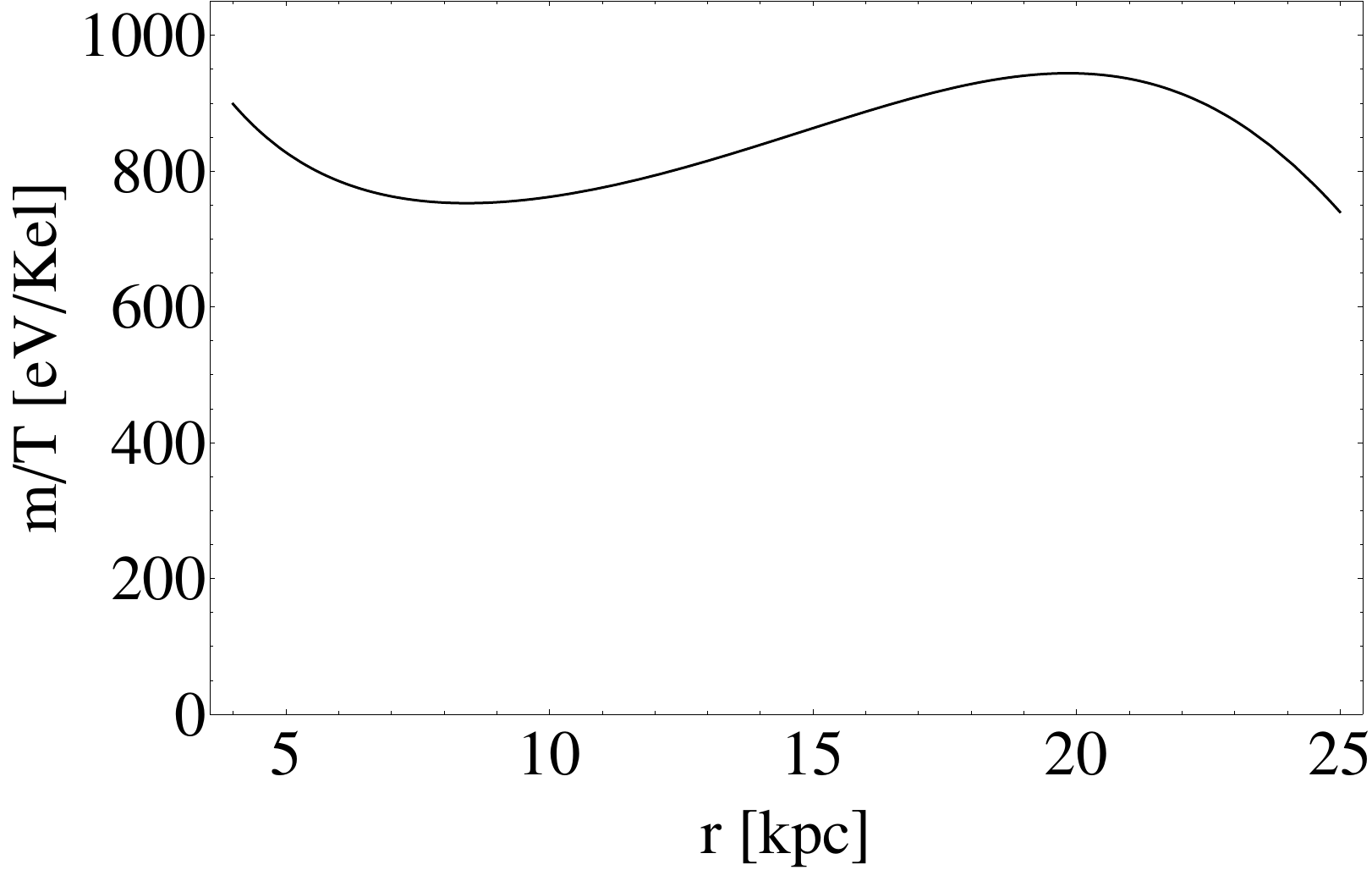}
\caption {$\;\;-\frac{d\log{{\rm \rho}}}{d\phi}$ versus $r$ for  NGC 3198. The plot ranges between $\sim 750$ eV/Kel  and  $\sim 950$ eV/Kel for the domain  $5<r<25$ kpc.   }
\end{center}

 \end{figure}
\newpage

Finally, we repeat the analysis for the dwarf galaxy  DD0 154.

\noindent 
\subsection{ DD0 154}

The morphology classification of  DD0 154 is IB(s)m.  It is a  gas-rich dwarf irregular galaxy with a negligible bulge.
 Mass models  (see  \cite{deBlok:2008wp}) give the disk  mass for  DD0 154 of  
 $2.63\times 10^{7}\; M_{\odot}$, with a  corresponding  distance scale of  $\sim .8$ kpc. \cite{Mannheim:2010xw}
  The HI gas mass for  DD0 154 is $3.58\times 10^{8} \;M_{\odot}$,\cite{Walter:2008wy} which is approximately $13.6$ times the disk  mass.

THINGS rotation velocity data are available at distances up to $8$ kpc, or ten times the  distance scale of  the disk.   The eight-parameter series fit of rotation velocity data from $r= .3$ to $ 8$ kpc appears in figure 24. The resulting  plot of $\log{({\rm \rho(r)/  constant})}$ versus $\phi(r)$ is given in figure 25.  An approximate straight line with negative slope is recovered upon restricting the radial coordinate to $1.5<r<4.5$ kpc. In figure 26 we plot  minus the derivative of  $\log{{\rm \rho}}$ with $\phi$ versus $r$.   It ranges between $\sim 7500$ eV/Kel  and  $\sim 10000$ eV/Kel  and has an average value of $\sim 8600$ eV/Kel for the domain  $1.5<r<6.5$ kpc.

 \begin{figure}[placement h]
\begin{center}
\includegraphics[height=2in,width=3in,angle=0]{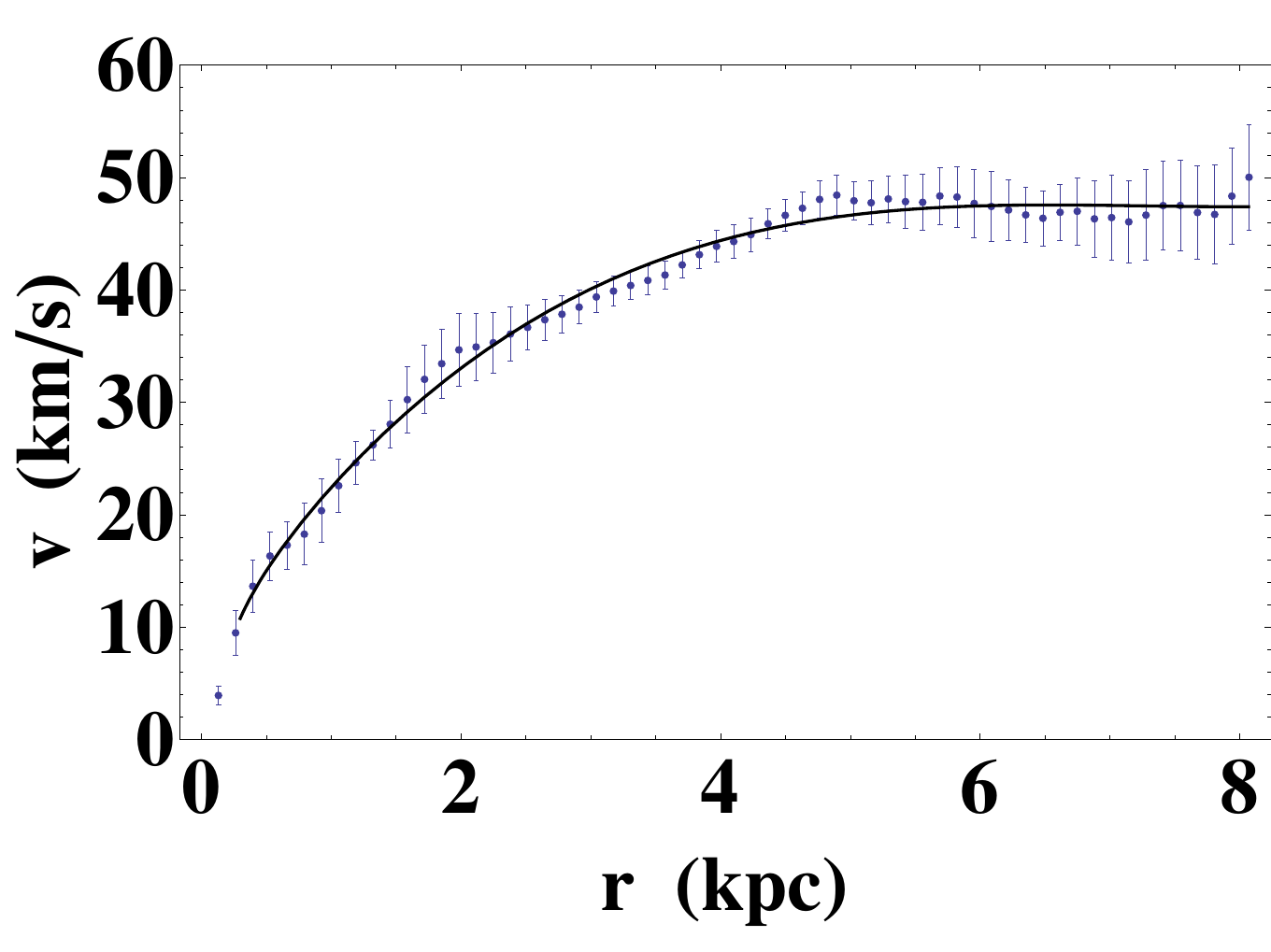}
\caption {Eight-parameter series fit of rotation velocity data for  DD0 154 from $r= .3$ to $ 8$ kpc.}
\end{center}

\begin{center}

\includegraphics[height=2in,width=3.3in,angle=0]{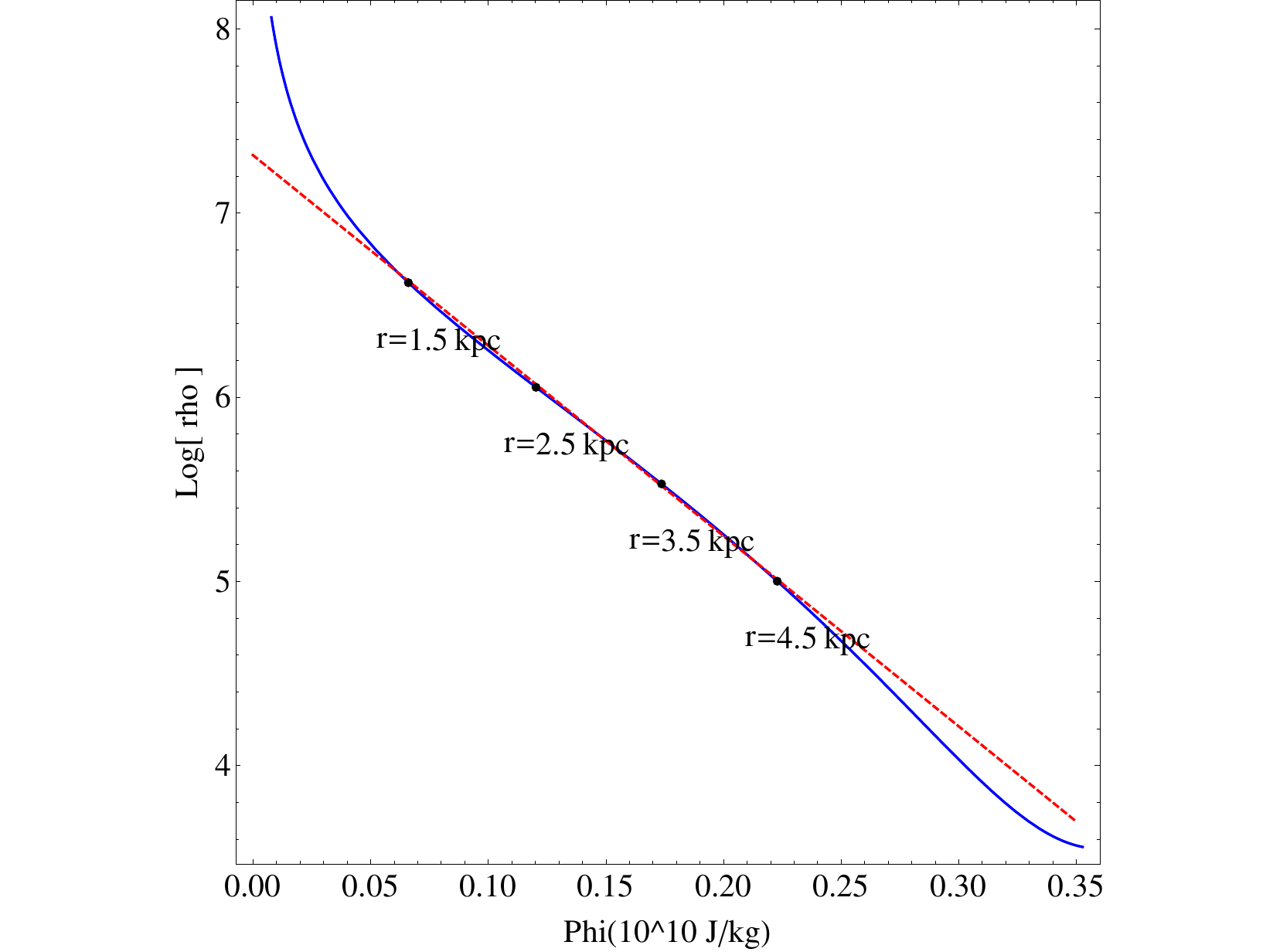}
\caption{$\log{({\rm density/  constant})}$ versus the gravitational potential for   DD0 154 for $r=.3$ to $8$ kpc.  It is compared to a straight line (red, dashed) with  slope $\approx - 8600$ eV/Kel.  The two plots approximately coincide for     $1.5<r<4.5$ kpc. }
\end{center}
\end{figure}

 \begin{figure}[placement h]
\begin{center}
\includegraphics[height=2in,width=2.5in,angle=0]{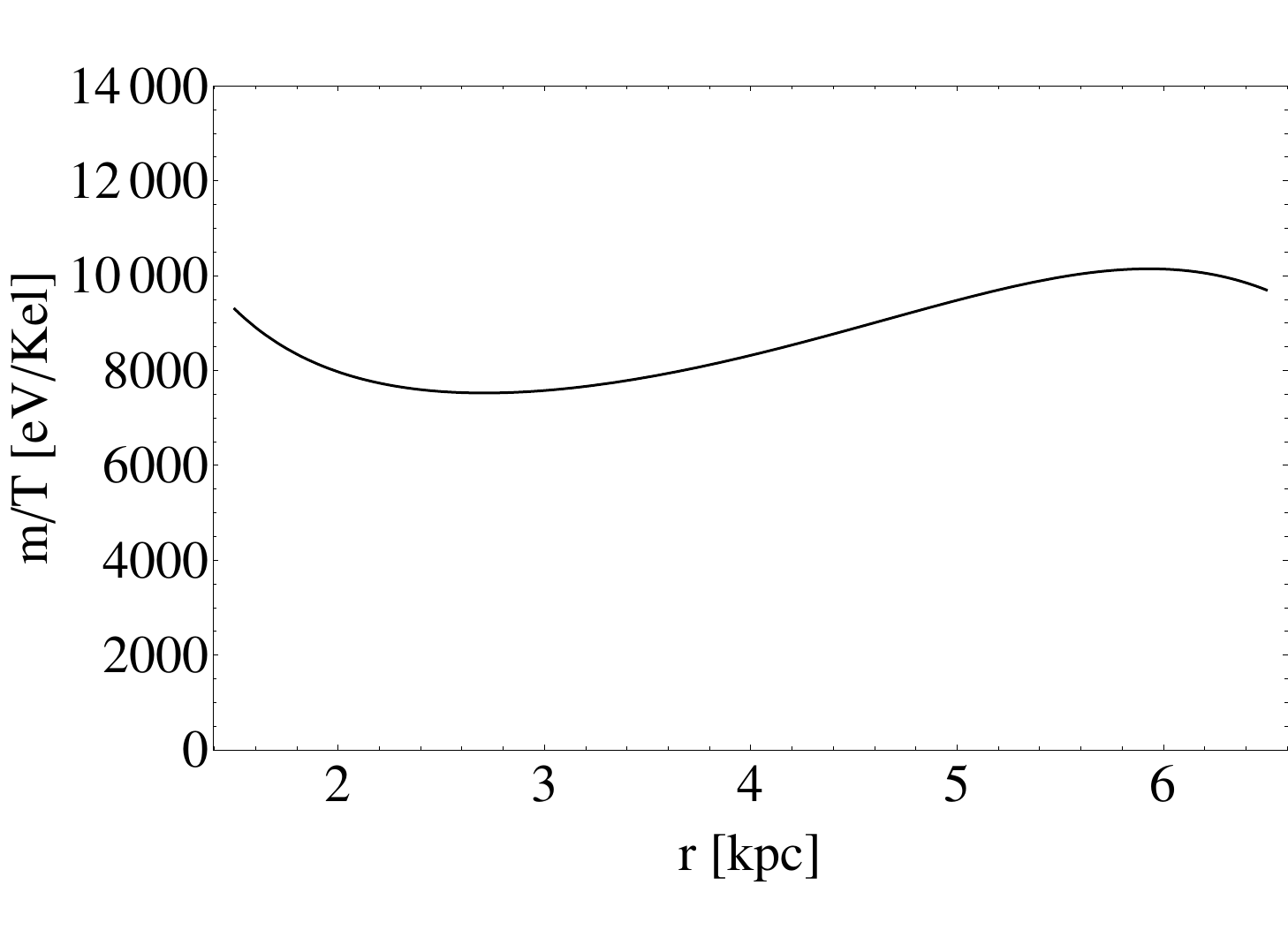}
\caption {$\;\;-\frac{d\log{{\rm \rho}}}{d\phi}$ versus $r$ for  DD0 154. The plot ranges between $\sim 7500$ eV/Kel  and  $\sim 10000$ eV/Kel over the domain  $1.5<r<6.5$ kpc.   }
\end{center}

 \end{figure}
\newpage
\section{An effective theory}

In the previous section we have presented evidence of Boltzmann-like   behavior in  regions of   haloes of eight galaxies in the THINGS survey.  
Here we model the Boltzmann behavior by  writing down the dynamics for a self-gravitating gas in thermodynamic equilibrium  in an external gravitational field.  The relevant domain for the model is a finite shell, associated with the regions where the exponential behavior was  found was found for $\rho$ versus $\phi$.  Here we  discuss the parameter space of the solutions of the model, along with consistency checks. Fits of the rotation curves   to the solutions  will be made in the following section. 

\subsection{Dynamics  and boundary conditions}

Here we write down an effective thermodynamic model for  regions, corresponding to spherical shells  $R_{\tiny {\tt G}}\le r\le R_{\tiny {\tt max}}$, where the  Boltzmann gas  is valid.   
More specifically, it is the  description of a self-gravitating isothermal gas, subject to an external gravitational potential. The  total gravitational potential is again denoted by $\phi $, which includes contributions from the gravitational self-interactions of the gas, as well as the external tidal force.     We shall  assume that the  gas consists of only one type of particle, with mass $m$, and that  $\phi $ is slowly  varying  with respect to the average  interparticle distance.  As in the previous section, we  assume that spherical symmetry holds, at least at lowest order, and so quantities of interest, such as $\phi$ and the  average  density  $\rho$ are functions only  of the radial variable.   It is desirable to have the radius $R_{\tiny {\tt G}}$ of the inner boundary sufficiently larger than the distance scale of  the disk.  (This was generally the case for the examples  considered in the previous section. For instance,   $R_{\tiny {\tt G}}$ was approximately twice the disk scale for NGC 2841.)
Then one expects that deviations from spherical symmetry and contributions from the
 baryonic matter to total density $\rho$ in the spherical shell can be
 ignored at lowest order.

We first impose boundary conditions at $r = R_{\tiny {\tt G}}$. The value of $\phi$  at the boundary  is arbitrary.  It is convenient to choose it to be zero.\footnote{Our fitting procedure
will not depend on this choice.  The fits determine values of  $ \rho( R_{\tiny {\tt G}})$ and $m/T$ in the density formula (\ref{natmr}).  The addition of a constant to the potential can be absorbed in $ \rho( R_{\tiny {\tt G}})$.}
 Then for a Boltzmann gas one has 
\be \rho(r) = \rho( R_{\tiny {\tt G}}) e^{- m\phi(r)/k_BT}\;\label{natmr}\ee
 On the other hand, the value of $\frac{d\phi}{dr}$   at  $r = R_{\tiny {\tt G}}$ must be  identified with the acceleration of gravity  at the inner boundary of the spherical shell.  Thus \be \phi(R_{\tiny {\tt G}})=0\qquad\qquad \frac{d\phi}{dr}\Big|_{r=R_{\tiny {\tt G}}}=  \frac{GM_{\tiny {\tt G}}}{R_{\tiny {\tt G}}^2}\;,\label{bndcnd}\ee $M_{\tiny {\tt G}}$ is the total mass inside the sphere of radius $R_{\tiny {\tt G}}$.  The model is thus defined by the two boundary conditions and the Poisson equation
\be\frac 1{r^2}\frac d{dr} \Bigl(r^2\frac{d\phi}{dr}\Bigr)=  4\pi G\rho( R_{\tiny {\tt G}}) e^{- m\phi(r)/k_BT}\;,\qquad\quad R_{\tiny {\tt G}}\le r\le R_{\tiny {\tt max}}\;\label{1.1}\ee

The pressure balance  implicitly holds everywhere within and beyond the isothermal region, including at the interfaces between the regions.  The pressure within the isothermal region is given by the ideal gas equation $ P(r) =\rho(r)kT/m$.
   
\subsection{Solutions}

 To examine the space of solutions it is convenient to do a rescaling to the dimensionless variables $ x=\frac r{R_{\tiny {\tt G}}} $ and $\chi = \frac{R_{\tiny {\tt G}}}{GM_{\tiny {\tt G}}}\,\phi$.  (\ref{1.1})  simplifies to
\be  \frac 1{x^2}\frac d{dx} \Bigl( x^2\frac{d\chi}{dx}\Bigr)=3\kappa\,e^{-\frac{\chi(x)} {\tau}}\;,\qquad\quad 1\le x\le x_{\tiny {\tt max}}\;,\label{sclddfeq}\ee where $x_{\tiny {\tt max}}=R_{\tiny {\tt max}}/R_{\tiny {\tt G}}$, and the boundary conditions become 
\be \chi(1)=0\qquad\qquad \frac{d\chi}{dx}\Big|_{x=1}=1 \;\label{bcsntrmsfx}\ee 
  $\kappa$  and $\tau$ are independent dimensionless parameters determining the solutions.  
  They depend, among other things, on $R_{\tiny {\tt G}}$ and $M_{\tiny {\tt G}}$.  $\kappa$ is the ratio of  $\rho(R_{\tiny {\tt G}})$ to the mean mass density in the interior of the sphere of radius $R_ {\tiny {\tt G}}$, i.e., $\rho_{\tiny {\tt G}}=\frac {M_{\tiny {\tt G}}}{\frac 43\pi  R_{\tiny {\tt G}}^3}$, while $\tau$ is a rescaled temperature
\be \kappa=\frac{ \frac 43\pi  R_{\tiny {\tt G}}^3\,\rho(R_{\tiny {\tt G}})}{M_{\tiny {\tt G}}}\;\qquad\quad \tau=\frac{k_B T}{mv_{x=1}^2 }\;, \label{kappatau}\ee where $ v_{x=1}$ is
 the speed   of objects undergoing  circular orbits at $r=R_{\tiny {\tt G}}$.  It is given by  $v_{x=1}^2=GM_{\tiny {\tt G}}/R_{\tiny {\tt G}}$.
From solutions to (\ref{sclddfeq}) and (\ref{bcsntrmsfx}) one can then determine the orbital
 speed $v(x)$ at any value of the radial coordinate, along with the total mass $M(x)$  enclosed in a sphere of radius $x$,  $1\le x\le x_{\tiny {\tt max}}$. They are given respectively by
\be \Bigl(\frac{v}{v_{x=1}}\Bigr)^2= x\frac{d\chi}{dx}\; \qquad{\rm and}\qquad \; \frac{M(x)}{ M_{\tiny {\tt G}}}= 3\kappa\int_1^x dx' x'^2 {e^{-\frac{\chi(x')}{\tau}}} \label{atmms}\ee

Eq. (\ref{sclddfeq}) is the spherically symmetric Lane-Emden
equation for infinite polytropic index, which is often referred to as the Emden equation.\cite{Emden}  Various techniques have been employed to obtain its solutions, and  application to galactic dynamics has been discussed. (See for 
example, \cite{Peebles:1994xt},\cite{Liu:1995db},\cite{Natarajan:1996ic},\cite{Mishchenko:2003in}.  Axially symmetric solutions with rotation have also been obtained.\cite{Christodoulou:2007br})  Boundary conditions for this equation are standardly imposed at the origin, where the gravitational force is required to vanish, i.e., $ \frac{d\chi}{dx}\Big|_{x=0}=0$.\footnote{$\chi$ is often set to zero, as well, at the origin, but this is a gauge condition which can be removed with a redefinition of $\kappa$.}
This boundary condition leads to a  solution known as the   isothermal sphere, which was first found by Emden.

In our case, the Boltzmann gas description is not valid  outside the domain $ 1\le x\le x_{\tiny {\tt max}}$, and it is then appropriate to impose boundary conditions at $x=1$, as we have done in (\ref{bcsntrmsfx}).
A two-parameter (i.e.,  $\kappa$ and $\tau$) family of solutions to the Emden equations result from the boundary conditions (\ref{bcsntrmsfx}).  

The solutions contain  Emden's  isothermal sphere as a special case.  That is, a subset of the solutions to (\ref{sclddfeq}) and (\ref{bcsntrmsfx})  can be consistently continued from $x=1$ to the origin, where $ \frac{d\chi}{dx}\Big|_{x=0}=0$.  
 $\kappa$ and $\tau$ are not independent for the subset. To see this one can   do another rescaling of the field, $\chi\rightarrow\tilde\chi=\chi/\tau$, and the coordinate,  $x\rightarrow\tilde x=\sqrt{\frac{3\kappa}\tau}\, x$, thereby removing the parameters from the differential equation (\ref{sclddfeq}), and putting them instead in the boundary conditions. Then
\be  \frac 1{\tilde x^2}\frac d{d\tilde x} \Bigl(\tilde  x^2\frac{d\tilde \chi}{d\tilde x}\Bigr)=e^{-{\tilde \chi(\tilde x)}}\;,\ee with \be \tilde\chi\Big|_{\tilde x=\sqrt{\frac{3\kappa}\tau}}=0\qquad\qquad \frac{d\tilde \chi}{d\tilde x}\Big|_{\tilde x=\sqrt{\frac{3\kappa}\tau}}=\frac1{\sqrt{3\kappa\tau}} \;\label{bcchitld}\ee  For generic values of $\kappa$ and $\tau$, the boundary conditions (\ref{bcchitld}) cannot  be generated by integrating the Emden equation from the origin, starting from $ \frac{d\tilde\chi}{d\tilde x}\Big|_{\tilde x=0}=0$ and any $\tilde\chi(0)$. [$\tilde\chi(0)$ must be negative since $\tilde\chi$ grows with  increasing $\tilde x$.]  If one were to impose the boundary condition  $ \frac{d\tilde\chi}{d\tilde x}\Big|_{\tilde x=0}=0$, along with some  $\tilde\chi(0)<0$, one can  integrate the Emden equations to find the intersection of $\tilde \chi(\tilde x)$ with the $\tilde x-$axis, along with the slope of the function at that point.  The result, along with (\ref{bcchitld}), can then be used    to numerically solve for specific values of  $\kappa$ and $\tau$, associated with the  subset of isothermal spheres.

Another special solution to the Emden equation is  the `singular' isothermal sphere.
It has the analytic expression 
\be \chi(x)=\tau \log{\Big(\frac{3\kappa x^2}{2\tau}\Bigl)}\label{exctsln}
\ee  
It is called singular  because it  is ill-defined at the origin (for any values of $\kappa$ and $\tau$). Rather than satisfying $\frac{d\chi}{dx}\Big|_{x=0}=0$, the acceleration of gravity diverges at the origin. On the other hand, (\ref{exctsln}) is mathematically and physically well defined  for the domain of interest here,  $ 1\le x\le x_{\tiny {\tt max}}$.  It satisfies the boundary conditions (\ref{bcsntrmsfx})  for particular values of  $\tau$ and $ \kappa$, namely $\tau=\frac 12 $ and $ \kappa=\frac 13 $, or in terms of dimensionfull quantities,
\be \rho(R_{\tiny {\tt G}}) =\frac{M_{\tiny {\tt G}}}{4\pi  R_{\tiny {\tt G}}^3}  \qquad\qquad \frac mT=\frac{2k_B }{v_{x=1}^2}\label{moverT}\;\ee    From $v_{x=1}^2=GM_{\tiny {\tt G}}/R_{\tiny {\tt G}}$,  one finds that  $m/T$ in (\ref {moverT}) is determined by the baryonic matter and dark matter contained in $r<R_{\tiny {\tt G}}$.  Using  (\ref{atmms}), one gets that
the  solution  (\ref{exctsln}) has an exactly flat rotation curve,  $v=v_{x=1}$ for all $x>0$, and a linearly increasing  mass function   $\;{M(x)}=(x-1) \,{ M_{\tiny {\tt G}}}$.   Since this behavior is roughly what is observed
  for all galactic haloes, 
 (\ref{moverT}) should provide  crude estimates for $\rho(R_{\tiny {\tt G}})$  and $m/T$.  In section 4 we shall compare these estimates   to  values  extracted from fits to the rotation curve data, and we can compare the fitted values  of  $\kappa$ and $\tau$ to this  special case.

Although eq. (\ref{exctsln})  is not a solution for arbitrary values of $\tau$ and $\kappa $, it  does  describe the  behavior of generic solutions  as $x\rightarrow \infty$.\footnote{ To see this we can do another change of coordinates and rewrite the differential equation in terms of $y=\log x$.  Upon assuming that $\frac{d^2 \chi}{dy^2}<<\frac{d \chi}{dy}$ when $y\rightarrow \infty$, one gets  $\frac{d \chi}{dy}\sim 3\kappa \exp{(2y -\chi(y)/\tau)}$, the solution of which is (\ref{exctsln}).}
So in the asymptotic region of the solution (which may be beyond the relevant  physical domain, $ 1\le x\le x_{\tiny {\tt max}}$),
 the orbital speed goes to a constant value, specifically  \be v_\infty=\sqrt{2\tau} \,v_{x=1}=\sqrt{\frac{2k_BT}m}\;,\label{vinft}\ee for any $\kappa>0$. (Of course, $v\rightarrow 0$ as $x\rightarrow \infty$ in the absence of any matter source in the region, i.e., $\kappa=0$.) 
 This is illustrated in figures 27 and 28.   There we plot $v/v_{x=1}$ for numerical solutions to  (\ref{sclddfeq}) and (\ref{bcsntrmsfx}) associated with various values of $\tau$ and $\kappa$ [including the values $\tau=\frac 12$ and $\kappa=\frac 13$ corresponding to the analytic solution  (\ref{exctsln})].

\begin{figure}[placement h]
\begin{center}
\includegraphics[height=2in,width=2.5in,angle=0]{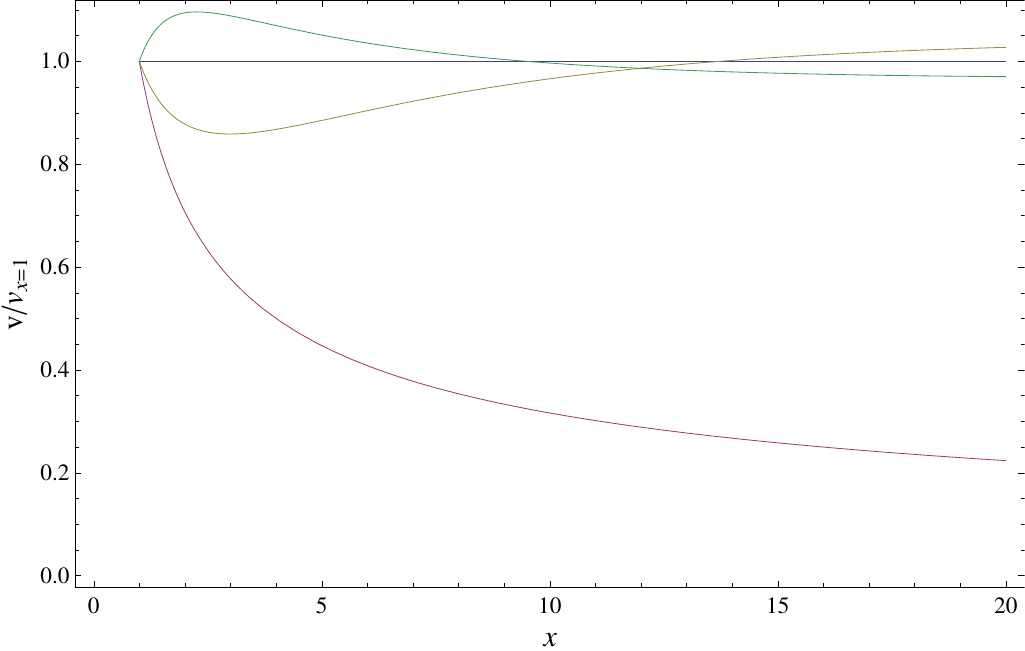}
\caption {$v/v_{x=1}$ is plotted versus $x$ for $\tau=\frac 12$ and $\kappa=0,\frac 16,\frac 13$ and $\frac 12$.  The  values of $\kappa$ are listed in ascending order of the slope at $x=1$, with $\kappa=0$ corresponding to no halo. }
\end{center}

\begin{center}
\includegraphics[height=2in,width=2.5in,angle=0]{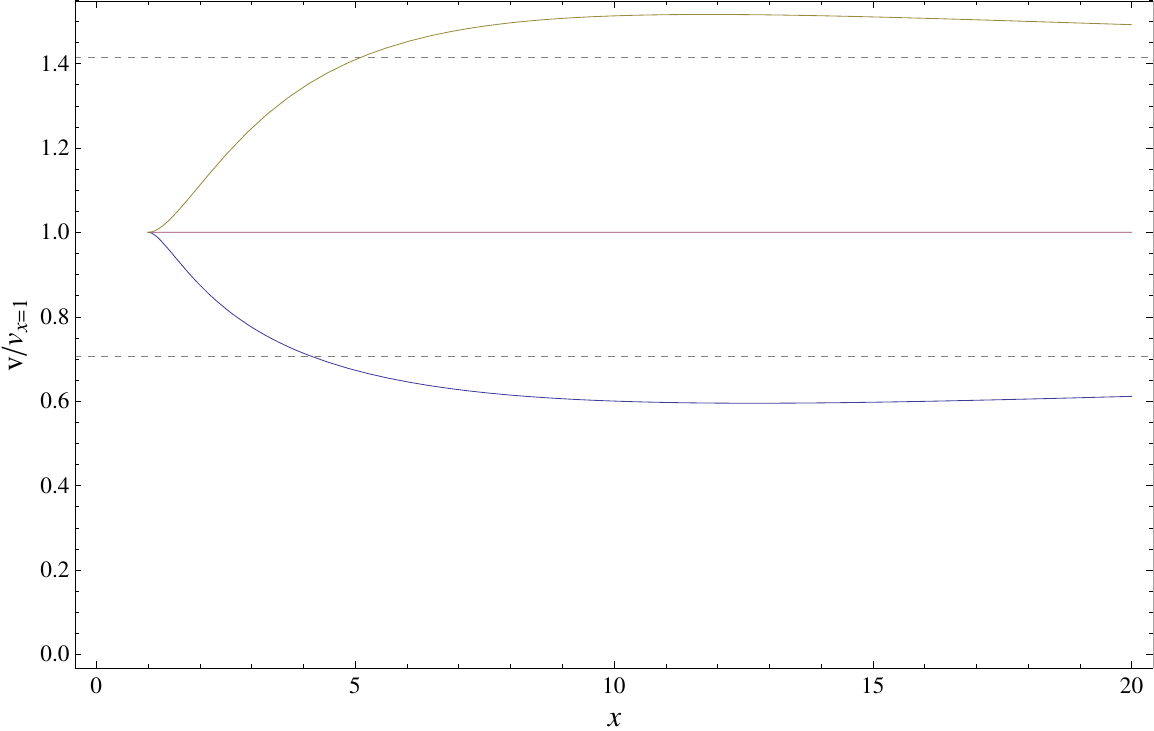}
\caption {$v/v_{x=1}$ is plotted versus $x$ for $\kappa=\frac 13$ and  $\tau=\frac 14,\frac 12$ and $1$. The values of  $\tau$ are listed in ascending order of the slope at $x=1$.  The limiting value of $v/v_{x=1}$ is $\sqrt{2\tau}$, which is indicated for the top and bottom curves by the dashed gridlines.}
\end{center}
\end{figure}

\subsection{Consistency checks}

Since we are dealing with an  Boltzmann gas description, it is reasonable to regard the effective temperature as a measure of the average kinetic energy of the dark matter particles. Then a simple application of the equipartition theorem gives an  rms speed of $u=\sqrt{3k_BT/m}$, assuming no internal structure for the particles.  This is of the same order as the average orbital speed of the HI gas in the halo.  From (\ref{vinft}), one gets $u=\sqrt{\frac 32} v_\infty$, where $v_\infty$ was the orbital speed in the limit  $r$ tends to infinity.
Since orbital speeds in haloes are nonrelativistic, the    particles in this Boltzmann gas description must also be nonrelativistic.  This provides one consistency check of the model, as no relativistic corrections need to be considered.

Another consistency check comes from examining the number  density $n(r)=\rho(r)/m$.  A  Boltzmann gas description necessitates that the particles behave classically.  This means that the de Broglie wavelength $\lambda$  is much less than the interparticle distance,\be\lambda\,  n^{1/3}(r)<<1\label{dbrlebnd}\ee
 As our analysis applies only for the domain $R_{\tiny {\tt G}}\le r\le R_{\tiny {\tt max}}$, the most stringent constraint will be  at the inner boundary of the shell, $r=R_{\tiny {\tt G}}$.  Applying this to the example of the  solution with $\tau=\frac 12$ and $\kappa=\frac13$,  which is associated with exactly flat rotation curves, one  has from (\ref{moverT}) that $n(R_{\tiny {\tt G}})= \frac {M_{\tiny {\tt G}}}{ 4\pi m R_{\tiny {\tt G}}^3}$, while $\lambda=\frac h{mu}=\sqrt{\frac23}\frac h{mv_\infty}$.  The condition (\ref{dbrlebnd}) becomes \be m^{8/3}>>\frac{ 8\pi }{3^{5/3}}\Bigl(\frac{\hbar\rho_{\tiny {\tt G}}^{1/3}}{v_{\infty}}\Bigr)^2\;,\label{consisty}\ee  
where $\rho_{\tiny {\tt G}}$ is again the mean mass density inside the  sphere of radius $R_ {\tiny {\tt G}}$.
For an order of magnitude estimate, let us take  $\rho_{\tiny {\tt G}}\sim10^{9} M_{\odot}/{\rm kpc}^3$ along with $v_{\infty}\sim 200$ km/s.   Substituting into the right hand side of (\ref{consisty}) gives $\sim ( 100\;$eV$)^{8/3}$.  Thus the classical approximation is valid provided that  particle mass is significantly greater than $\sim 100$ eV. In other  words, corrections due to quantum statistics can only be significant in the spherical shell,  $R_{\tiny {\tt G}}\le r\le R_{\tiny {\tt max}}$,  if the  dark matter particle mass is  $\sim 100$ eV or  less (smaller  bounds result from typical densities of dwarf galaxies).  However, since such  hot dark matter scenarios are currently disfavored, classical statistics is all that is needed for this model.

\section{Boltzmann fits}

In section 2 we fit the rotation velocity data for galactic haloes to an eight-parameter series and found  significant regions $R_{\tiny {\tt G}}\le r\le R_{\tiny {\tt max}}$ in the haloes   where $\rho(\phi)$ decreases exponentially.   
 Here we directly compare the  data in this region to the model  of the   self-gravitating isothermal  Boltzmann gas constructed in the previous section. More specifically, we  fit the rotational velocity  data from THINGS\cite{deBlok:2008wp} to solutions of (\ref{bndcnd}) and (\ref{1.1}). As stated in the introduction, our procedure is very different from the  common practice of fitting the data using  some particular density profile for  the dark matter, such as NFW.   The latter requires detailed knowledge of all the baryonic contributions, which are deduced from mass models.    The results of  such fits are highly dependent on the choice of mass models.  On the other hand,  precise density profiles for the baryonic contributions are not required in the analysis we do here.   The baryonic component, nevertheless, plays an important role in our approach.  It, along with the dark matter contribution to the mass $M(R_{\tiny {\tt G}})$ in the interior region $r\le R_{\tiny {\tt G}}$,  determine the second boundary condition in (\ref{bndcnd}) at $R_{\tiny {\tt G}}$, which gives the strength of the  external  gravitational field in the spherical shell $R_{\tiny {\tt G}}\le r\le R_{\tiny {\tt max}}$.  (In practice,  we do the inverse.  We determine the boundary condition from the fits, i.e., we obtain  $M(R_{\tiny {\tt G}})$  from the data.)  As the purpose of this section is to test the validity of the model, it makes sense to restrict the fit to the region  $R_{\tiny {\tt G}}\le r\le R_{\tiny {\tt max}}$ where the Boltzmann-like behavior was previously found.

The Boltzmann fits allow us to determine the  three dimensionfull parameters   of the model, namely, $m/T$, $\rho(R_{\tiny {\tt G}})$ and $M(R_{\tiny {\tt G}})$, for the given $R_{\tiny {\tt G}}$.  From them,

\noindent
a) We compare the results for $m/T$ to those found in section 2.

\noindent b)
 Using the Boltzmann form for the density (\ref{natmr}), we  numerically estimate the mass $M_{(R_{\tiny {\tt G}}, R_{\tiny {\tt max}})}$ of the region, 
 $R_{\tiny {\tt G}}\le r\le R_{\tiny {\tt max}}$.   The mass of the Boltzmann-like region  is simply
\be M_{(R_{\tiny {\tt G}}, R_{\tiny {\tt max}})}=4\pi \rho( R_{\tiny {\tt G}}) \int_{R_{\tiny {\tt G}}}^{ R_{\tiny {\tt max}}} dr\, r^2 e^{- m\phi(r)/k_BT}\label{regnmas}\ee
In order to understand what fraction of the total mass of the galaxy it represents, we compare  $ M_{(R_{\tiny {\tt G}}, R_{\tiny {\tt max}})}$  to the mass  $M(R_{\tiny {\tt G}})$ in the interior region  $r<R_{\tiny {\tt G}}$, and to the total  baryonic mass of the galaxy, as  was determined previously from a mass model.\cite{deBlok:2008wp}

\noindent c) Lastly, we obtain fitted values for  the dimensionless parameters $\kappa$ and $\tau$ and  compare the results to  the flat solution $(\tau,\kappa)=   (\frac 12,\frac 13). $

We again consider the  eight galaxies in the order: unbarred, weakly barred, barred  and  dwarf galaxies.

\subsection{NGC 2841}

In subsection 2.1 the exponential behavior  for $\rho(\phi)$ was found in the region $ 7 \;{\rm kpc}\le r\le 22\;{\rm kpc}$ of the NGC 2841 halo.  Here we perform the fit in  this region of  the rotation velocity data to solutions of (\ref{bndcnd}) and (\ref{1.1}).  It is shown in figure 29.
The   fit is   done to the midpoints [corresponding to the
dots] of the error bars.
The values  found for  the three dimensionfull parameters of the Boltzmann gas are: $m/T\approx205$ eV/Kel,   $\rho(R_{\tiny {\tt G}})\approx 4.6\times 10^7 M_{\odot}/{\rm kpc}^3 $ and $M(R_{\tiny {\tt G}})\approx 1.7\times 10^{11} M_{\odot} $.
The result for $m/T$ is close to the value of  $\sim 190$ eV/Kel  obtained in section 2.  The latter corresponds to the average of $\;\;-\frac{d\log{{\rm \rho}}}{d\phi}$ over the domain  $7<r<22$ kpc in figure 5.  One can also compare the results for  $\rho(R_{\tiny {\tt G}})$ and  $M(R_{\tiny {\tt G}}) $ with those of the eight-parameter fit in section 2.  For $\rho(7$ kpc $)$,  the latter gives  $\sim4.4\times 10^7  M_{\odot}/{\rm kpc}^3$, which is close to the result obtained from the Boltzmann fit.  More generally, in figure 30 we plot the  Boltzmann gas density for the fitted values of the parameters and compare it with the density  in figure 3, which was obtained from the eight-parameter series fit.  Agreement is quite close beyond $10$ kpc.  (The analogous comparisons for the other seven galaxies [not shown here] are reasonably similar, with one exception.)  To calculate   $M(R_{\tiny {\tt G}}) $ from the eight-parameter fit, one can use $M(R_{\tiny {\tt G}})=v(R_{\tiny {\tt G}})^2R_{\tiny {\tt G}}/G$.  The result is  $M(7$ kpc $)\sim 1.67\times 10^{11} M_{\odot} $.  This value is  also close to the results we get for the Boltzmann fit.

We next numerically estimate the mass of the Boltzmann-like region  $7<r<22$ kpc using (\ref{regnmas}).
We get $M_{(7 \,{\rm kpc}, 22 \,{\rm kpc})}\approx 2.9 \times 10^{11} M_{\odot} $.  It is significantly greater than  the mass $M(R_{\tiny {\tt G}})$ of the inner region, $r<7 \,{\rm kpc}$.
Furthermore, it is also significantly greater than the sum of the baryonic mass contributions   for   NGC 2841 which were  obtained from a  mass model. \cite{deBlok:2008wp}  Summing the disk,  bulge and HI gas masses cited in  subsection 2.1 gives
 $\sim 1.43  \times 10^{11} M_{\odot} $ [which in this example is close to the estimate for the mass  $M(R_{\tiny {\tt G}})$ in the inner region].    From these values one gets that the region exhibiting Boltzmann-like   behavior is  more than twice as massive as  the baryonic component of  the galaxy, and it therefore must consist mainly of dark matter. 

 From (\ref{kappatau}) we compute the fitted values for the dimensionless parameters $\tau$ and $\kappa$.  We get $\tau\approx .37$ and $\kappa\approx .42$.  This in contrast to the trivial isothermal solution $(\tau,\kappa)=   (\frac 12,\frac 13) $, associated  with   an exactly flat rotation curve.  Since $v(r)$ is far from being flat, it is  a nontrivial   result that the region  $7<r<22$ kpc is well described by a Boltzmann gas.

\begin{figure}[placement h]
\centering
  \includegraphics[height=2.25in,width=2.5in,angle=0]{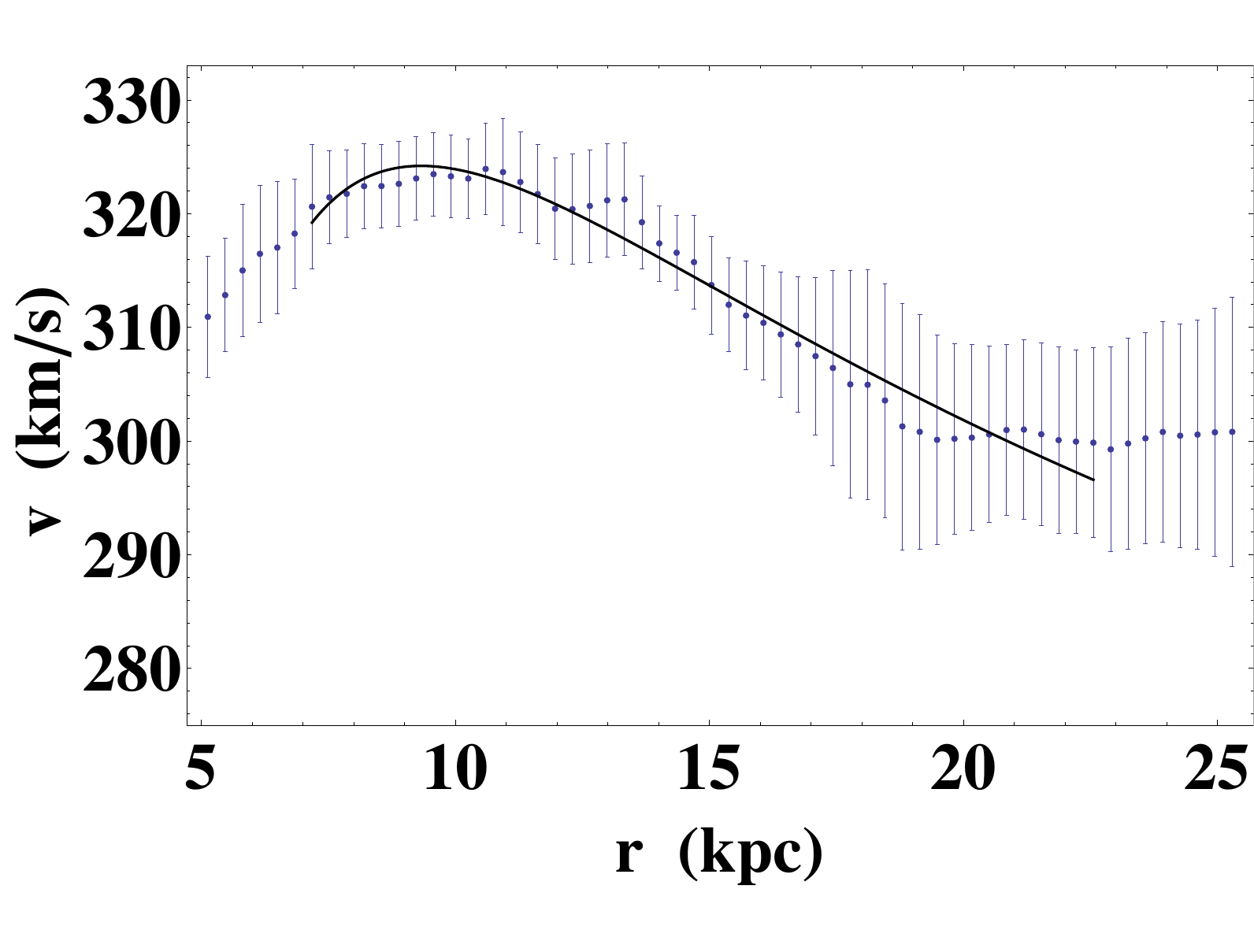}
  \caption {Fit of  the rotation velocity data  for NGC 2841 to a Boltzmann gas  in  the region $ 7 \;{\rm kpc}\le r\le 22\;{\rm kpc}$.  The fitted value for  the three dimensionfull parameters of the Boltzmann gas are: $m/T\approx205$ eV/Kel,   $\rho(R_{\tiny {\tt G}})\approx 4.6\times 10^7 M_{\odot}/{\rm kpc}^3 $ and $M(R_{\tiny {\tt G}})\approx 1.7\times 10^{11} M_{\odot} $. }
\label{fig:test}

\begin{center}
\includegraphics[height=2in,width=2.5in,angle=0]{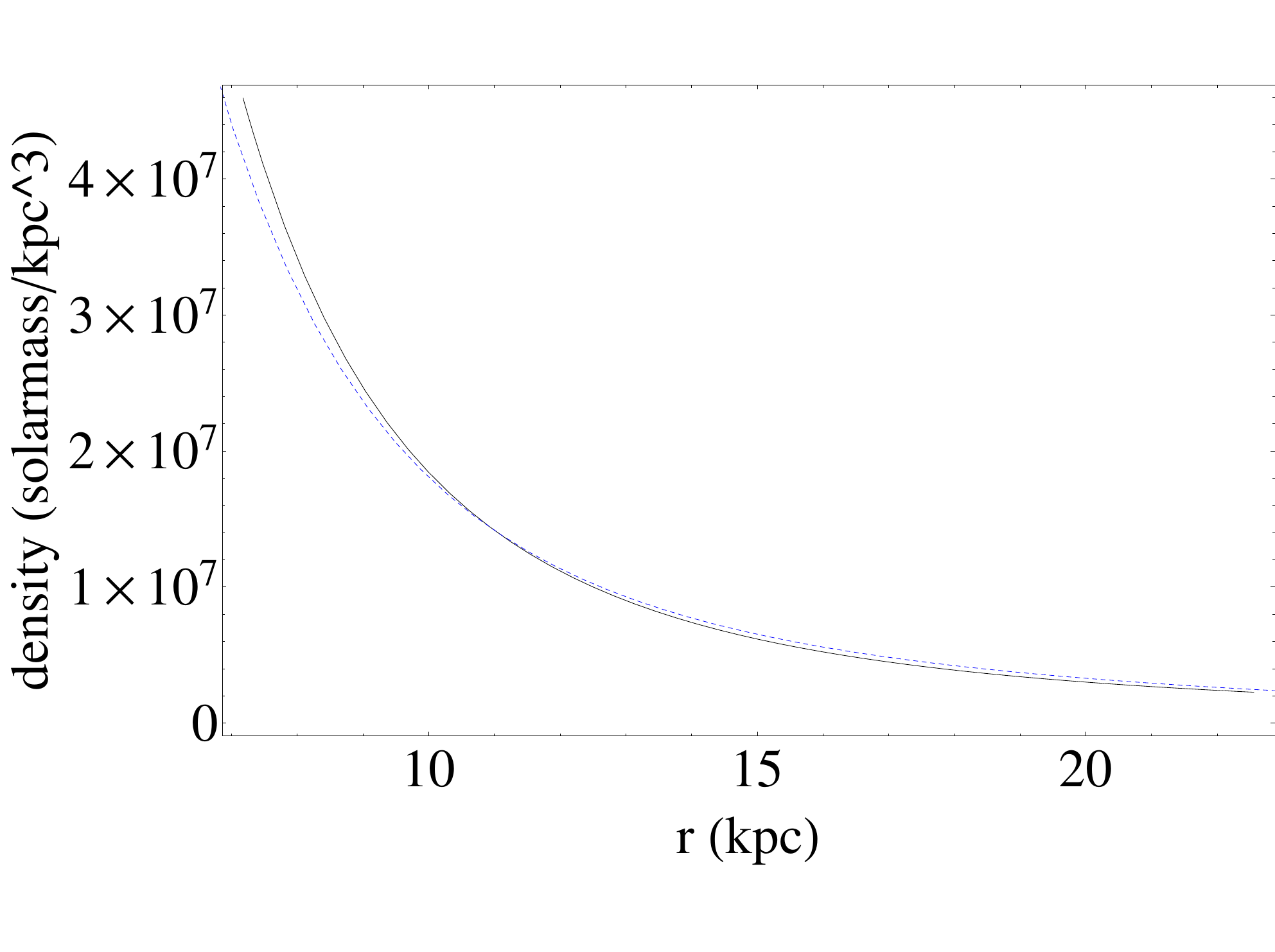}
\caption {Plot of the Boltzmann gas density for the fitted values of the parameters   (solid) compared to the density resulting from the eight-parameter series fit in section 2     which was obtained from the eight-parameter series fit restricted to  the region $ 7 \;{\rm kpc}\le r\le 22\;{\rm kpc}$ (dashed).}
\end{center}
\end{figure}

\newpage

\subsection{ NGC 5055}

In subsection 2.2 the exponential behavior for $\rho(\phi)$  was found in the region $10<r<25$ kpc of the NGC 5055 halo. A fit  in  this region of  the rotation velocity data to the solutions of (\ref{bndcnd}) and (\ref{1.1})  is shown in figure  31.  The fitted value for  the three dimensionfull parameters of the Boltzmann fit are: $m/T\approx 300$ eV/Kel,   $\rho(R_{\tiny {\tt G}})\approx 5.7\times 10^6 M_{\odot}/{\rm kpc}^3 $ and $M(R_{\tiny {\tt G}})\approx 1.05\times 10^{11} M_{\odot} $.   The fit for $m/T$ is a bit below the average value    of $\sim 440$ eV/Kel found for this region from figure 8, but is somewhat sensitive to the choice of end points.  
Using  (\ref{regnmas}), we can numerically estimate the mass of the region  $10<r<25$ kpc exhibiting Boltzmann-like behavior.
We get $M_{(10 \,{\rm kpc}, 25 \,{\rm kpc})}\approx 1.29 \times 10^{11} M_{\odot} $. This  is approximately the same as the estimated value of the mass $M(R_{\tiny {\tt G}})$ in the interior region $r<R_{\tiny {\tt G}}$, and also approximately the same as the sum of the baryonic mass contributions for   NGC 5055  quoted in subsection 2.2. The latter  is $\sim  1.34\times 10^{11} M_{\odot} $.
From (\ref{kappatau})  the fitted values for $\tau$ and $\kappa$ are $\tau\approx 0.55$ and $\kappa\approx 0.21$, as compared with the flat  solution $(\tau,\kappa)=   (\frac 12,\frac 13) $.  
 
\begin{figure}[placement h]
\centering
  \includegraphics[height=2.5in,width=2.5in,angle=0]
{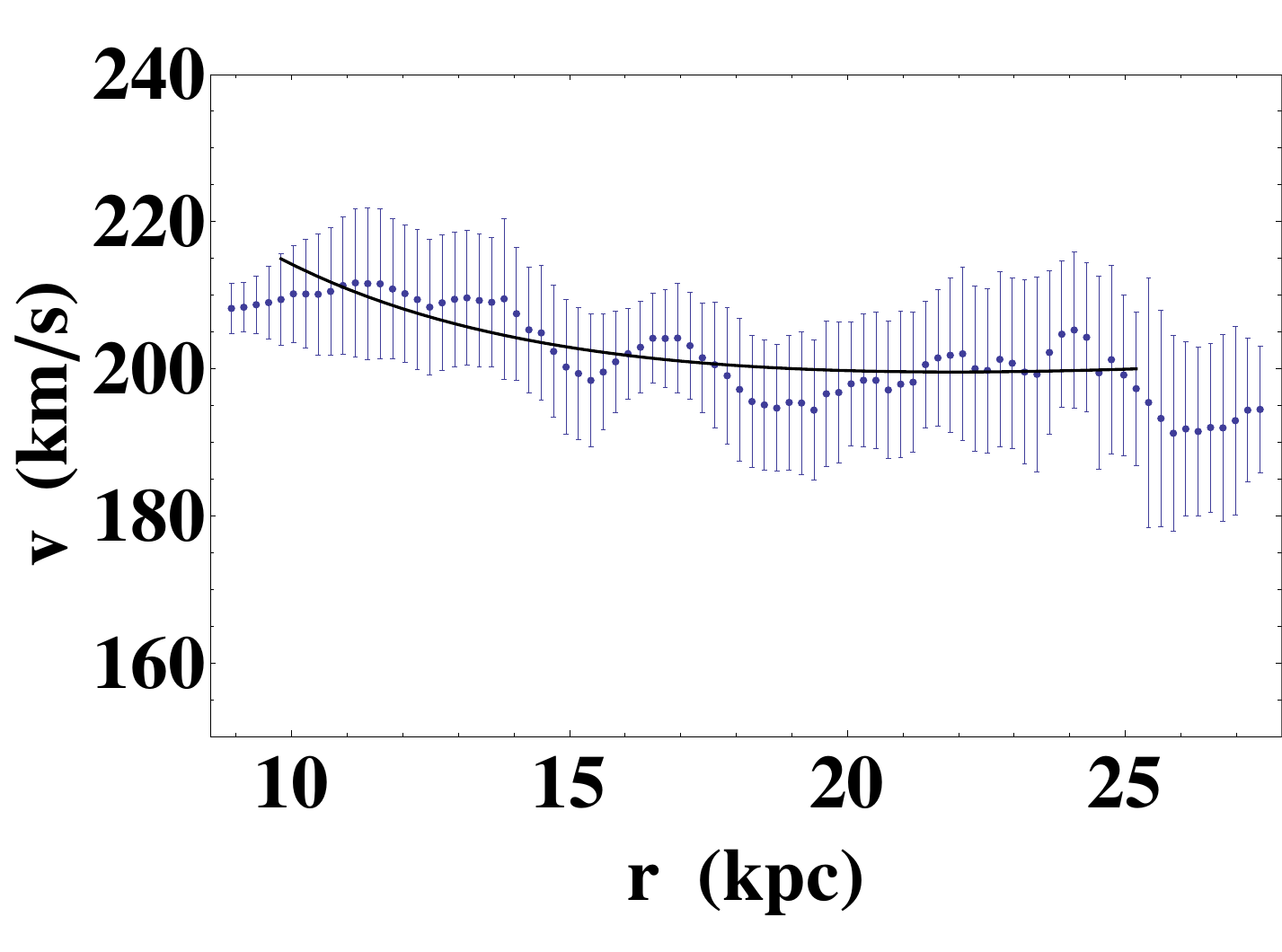}
\caption {Fit of  the rotation velocity data  for NGC 5055 to  a Boltzmann gas  in  the region  $ 10 \;{\rm kpc}\le r\le 25\;{\rm kpc}$. The fitted values for  the three dimensionfull parameters of the Boltzmann gas  are:  $m/T\approx 300$ eV/Kel,   $\rho(R_{\tiny {\tt G}})\approx 5.7\times 10^6 M_{\odot}/{\rm kpc}^3 $ and $M(R_{\tiny {\tt G}})\approx 1.05\times 10^{11} M_{\odot} $.  }
\label{fig:test}
\end{figure}

\subsection{ NGC 7331 }

In subsection  2.3 the exponential behavior for $\rho(\phi)$  was found in the region  $8<r<16$ kpc
 of the NGC  7331 halo. 
A fit  in  this region of  the rotation velocity data to the solutions of (\ref{bndcnd}) and (\ref{1.1})  is shown in figure  32.  The fitted value for  the three dimensionfull parameters of the Boltzmann fit are: $m/T\approx 225$ eV/Kel,   $\rho(R_{\tiny {\tt G}})\approx 1.3\times 1 0^7 M_{\odot}/{\rm kpc}^3 $ and $M(R_{\tiny {\tt G}})\approx  1.3 \times 10^{11} M_{\odot} $.   The fit for $m/T$ is to be compared with the average value  of $ \sim  260$ eV/Kel found for this region from figure 10. 
Using (\ref{regnmas}), we can numerically estimate the mass of the region  $8<r<16$ kpc.
We get $M_{(8 \,{\rm kpc}, 16 \,{\rm kpc})}\approx .90 \times 10^{11} M_{\odot} $, which is a little less than the mass  $M(R_{\tiny {\tt G}})$ in the interior region $r<R_{\tiny {\tt G}}$.
The sum  of the  quoted baryonic mass contributions for    NGC  7331 in subsection 2.3, which were obtained from a mass model is $\sim 1.925 \times 10^{11} M_{\odot} $. \cite{deBlok:2008wp}  From (\ref{kappatau})  the fitted values for $\tau$ and $\kappa$ are $\tau\approx 0.50$ and $\kappa\approx 0.21$.  

\begin{figure}[placement h]
\centering
  \includegraphics[height=2.25in,width=2.5in,angle=0]{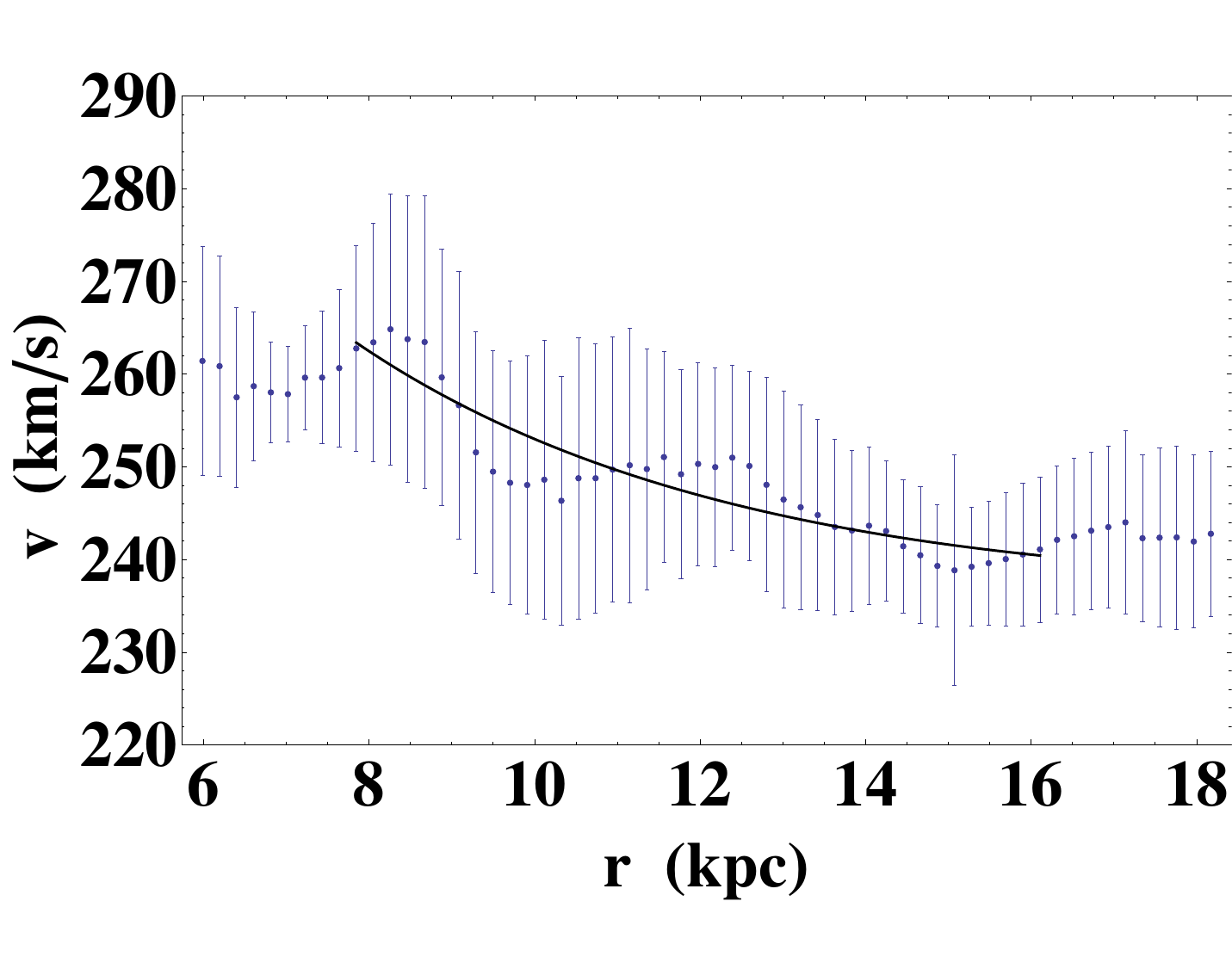}

\caption {Fit of  the rotation velocity data  for NGC 7331 to  a Boltzmann gas  in  the region  $ 8 \;{\rm kpc}\le r\le 16\;{\rm kpc}$. The fitted values for  the three dimensionfull parameters of the Boltzmann gas  are:  $m/T\approx 225$ eV/Kel,   $\rho(R_{\tiny {\tt G}})\approx  1.3  \times 10^7 M_{\odot}/{\rm kpc}^3 $ and $M(R_{\tiny {\tt G}})\approx 1.3 \times 10^{11} M_{\odot} $.  }
\label{fig:test}
\end{figure}

\subsection{ NGC 2403}

In subsection 2.4 we found exponential behavior for $\rho(\phi)$ in the region   $3<r<10$ kpc 
 of the halo  for weakly barred spiral galaxy NGC  2403 . 
A fit of  the rotation velocity data to the solutions of (\ref{bndcnd}) and (\ref{1.1})  in  the region  is shown in figure 33.     The fitted value for  the three dimensionfull parameters of the Boltzmann fit are: $m/T\approx 1140$ eV/Kel,   $\rho(R_{\tiny {\tt G}})\approx 3.5\times 1 0^7 M_{\odot}/{\rm kpc}^3 $ and $M(R_{\tiny {\tt G}})\approx  5.9 \times 10^{9} M_{\odot} $.   The fit for $m/T$ is close to the average value of $\sim 1100$ eV/Kel,  found for this region from figure 13. 
Using (\ref{regnmas}), we can numerically estimate the mass of the region  $3<r<10$ kpc.
We get $M_{(3 \,{\rm kpc}, 10 \,{\rm kpc})}\approx 3.25 \times 10^{10} M_{\odot} $, which is over five times the mass $M(R_{\tiny {\tt G}})$ in the interior region $r<R_{\tiny {\tt G}}$.
It is also much greater than the combined baryonic mass contributions from a mass model for      NGC  2403, as was cited in subsection  2.4.   The latter is $\sim  7.7 \times 10^{9} M_{\odot} $.
  From (\ref{kappatau})  the fitted values for $\tau$ and $\kappa$ are $\tau\approx 0.81$ and $\kappa\approx 0.67 $.  

\begin{figure}[placement h]
\centering

   \includegraphics[height=2.25in,width=2.5in,angle=0]{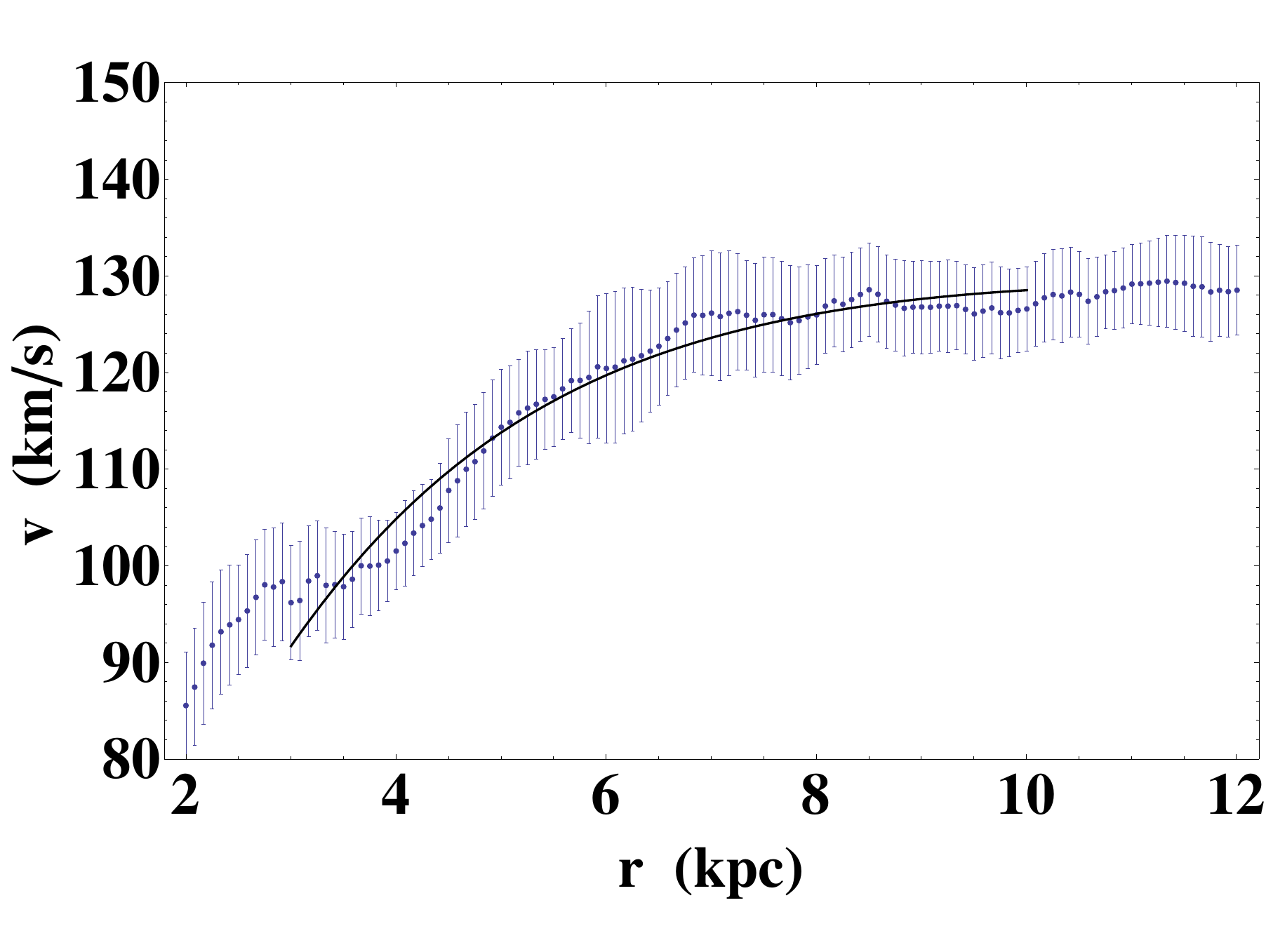}

\caption {Fit of  the rotation velocity data  for NGC 2403 to a Boltzmann gas  in  the region $3 \;{\rm kpc}\le r\le 10\;{\rm kpc}$. The fitted values for  the three dimensionfull parameters of the Boltzmann gas  are: $m/T\approx 1140$ eV/Kel,   $\rho(R_{\tiny {\tt G}})\approx  3.5 \times 10^7 M_{\odot}/{\rm kpc}^3 $ and $M(R_{\tiny {\tt G}})\approx 5.9 \times 10^{9} M_{\odot} $. }
\label{fig:test}
\end{figure}
\newpage
\subsection{ NGC 2903}

A fit of  the rotation velocity data to the solutions of (\ref{bndcnd}) and (\ref{1.1})  in  the region $3<r<9$ kpc  is shown in figure 34.    The fitted value for  the three dimensionfull parameters of the Boltzmann fit  are: $m/T\approx 580$ eV/Kel,   $\rho(R_{\tiny {\tt G}})\approx 1.6  \times 10^8 M_{\odot}/{\rm kpc}^3 $ and $M(R_{\tiny {\tt G}})\approx 2.4 \times 10^{10} M_{\odot} $.   The result for $m/T$ is compared with average  value  of $\sim  490$ eV/Kel found in figure 16.
Using (\ref{regnmas}), we can numerically estimate the mass of the region  $3<r<9$ kpc.
We get $M_{(3 \,{\rm kpc}, 9 \,{\rm kpc})}\approx 6.3\times 10^{10} M_{\odot} $.  This is significantly greater than the mass $M(R_{\tiny {\tt G}})$ in the interior region $r<R_{\tiny {\tt G}}$, and also the combined baryonic mass contributions from a mass model for       NGC  2903.  Using the values quoted in subsection 2.5, the latter is  $\sim 2.06 \times 10^{10} M_{\odot} $.   
  From (\ref{kappatau})  the fitted values for $\tau$ and $\kappa$ are $\tau\approx 0.41$ and $\kappa\approx  0.90 $.

\begin{figure}[placement h]
\centering
  \includegraphics[height=2.25in,width=2.5in,angle=0]{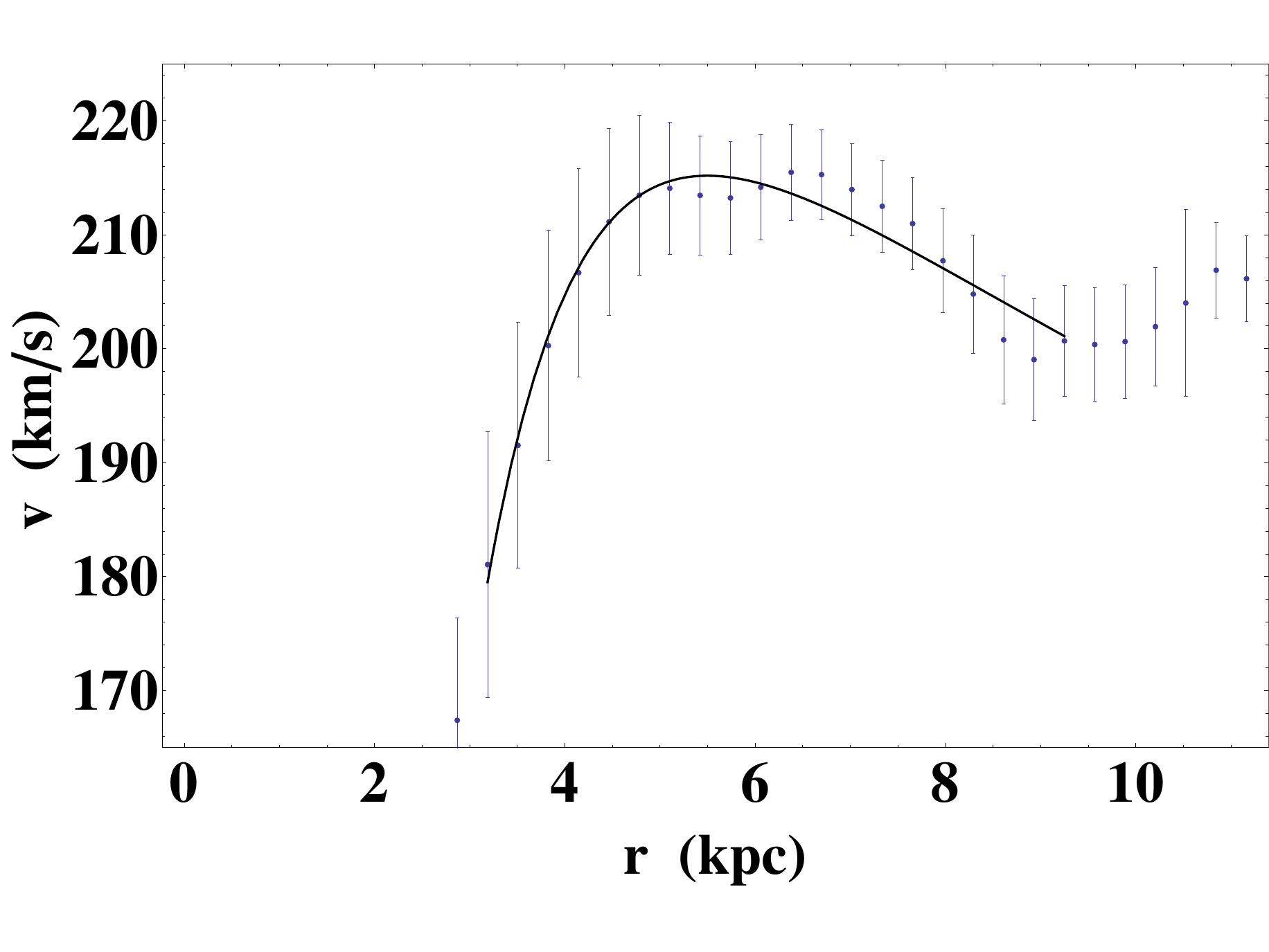}
  
\caption {Fit of  the rotation velocity data  for NGC 2903 to  a Boltzmann gas  in  the region $3 \;{\rm kpc}\le r\le 9\;{\rm kpc}$. The fitted values for  the three dimensionfull parameters of the Boltzmann gas  are: $m/T\approx 580$ eV/Kel,   $\rho(R_{\tiny {\tt G}})\approx 1.6  \times 10^8 M_{\odot}/{\rm kpc}^3 $ and $M(R_{\tiny {\tt G}})\approx 2.4 \times 10^{10} M_{\odot} $. }
\label{fig:test}
\end{figure}

\subsection{ NGC 3521}

A fit of  the rotation velocity data to the solutions of (\ref{bndcnd}) and (\ref{1.1})  in  the region $7<r<15$ kpc  is shown in figure 35.   The fitted value for  the three dimensionfull parameters of the Boltzmann fit are:  $m/T\approx 370$ eV/Kel,   $\rho(R_{\tiny {\tt G}})\approx 2.1 \times 10^7 M_{\odot}/{\rm kpc}^3 $ and $M(R_{\tiny {\tt G}})\approx  8.9\times 10^{10} M_{\odot} $. The fit for $m/T$ is close to the average value of $ 380$ eV/Kel,  found for this region in figure 19.   Using (\ref{regnmas}), we can numerically estimate the mass of the region  $7<r<15$ kpc.
We get $M_{(7 \,{\rm kpc},15 \,{\rm kpc})}\approx 8.8\times 10^{10} M_{\odot} $, which is approximately the same as the mass $M(R_{\tiny {\tt G}})$ in the interior region $r<R_{\tiny {\tt G}}$.  This is compared with the sum of the quoted baryonic mass contributions in subsection 2.6 for  
 NGC  3521 of  $\sim 1.31\times 10^{11} M_{\odot} $.   From (\ref{kappatau})  the fitted values for $\tau$ and $\kappa$ are $\tau\approx 0.39$ and $\kappa\approx  0.36 $.  

\begin{figure}[placement h]
\centering
  \includegraphics[height=2.25in,width=2.5in,angle=0]{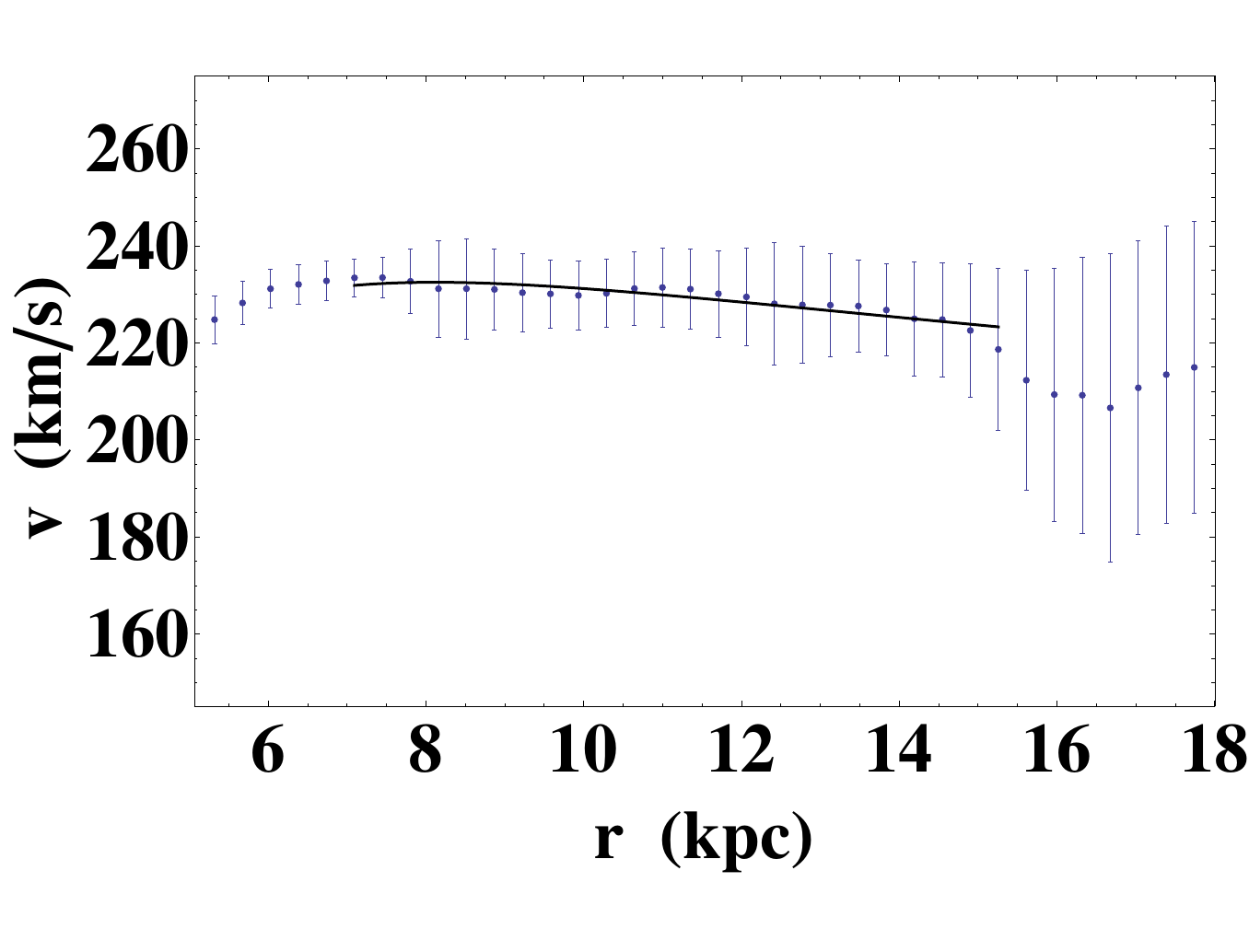}

\caption {Fit of  the rotation velocity data  for NGC 3521 to  a Boltzmann gas  in  the region $ 7 \;{\rm kpc}\le r\le 15\;{\rm kpc}$. The fitted values for  the three dimensionfull parameters of the Boltzmann gas  are: $m/T\approx 370$ eV/Kel,   $\rho(R_{\tiny {\tt G}})\approx 2.1 \times 10^7 M_{\odot}/{\rm kpc}^3 $ and $M(R_{\tiny {\tt G}})\approx  8.9\times 10^{10} M_{\odot} $.}
\label{fig:test}
\end{figure}

\newpage
\subsection{ NGC  3198}

For the case of the barred spiral  galaxy NGC  3198, a fit of  the rotation velocity data to the solutions of (\ref{bndcnd}) and (\ref{1.1})  in  the region $5<r<20$ kpc  is shown in figure 36.  The fitted value for  the three dimensionfull parameters of the Boltzmann fit  are:  $m/T\approx 800$ eV/Kel,   $\rho(R_{\tiny {\tt G}})\approx 2.2 \times 10^7 M_{\odot}/{\rm kpc}^3 $ and $M(R_{\tiny {\tt G}})\approx  1.9\times 10^{10} M_{\odot} $. 
The fit for $m/T$ is within the range of values for this region  found from figure 23. 
 Using (\ref{regnmas}), we can numerically estimate the mass of the region  $5<r<20$ kpc.
We get $M_{(5 \,{\rm kpc},20 \,{\rm kpc})}\approx 8.9\times 10^{10} M_{\odot} $,  which is much greater than the mass $M(R_{\tiny {\tt G}})$ in the interior region $r<R_{\tiny {\tt G}}$. 
It is also greater than  the sum of the stated baryonic masses  for   NGC  3198 in subsection 2.7.  The latter is  $4.12 \times 10^{10} M_{\odot} $.    From (\ref{kappatau})  the fitted values for $\tau$ and $\kappa$ are $\tau\approx 0.58$ and $\kappa\approx  0.59 $.

\begin{figure}[placement h]
\centering

  \includegraphics[height=2.25in,width=2.5in,angle=0]{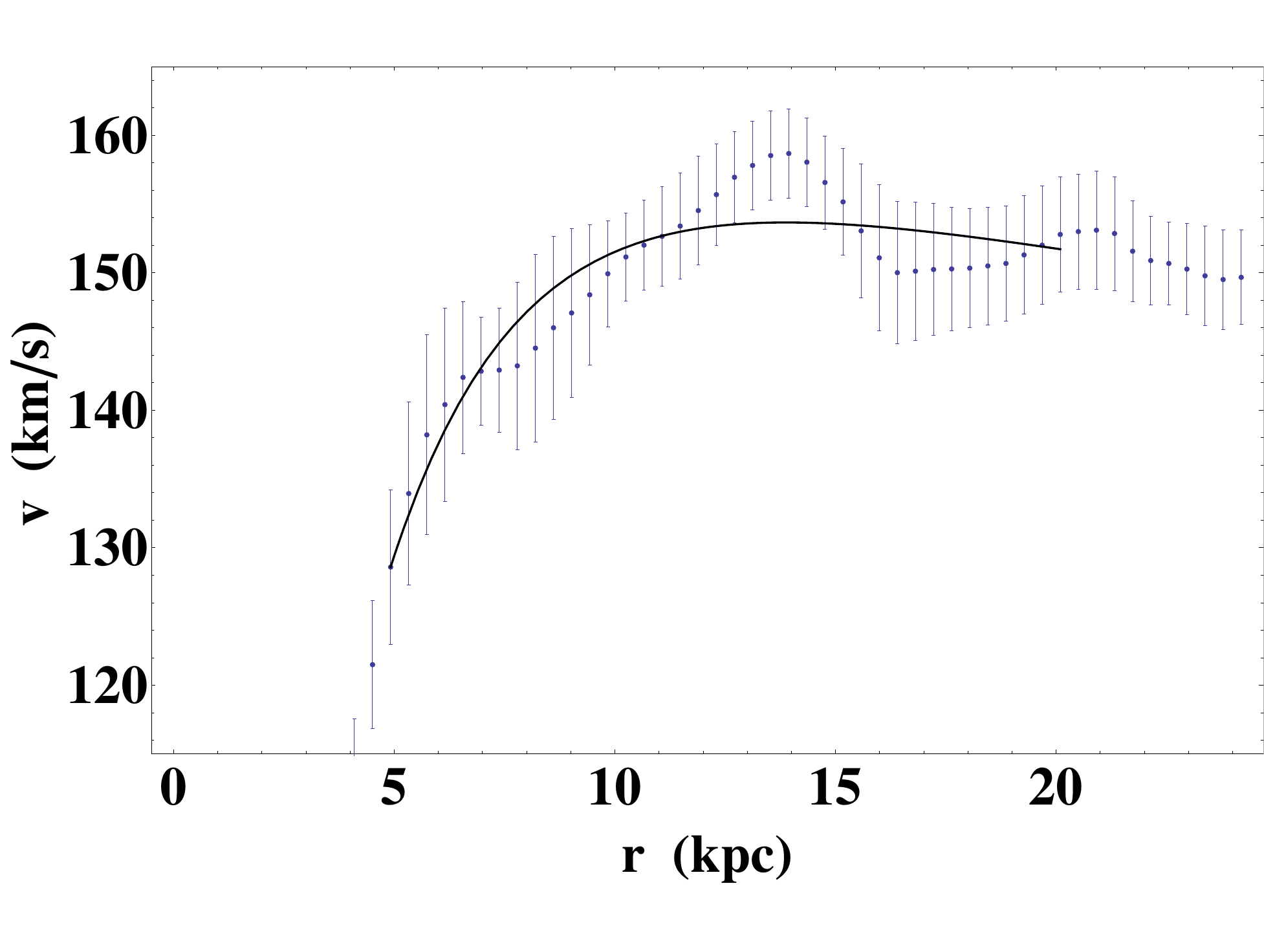}
 
\caption {Fit of  the rotation velocity data  for NGC 3198 to a Boltzmann gas  in  the region $ 5 \;{\rm kpc}\le r\le 20\;{\rm kpc}$. The fitted values for  the three dimensionfull parameters of the Boltzmann gas  are:  $m/T\approx 800$ eV/Kel,   $\rho(R_{\tiny {\tt G}})\approx 2.2 \times 10^7 M_{\odot}/{\rm kpc}^3 $ and $M(R_{\tiny {\tt G}})\approx  1.9\times 10^{10} M_{\odot} $.}
\label{fig:test}
\end{figure}

\subsection{ DD0 154}

Lastly, for the case of the dwarf   galaxy  DD0 154, a fit of  the rotation velocity data to the solutions of (\ref{bndcnd}) and (\ref{1.1})  in  the region $1.5<r<6.5$ kpc  is shown in figure 37.    The fitted value for  the three dimensionfull parameters of the Boltzmann fit  are: 
 $m/T\approx 7500$ eV/Kel,   $\rho(R_{\tiny {\tt G}})\approx 1.2 \times 10^7 M_{\odot}/{\rm kpc}^3 $ and $M(R_{\tiny {\tt G}})\approx  3.4\times 10^{8} M_{\odot} $. 
The fit for $m/T$ is within the range found for this region from figure 26.   Using (\ref{regnmas}), we can numerically estimate the mass of the region  $1.5<r<6.5$ kpc.
We get $M_{(1.5 \,{\rm kpc},6.5 \,{\rm kpc})}\approx 3.2\times 10^{9} M_{\odot} $, which is an order of magnitude bigger than the mass $M(R_{\tiny {\tt G}})$ in the interior region $r<R_{\tiny {\tt G}}$.  It  is also  approximately an order of magnitude bigger than   the sum of the stated baryonic masses   for  DD0 154 in  subsection 2.8  ($\sim    3.84\times 10^{8} M_{\odot} $).  This is not surprising since the dark matter component is dominant in dwarf galaxies.  From (\ref{kappatau})  the fitted values for $\tau$ and $\kappa$ are $\tau\approx 1.13$ and $\kappa\approx 0.62 $.  

\begin{figure}[placement h]
\centering

  \includegraphics[height=2.25in,width=2.5in,angle=0]{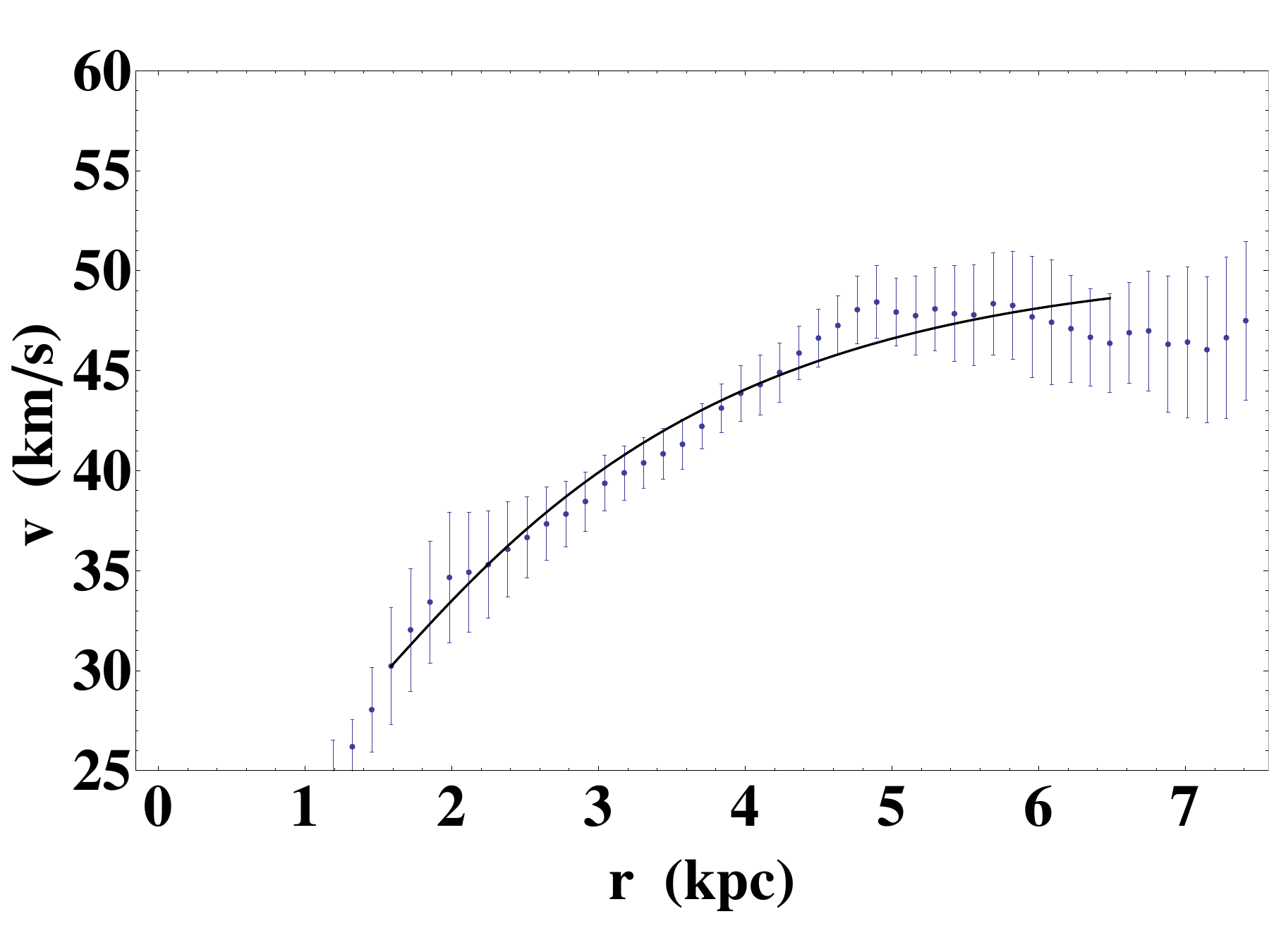}
 
\caption {Fit of  the rotation velocity data  for DD0 154 to  a Boltzmann gas  in  the region  $1. 5 \;{\rm kpc}\le r\le 6.5\;{\rm kpc}$. The fitted values for  the three dimensionfull parameters of the Boltzmann gas  are: $m/T\approx 7500$ eV/Kel,   $\rho(R_{\tiny {\tt G}})\approx 1.2 \times 10^7 M_{\odot}/{\rm kpc}^3 $ and $M(R_{\tiny {\tt G}})\approx  3.4\times 10^{8} M_{\odot} $ .}
\label{fig:test}
\end{figure}

\section{Summary and Concluding Remarks} 

In section two we examined  the rotation curve data of eight galaxies  from THINGS\cite{deBlok:2008wp}, and searched for universal behavior for the function  $\rho(\phi)$.  We found indications that it has an exponential behavior in  regions  $R_{\tiny {\tt G}}\le r\le R_{\tiny {\tt max}}$ of the haloes.  We modeled the regions with an effective theory in section three.    The dynamics in the spherical shell  $R_{\tiny {\tt G}}\le r\le R_{\tiny {\tt max}}$ was described  by a self-gravitating isothermal gas, subject to an external gravitational field due to   the inner-most region $r\le R_{\tiny {\tt G}}$.    General solutions are specified by two dimensionless parameters, which were denoted by $\tau$ and $\kappa$, while three dimensionfull parameters  are required for a fit to the data.    From typical rotation velocities, we checked that  the (dark matter) particles responsible for flat rotation curves move nonrelativistically, and that quantum statistics is negligible   in  typical regions where the model has application. In section four, the rotation curve data was fit to solutions to the model in the regions   $R_{\tiny {\tt G}}\le r\le R_{\tiny {\tt max}}$ where Boltzmann-like behavior was found.  The fits rely on the gravitational attraction to the baryonic matter in the inner region  $r\le R_{\tiny {\tt G}}$, but do not require any detailed knowledge of the various density profiles; i.e., they do not require the use of mass models.
 
Our  results for the fits and searches for Boltzmann-like behavior of the eight galaxies are summarized in tables 1 and 2 below.  
Table 1 contains two different estimates for $m/T$ for each spherical shell region (columns four and five), along with distance scales for the various galaxies.
The second column of table 1 lists the distances to the galaxies used in this article, while the third column contains the distance scale of the disks.\cite{Mannheim:2010xw}
In column four we give the result for the mean values  of $\;-\frac{d\log{{\rm \rho}}}{d\phi}$ in the  regions (indicated below in parenthesis) where  $\rho(\phi)$ exhibited exponential behavior. (Recall,  $\;-\frac{d\log{{\rm \rho}}}{d\phi}$ is a constant, namely  $m/T$, for the Boltzmann gas.)  The results were obtained using the eight-parameter series fits of the rotation  velocity data described in section 2.    The table  shows  that the regions exhibiting  Boltzmann-like behavior are generally much larger than the disk scale.  
 For the examples of  NGC 5055, as well as for NGC 2841, these regions are $\sim 15 $ kpc, corresponding to roughly four times the distance scale of  the disk.  Also, the inner boundary $r=R_{\tiny {\tt G}}$ of the spherical shell is  greater  than the disk scale. Baryonic matter is expected to be subdominant at such distances.  {\it Thus,  dark matter should give the  dominant contribution to the total mass density  in the spherical shell  $R_{\tiny {\tt G}}\le r\le R_{\tiny {\tt max}}$, and it should therefore be responsible for the apparent Boltzmann-like behavior.}

 As stated above, column four is an estimate of  $m/T$ in the region $R_{\tiny {\tt G}}\le r\le R_{\tiny {\tt max}}$, which was obtained utilizing the eight-parameter fit.  Column five   of table 1 gives another determination for $m/T$ in the same region.  These values were obtained in section 4 by making a fit of the rotation curve data to solutions  of the  equation for a self-gravitating Boltzmann gas.  The two different determinations of $m/T$  are
generally in agreement with each other for the eight galaxies.  We note that after  assuming that the (dark matter) mass $m$ is a universal constant, we can compare the relative effective temperatures of (dark matter in) the spherical shells of the eight dark matter haloes.  We get
$$ T_{{\rm  NGC}\, 2841}> T_{{\rm  NGC}\, 7331}>T_{{\rm  NGC}\, 3521}\sim T_{ {\rm NGC }\,5055} > T_{ {\rm NGC}\, 2903}  > T_{ {\rm NGC}\, 3198}  > T_{ {\rm NGC}\, 2403}  > T_{ {\rm DD0}\, 154} $$

\begin{table}[h]
\centering

\begin{tabular}{|l|l|l|l|l|}

\hline
Galaxy& distance&  $R_{\tiny {\tt disk}}$  &   $\;-\Big<\frac{d\log{{\rm \rho}}}{d\phi}\Big>$   & $m/T$ (Boltzmann\\
&[Mpc]
&[ kpc]& [ eV/Kel]& fit)$\;\;$ [ eV/Kel]\\
\hline
NGC 2841&14.1 &$3.5$   &$ 190$  & $205$ \\
& & &($7\le r\le22$ kpc) &($7\le r\le22$ kpc) \\
\hline
NGC 5055&9.2 &$3.622$ & $440$ & $ 300$\\
 &&&($10\le r\le 25$ kpc) & ($10\le r\le 25$ kpc) \\
\hline
NGC  7331&14.2 &$3.2$ & $260$ & $ 225$\\
& && ($8\le r\le 16$ kpc) & ($8\le r\le 16$ kpc)  \\
\hline
NGC 2403&4.3 &$2.7$ & $ 1100$& $1140$\\
&& &($3\le r\le 10$ kpc)&($3\le r\le 10$ kpc)\\

\hline
NGC 2903 &9.4&$3.0$& $ 490$& $ 580$\\
&& &($3\le r\le 9$ kpc)&($3\le r\le 9$ kpc)\\

\hline
NGC 3521&12.2 &$3.3$ &$380$  &  $370$ \\
&& & ($7\le r\le 15$ kpc)& ($7\le r\le 15$ kpc)\\

\hline
NGC 3198 &14.1&$4.0$ &$820$  &  $800$ \\
&& & ($5\le r\le 25$ kpc)& ($5\le r\le 20$ kpc) \\

\hline
DD0 154&4.2&$ .8$ &$ 8600$  &  $7500$ \\
&& & ($1.5\le r\le 6.5$ kpc)& ($1.5\le r\le 6.5$ kpc) \\

\hline

\end{tabular}
\caption{Summary of estimates for $m/T$ in the haloes of eight galaxies. The second column lists the distances to the galaxies used in this article.  The third column quotes the distance scale of the disks from \cite{Mannheim:2010xw}. In column 4 we give the mean values  of $\;-\frac{d\log{{\rm \rho}}}{d\phi}$ in regions (indicated below in parenthesis) where the derivatives are approximately constant.  The values for $m/T$ obtained from the fits of  to the Boltzmann gas (for the region  in the parenthesis below) is  given in column 5. }

\label{tab:template}

\end{table}

Table 2 reports on the mass estimates for the spherical shell $R_{\tiny {\tt G}}\le r\le R_{\tiny {\tt max}}$ and the inner region $r\le R_{\tiny {\tt G}}$.
Table 2 also gives results for the dimensionless parameters ($\tau,\kappa)$ for  the Boltzmann fits.  Column two lists the sum $M_{\tiny {\tt baryons}}$ of all the  baryonic contributions to the mass of the galaxy reported in the literature.  The disk and bulge contributions were obtained from a mass model in \cite{deBlok:2008wp}  (specifically, the model with fixed values of the mass to luminosity ratio and the diet Salpeter stellar initial mass function), while the HI gas mass estimates come from \cite{Walter:2008wy}.   Column three gives the result for the mass $M(R_{\tiny {\tt G}})$ in the interior region $r\le R_{\tiny {\tt G}}$ obtained from the Boltzmann fits, while column four
lists   estimates of the total  matter content in the spherical shell  $R_{\tiny {\tt G}}\le r\le R_{\tiny {\tt max}}$ using (\ref{regnmas}).
 The table   shows that for the majority of the galaxies, the regions exhibiting Boltzmann-like behavior are at least as massive, if not more, than the reported values for the sum of all the baryonic  components of the galaxies. This is another indication that the region is strongly dominated by dark matter.  Also, the spherical shells are, in general, more massive than the interior regions  $r\le R_{\tiny {\tt G}}$.  Two exceptional cases are NGC 7331 and NGC 3521, where the mass for the spherical shell ${ M_{(R_{\tiny {\tt G}}, R_{\tiny {\tt max}})}}$ is less than $M_{\tiny {\tt baryons}}$.  On the other hand, for the case of the dwarf galaxy DD0 154, the region associated with the exponential behavior for $\rho(\phi)$  is  almost ten times as massive as the baryonic  component of the galaxy. 

 Finally, column five of  table 2 gives the fitted values for the dimensionless parameters $\tau$ and $\kappa$ defining the Boltzmann gas. The results  vary from one galaxy to another,  and generally differ from the special case of $(\tau,\kappa)=(\frac 12,\frac 13) $, which  corresponds to an exactly flat rotation curve. As was discussed in the introduction, isothermal behavior is obvious  for the ideal  case of  exactly flat rotation curves.  On the other hand,  since from column five,  the results are, in  general, not in agreement with the exactly flat solution, the apparent isothermal behavior could not have been guessed from the outset. 
\begin{table}[h]
\centering

\begin{tabular}{|l|l|l|l|l|l|l|}

\hline
Galaxy&$M_{\tiny {\tt baryons}}$&$M(R_{\tiny {\tt G}})$&
${ M_{(R_{\tiny {\tt G}}, R_{\tiny {\tt max}})}}$  & ($\tau,\kappa)$\\
&$[10^{10} M_{\odot} ] $&$[10^{10} M_{\odot} ] $ &$[10^{10} M_{\odot} ] $   &\\
\hline
NGC 2841   &$14.3$&$17$  &$29$&$  ( .37,.42) $ \\
&&($r\le 7$ kpc)&($7\le r\le22$ kpc)&  \\
\hline
NGC 5055&$13.4$&$ 10.5$&$ 12.9$ &$(.55,.21)$\\ & & ($r\le 10$ kpc) 
& ($10\le r\le 25$ kpc) &  \\
\hline
NGC  7331&$ 19.25$&$13$&$9$ &$(.50,.21)$\\&& ($r\le 8$ kpc)
& ($8\le r\le 16$ kpc) & \\
\hline
NGC 2403&$.77$&$.59$& $3.25$ & $(.81,.67 )$\\&&($r\le 3$ kpc)
&($3\le r\le 10$ kpc)& \\

\hline
NGC 2903&$ 2.06$ &$ 2.4 $&$ 6.3$ & $(.41,.90 )$\\&&($r\le 3$ kpc)
&($3\le r\le 9$ kpc)& \\

\hline
NGC 3521&$ 13.1$&$ 8.9$& $ 8.8$ & $( .39,.36 )$ \\&& ($r\le 7$ kpc)
& ($7\le r\le 15$ kpc)& \\

\hline
NGC 3198 & $4.12$& $ 1.9$& $8.9$ & $(.58,.59   )$ \\&& ($r\le 5$ kpc)
& ($5\le r\le 20$ kpc)&  \\

\hline
DD0 154&$.0384$&$.034$&  $.32$ & $( 1.13, .62   )$ \\&& ($r\le 1.5$ kpc)
& ($1.5\le r\le 6.5$ kpc)& \\

\hline

\end{tabular}
\caption{Summary of mass estimates and ($\tau,\kappa)$ for eight galaxies.  Column 2 lists  the total mass baryonic mass of the galaxy, as determined from a mass model. \cite{deBlok:2008wp}  In column 3, we give the value of the mass $M(R_{\tiny {\tt G}})$ in the interior region ($r\le R_{\tiny {\tt G}}$), obtained from the fit to the Boltzmann gas.    The mass $ M_{(R_{\tiny {\tt G}}, R_{\tiny {\tt max}})}$ of the Boltzmann-like region,  estimated using  (\ref{regnmas}), appears in column  4.  Column 5 lists the fitted values of the dimensionless parameters $\tau$ and $\kappa$.}

\label{tab:template}

\end{table}

 We mention a few words about some of the  galaxies in  THINGS that were not analyzed here.  Two galaxies, NGC 2976 and  NGC 3627, were not considered at all because they contained too few data points for an eight-parameter fit.  NGC 3031 and NGC 4736 may not be ideal candidates because they are reported as having non-circular motions. Also,  at sufficient distances from the origin,  the former is affected by tidal interactions from other members of the group (the M81 group, for the case of NGC 3031).\cite{deBlok:2008wp}   There are other galaxies in the survey with features that may not make them ideal candidates for an analysis of this sort as well, such as  NGC 3627, which has  a pronounced bar and  an asymmetric spiral structure.  The data for the remaining five galaxies in the survey did not exhibit any signs of universal behavior after doing the eight-parameter fit used in section 2.  More specifically, we did not find convincing evidence of  Boltzmann-like regions from the data for those  galaxies.

In summary, from the rotation velocity data  for 19 galaxies from THINGS, we found evidence of Boltzmann-like behavior  in the haloes of eight  galaxies,  corresponding to approximately $40\%$ of the sample. More specifically, after making the usual assumptions of spherical symmetry and that the rotation curves are generated by circular orbits, we showed that there were substantial regions of the haloes where the density decreased exponentially with the gravitational potential.   For  NGC 2841 and  NGC 5055, agreement with Boltzmann-like behavior was found  over a distance which was roughly four times the distance scale of  the disk, and the relevant regions were estimated to be  much more massive than the  total baryonic masses of the galaxies.
 For the case of the dwarf galaxy  DD0 154, the region exhibiting  exponential behavior for $\rho(\phi)$ was approximately six times larger than the disk scale  and eight times  more  massive  than the baryonic component, indicating  that dark matter  dominates the region.
In all cases, the  exponential behavior  breaks down at short and  large distances.    The breakdown at extremely large distances
 may  indicate that the dynamics described by   (\ref{1.1}) is not valid when densities  go below a certain threshold.
  It
is not hard to understand the breakdown at short distances, since the assumptions of   symmetry and a single-component gas are not valid approximations close to the disk.

A number of improvements can be performed to the model presented in section 3 and its fits of galactic haloes in sections 4.  The most obvious are to drop the assumption of spherical symmetry, and to take into account effects of the HI gas present in the haloes.  Furthermore, one can try to  extend the analysis  to include the interior region.  As densities  grow in the interior, it may  then be of interest  to consider the effect of quantum statistics.

Finally, the fact that we have evidence of some  Boltzmann-like behavior in substantial regions of eight galactic haloes may not be accidental, as it may  signal the presence of (non gravitational) dark matter self-interactions.  Moreover, it is an indication that such  interactions could be sufficient in strength and number for the system to settle down to an equilibrium  configuration. 

\bigskip
{\Large {\bf Acknowledgments} }

\noindent
We are very grateful to  P. Biermann,  R. Buta, N. Okada and J. Reichenbacher for valuable discussions.   We are  also grateful to Erwin de Blok for providing us with rotation curve data from THINGS. 
A.S. was supported in part by the DOE,
Grant No. DE-FG02-10ER41714.

\end{document}